\documentclass[10pt,journal,compsoc]{IEEEtran}

\ifCLASSOPTIONcompsoc
  \usepackage[nocompress]{cite}
\else
  \usepackage{cite}
\fi
\ifCLASSINFOpdf
\else
\fi

\usepackage{epsfig}
\usepackage{graphicx}
\usepackage{amsmath}
\usepackage{amssymb}
\usepackage{enumitem}

\usepackage[pagebackref=true,breaklinks=true,letterpaper=true,colorlinks,bookmarks=false,urlcolor=red,citecolor=cyan]{hyperref}
\usepackage{gensymb,graphics,subfigure,inputenc}
\usepackage[linesnumbered,algo2e,boxed]{algorithm2e}
\graphicspath{{Figure/}}
\usepackage[figuresright]{rotating}
\usepackage{amsmath,amssymb,textcomp,subfigure,multirow,upgreek}
\usepackage{booktabs}
\usepackage{color,mdwlist}

\usepackage{ragged2e}

\usepackage{review}
\usepackage{libertine} 

\usepackage{graphicx,fontawesome}
\graphicspath{{Figures/}}

\usepackage{caption}
\usepackage{wrapfig}

\usepackage{enumitem}
\setitemize{noitemsep,topsep=0pt,parsep=0pt,partopsep=0pt}

\usepackage{times}
\usepackage{soul}
\usepackage{url}
\usepackage{caption}

\usepackage{graphicx}
\usepackage{amsmath}
\usepackage{amsthm}
\usepackage{booktabs}
\usepackage{algorithm}
\usepackage{algorithmic}
\usepackage[switch]{lineno}
\usepackage{booktabs}
\usepackage{fontawesome}
\usepackage{tikz}
\usepackage{inconsolata}
\usepackage{pifont}
\usepackage[edges]{forest}
\usepackage{makecell}
\usepackage{cleveref}
\usepackage{setspace}
\usepackage[numbers,sort&compress]{natbib}
\definecolor{hidden-draw}{RGB}{205, 44, 36}
\definecolor{hidden-blue}{RGB}{194,232,247}
\definecolor{hidden-orange}{RGB}{243,202,120}
\definecolor{hidden-yellow}{RGB}{255,229,204}
\definecolor{hidden-red}{RGB}{255,204,204}
\definecolor{hidden-draw}{RGB}{20,68,106}
\definecolor{hidden-pink}{RGB}{255,245,247}

\begin{document}
	\pdfoutput=1
%
\title{Dynamic Scene Reconstruction: Recent Advance in Real-time Rendering and Streaming}
\author{Jiaxuan Zhu \quad Hao Tang$^*$
	\IEEEcompsocitemizethanks{
	    \IEEEcompsocthanksitem Jiaxuan Zhu is with the School of Computer Science and Engineering, Southeast University, Nanjing 211189, China. E-mail: zjxx@seu.edu.cn \protect
	    \IEEEcompsocthanksitem Hao Tang is with the School of Computer Science, Peking University, Beijing 100871, China. E-mail: haotang@pku.edu.cn \protect
        }
	\thanks{$^*$Corresponding author: Hao Tang.}
}

%
%

\markboth{IEEE Transactions on Pattern Analysis and Machine Intelligence}%
{Shell \MakeLowercase{\textit{et al.}}: Bare Demo of IEEEtran.cls for Computer Society Journals}
%



\IEEEtitleabstractindextext{%
\justify
\begin{abstract}

Representing and rendering dynamic scenes from 2D images is a fundamental yet challenging problem in computer vision and graphics. This survey provides a comprehensive review of the evolution and advancements in dynamic scene representation and rendering, with a particular emphasis on recent progress in Neural Radiance Fields based and 3D Gaussian Splatting based reconstruction methods. We systematically summarize existing approaches, categorize them according to their core principles, compile relevant datasets, compare the performance of various methods on these benchmarks, and explore the challenges and future research directions in this rapidly evolving field. In total, we review over 170 relevant papers, offering a broad perspective on the state of the art in this domain.
\end{abstract}

}

\maketitle

\IEEEdisplaynontitleabstractindextext

%
\IEEEpeerreviewmaketitle


%
%
%
%


\section{Introduction}
\tikzstyle{my-box}=[
    rectangle,
    draw=hidden-draw,
    rounded corners,
    text opacity=1,
    minimum height=1.5em,
    minimum width=5em,
    inner sep=2pt,
    align=left,
    fill opacity=.5,
    line width=0.8pt,
]
\tikzstyle{leaf}=[my-box, minimum height=1.5em,
    text=black, align=left,font=\normalsize,
    inner xsep=2pt,
    inner ysep=4pt,
    line width=0.8pt,
]
\definecolor{c1}{HTML}{6d8ac3}
\definecolor{c2}{HTML}{c36d8a}
\definecolor{c3}{HTML}{8ac36d}

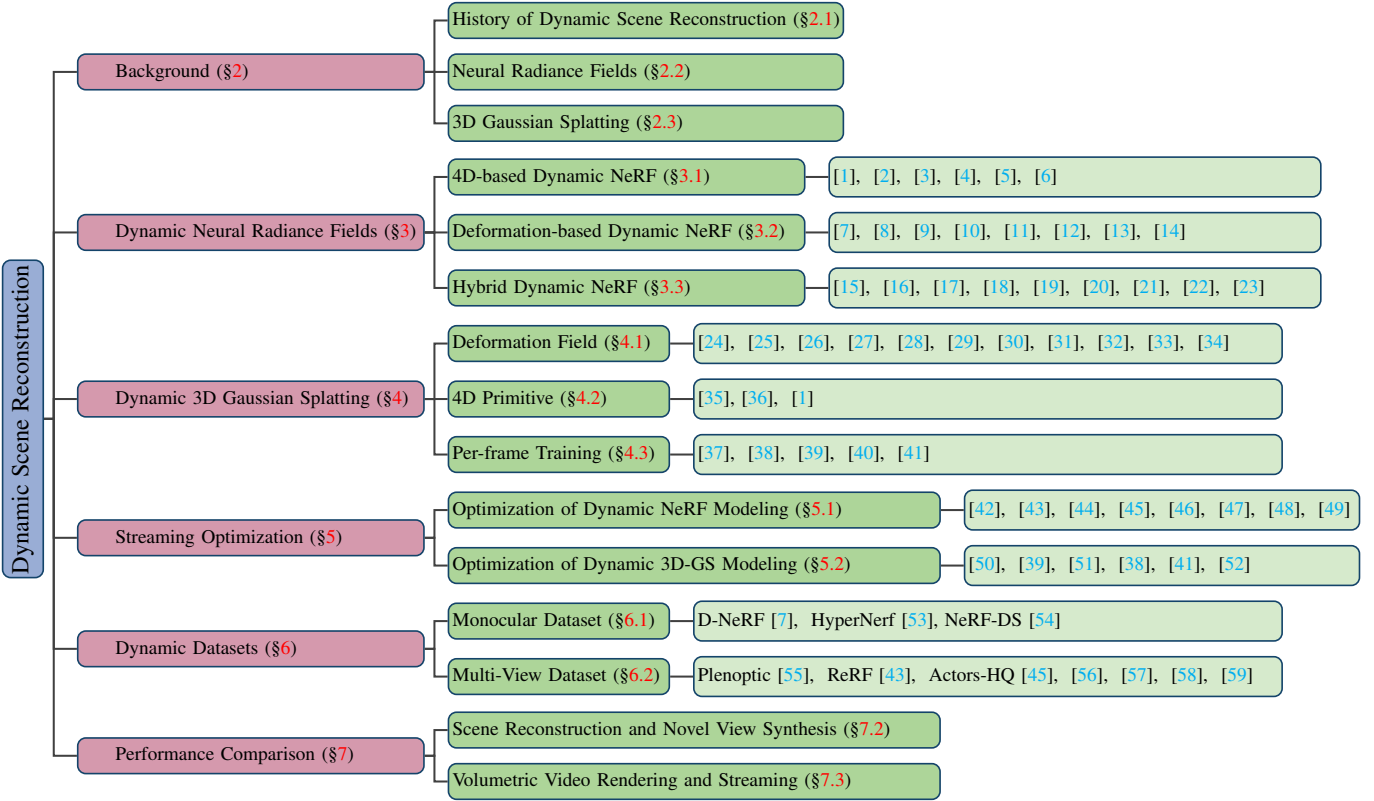
\begin{figure*}[ht]
    
    \resizebox{\textwidth}{!}{
        \begin{forest}
            forked edges,
            for tree={
                grow=east,
                reversed=true,
                anchor=base west,
                parent anchor=east,
                child anchor=west,
                base=left,
                font=\large,
                rectangle,
                draw=hidden-draw,
                rounded corners,
                align=left,
                minimum width=4em,
                edge+={darkgray, line width=1pt},
                s sep=7pt,
                inner xsep=2pt,
                inner ysep=3pt,
                line width=0.8pt,
                ver/.style={rotate=90, child anchor=north, parent anchor=south, anchor=center},
            },
            where level=1{text width=14em,font=\normalsize,}{},
            where level=2{text width=11em,font=\normalsize,}{},
            where level=3{text width=11em,font=\normalsize,}{},
            where level=4{text width=10em,font=\normalsize,}{},
            [
                Dynamic Scene Reconstruction,fill=c1!70,ver
                [
                    Background   (\S \ref{sec:background}),fill=c2!70, minimum width=6cm
                    [
                        History of Dynamic Scene Reconstruction  (\S \ref{subsec:history}),text width=20em,fill=c3!70
                    ]
                    [
                        Neural Radiance Fields (\S \ref{sec:nerf}),text width=20em,fill=c3!70
                    ]
                    [
                        3D Gaussian Splatting  (\S \ref{sec:3dgs}),text width=20em,fill=c3!70
                    ]
                ] 
                [
                    Dynamic Neural Radiance Fields   (\S \ref{sec:dynamicnerf}),fill=c2!70, minimum width=6cm
                    [
                        4D-based Dynamic NeRF  (\S \ref{subsec:4dnerf}),text width=18em,fill=c3!70
                        [
                            \cite{liSpacetimeGaussianFeature2024}{, }
                            \cite{liNeuralSceneFlow2021}{, }
                            \cite{Gao-ICCV-DynNeRF}{, }
                            \cite{duNeuralRadianceFlow2021}{, }
                            \cite{wangNeuralTrajectoryFields2021}{, }
                            \cite{sunDyBluRFDynamicNeural2024},
                    leaf, text width=25em,fill=c3!70
                        ]
                    ]
                    [
                        Deformation-based Dynamic NeRF (\S \ref{subsec:deformnerf}),text width=18em,fill=c3!70
                        [
                            \cite{pumarola2020d}{, }
                            \cite{Park_2021_ICCV}{, }
                            \cite{parkHyperNeRFHigherDimensionalRepresentation2021}{, }
                            \cite{tretschkNonRigidNeuralRadiance2021a}{, }
                            \cite{fangFastDynamicRadiance2022}{, }
                            \cite{liNeural3DVideo2022a}{, }
                            \cite{yanNeRFDSNeuralRadiance2023}{, }
                            \cite{maDeformableNeuralRadiance2023}
                            ,
                    leaf, text width=25em,fill=c3!70
                        ]
                    ]
                    [
                        Hybrid Dynamic NeRF  (\S \ref{subsec:hybridnerf}),text width=18em,fill=c3!70
                        [
                            \cite{fridovich-keilKPlanesExplicitRadiance2023a}{, }
                            \cite{caoHexPlaneFastRepresentation2023}{, }
                            \cite{jangDTensoRFTensorialRadiance2022}{, }
                            \cite{louDaReNeRFDirectionawareRepresentation2024}{, }
                            \cite{10049689}{, }
                            \cite{chenMultiResolutionHybridExplicit2024}{, }
                            \cite{qiaoDynamicMeshAwareRadiance2023}{, }
                            \cite{ganV4DVoxel4D2024}{, }
                            \cite{NEURIPS2023_df311263}
                            ,
                    leaf, text width=25em,fill=c3!70
                        ]
                    ]
                ]
                [
                    Dynamic 3D Gaussian Splatting  (\S \ref{sec:dynamicRepresentation}),fill=c2!70, minimum width=6cm
                    [
                        Deformation Field  (\S \ref{sec:deformation field}),fill=c3!70
                            [
                                \cite{yangDeformable3DGaussians2023}{, } 
                                \cite{liangGauFReGaussianDeformation2023}{, } 
                                \cite{huangSCGSSparseControlledGaussian2023}{, } 
                                \cite{linGaussianFlow4DReconstruction2024}{, } 
                                \cite{kratimenosDynMFNeuralMotion2023}{, } 
                                \cite{wan2024superpointgaussiansplattingrealtime}{, } 
                                \cite{wu4DGaussianSplatting2024}{, } 
                                \cite{zhengGPSGaussianGeneralizablePixelwise2023}{, } 
                                \cite{waczyńska2024dmiso}{, } 
                                \cite{7780814}{, } 
                                \cite{shin2024enhancing},  leaf, text width=30em,fill=c3!70
                            ]
                    ]
                    [
                        4D Primitive  (\S \ref{sec:4D primitive}),fill=c3!70
                        [
                            \cite{yangRealtimePhotorealisticDynamic2023}{, }\cite{duan4DRotorGaussianSplatting2024}{, } \cite{liSpacetimeGaussianFeature2024},  leaf, text width=30em,fill=c3!70
                        ]
                    ]
                    [
                        Per-frame Training  (\S \ref{sec:per-frame training}),fill=c3!70
                        [
                            \cite{luitenDynamic3DGaussians2023}{, } 
\cite{sun3DGStreamFlyTraining}{, } 
\cite{jiang2024robust}{, } 
\cite{dasNeuralParametricGaussians2024}{, } 
\cite{wang2024v3viewingvolumetricvideos} ,  leaf, text width=30em,fill=c3!70
                        ]
                    ]
                ]
                [
                    Streaming Optimization  (\S \ref{sec:streamopt}),fill=c2!70, minimum width=6cm
                    [
                        Optimization of Dynamic NeRF Modeling  (\S \ref{subsec:streamnerf}),text width=25em,fill=c3!70
                        [
                            \cite{zhangEfficientDynamicNeRFBased2024a}{, }
                            \cite{wangNeuralResidualRadiance2023}{, }
                            \cite{zhangRateawareCompressionNeRFbased2024}{, }
                            \cite{10.1145/3592415}{, }
                            \cite{zhengJointRFEndtoEndJoint2024}{, }
                            \cite{guoCompactNeuralVolumetric2023}{, }
                            \cite{wuTeTriRFTemporalTriPlane2024}{, }
                            \cite{wangVideoRFRenderingDynamic2024}
                        , leaf, text width=20em,fill=c3!70
                        ]
                    ]
                    [
                        Optimization of Dynamic 3D-GS Modeling (\S \ref{subsec:stream3dgs}),text width=25em,fill=c3!70
                        [
                            \cite{xiaoBridging3DGaussian2024}{, }
                            \cite{jiang2024robust}{, }
                            \cite{sunMultiframeBitrateAllocation2024a}{, }
                            \cite{sun3DGStreamFlyTraining}{, }
                            \cite{wang2024v3viewingvolumetricvideos}{, }
                            \cite{10.1145/3687919}
                            , leaf, text width=20em,fill=c3!70
                        ]
                    ]
                ]
                [
                    Dynamic Datasets  (\S \ref{sec:datasets}),fill=c2!70, minimum width=6cm
                    [
                        Monocular Dataset  (\S \ref{sec:monocular_datasets}),fill=c3!70
                            [
                                D-NeRF~\cite{pumarola2020d}{, } HyperNerf~\cite{park2021hypernerf}{, }NeRF-DS~\cite{yan2023nerf}
                                , leaf, text width=30em,fill=c3!70
                            ]
                    ]
                    [
                        Multi-View Dataset (\S \ref{sec:multi-view_datasets}),fill=c3!70
                            [
                                Plenoptic~\cite{li2022neural3dvideosynthesis}{, } 
                                ReRF~\cite{wangNeuralResidualRadiance2023}{, }
                                Actors-HQ~\cite{10.1145/3592415}{, }
                                \cite{broxton2020immersive}{, } 
                                \cite{li2022streaming}{, } 
                                \cite{Jiang_2024_CVPR}{, } 
                                \cite{abou2022particlenerf}
                                , leaf, text width=30em,fill=c3!70
                            ]
                    ]
                ]
                [
                    Performance Comparison (\S \ref{sec:performance}),fill=c2!70, minimum width=6cm
                    [
                        Scene Reconstruction and Novel View Synthesis  (\S \ref{sec:scene_reconstruction}),text width=25em,fill=c3!70
                    ]
                    [
                        Volumetric Video Rendering and Streaming (\S \ref{sec:fvv}),text width=25em,fill=c3!70
                    ]
                ]
            ] 
        \end{forest}
    }
    \centering
  \centering
  \captionsetup{justification=centering}
    \caption{\textbf{The structure of the survey.}}
    \label{fig:structure}
\end{figure*}
\IEEEPARstart{R}{epresenting} and rendering dynamic scenes from 2D images has always been a significant but difficult task in computer vision and graphics, as the real world is dynamic and ever-changing~\cite{wu2024cl}.
Its widely used applications in AR/VR~\cite{yangRealtimePhotorealisticDynamic2023} have gathered increasing attention in recent years.
The complexity of dynamic scene representation and rendering lies in the fact that the scene changes not only in appearance but also in geometry and motion~\cite{xu2019flyfusion}.
In the past few decades, dynamic scene rendering has witnessed a remarkable evolution, with various traditional methods being developed and refined.

Early techniques relied on dense camera arrays or multi-view video inputs to reconstruct dynamic scenes. Methods like Structure from Motion (SfM) and Multi-View Stereo (MVS)~\cite{seitzComparisonEvaluationMultiView2006} extended to dynamic scenarios by processing per-frame geometry independently, effectively treating dynamic scenes as sequences of static reconstructions~\cite{leroyMultiViewDynamicShape2017}. Non-Rigid SfM (NRSfM) further advanced this by decomposing temporal shape variations into low-rank basis representations~\cite{paladini2012optimal,garg2013dense}, enabling motion estimation from 2D correspondences. Although these methods achieved reasonable geometric reconstruction, they often struggled with complex topology changes and suffered from temporal inconsistencies due to their frame-wise processing paradigm. Moreover, the requirement for extensive camera setups or precisely calibrated multi-view systems limited their practical applicability.

Template-based approaches addressed some limitations by incorporating prior shape knowledge. The shape-from-template (SfT) methods~\cite{bartoliTemplatebasedReconstructionSingle2012,liRobustSingleviewGeometry2009} leveraged predefined 3D meshes as references, deforming them to match the observed frames through physical or geometric constraints. These methods excelled in scenarios with known object categories, such as human performance capture~\cite{douFusion4DRealtimePerformance2016}, but faltered when faced with topological changes or unseen objects. 

The advent of consumer-grade RGB-D sensors like Microsoft Kinect catalyzed breakthroughs in real-time dynamic reconstruction. Seminal works such as DynamicFusion~\cite{newcombeDynamicFusionReconstructionTracking2015} pioneered volumetric non-rigid alignment, enabling incremental surface updates from depth streams. Subsequent innovations like Fusion4D~\cite{douFusion4DRealtimePerformance2016} integrated multi-view RGB-D inputs with deformation graphs, improving robustness to rapid motions. Although these sensor-driven methods achieved unprecedented real-time performance, they remained fundamentally geometry-centric, lacking inherent mechanisms for photorealistic appearance modeling and view synthesis.

Neural Radiance Fields (NeRF)~\cite{mildenhall2020nerfrepresentingscenesneural} is a milestone work that can render novel views of a static scene from 2D images using implicit 3D representations, which demands extensive queries to a Multi-Layer Perceptron (MLP) to get the radiance value of a point in the scene~\cite{xiaoBridging3DGaussian2024}.
Due to its high costs in training~\cite{10296239} and rendering~\cite{chen2024far}, the original NeRF method is far from real-time rendering and unsuitable for dynamic scenes.
Consequently, numerous follow-up works have aimed to enhance the efficiency and quality of dynamic scene representation and rendering based on the original NeRF.

3D Gaussian Splatting(3D-GS)~\cite{kerbl20233d} has emerged as a paradigm-shifting approach to static scene representation and real-time rendering by representing objects as gaussians in a more flexible and adaptive way~\cite{chen2024survey3dgaussiansplatting}. Recent advances in 3D-GS have shown its potential to render dynamic scenes~\cite{wu4DGaussianSplatting2024,duan4DRotorGaussianSplatting2024,huangSCGSSparseControlledGaussian2023,kratimenosDynMFNeuralMotion2023,liangGauFReGaussianDeformation2023}.
Generally, these methods can be divided into three categories: 1) methods that implement a deformation field to modify Gaussians across frames~\cite{liangGauFReGaussianDeformation2023}; 2) methods that initialize Gaussians at a certain frame and conduct per-frame training~\cite{sun3DGStreamFlyTraining}; 
3) methods that extend 3D-GS to 4D-GS to represent dynamic scenes~\cite{wu4DGaussianSplatting2024,duan4DRotorGaussianSplatting2024}.

Although some existing surveys have also reviewed dynamic scene reconstruction~\cite{ingaleRealtime3DReconstruction2021d,tretschkStateArtDense2023}, some of them are outdated and lack the latest advances in neural rendering and 3D-GS based dynamic scene reconstruction methods. ~\citet{yunusRecentTrends3D2024} provides a comprehensive review of recent advancements in 3D reconstruction techniques for general non-rigid scenes but does not cover volumetric video rendering. ~\citet{entezamiAIDrivenInnovationsVolumetric2024} is closely related to ours as it also investigates the challenges in real-time rendering and streaming of volumetric video, whereas our article provides a broader examination of dynamic scene reconstruction with a focus on advancements in rendering techniques and streaming strategies.

The structure of this survey is shown in Figure~\ref{fig:structure}.
In this survey, we review the history and development of dynamic scene representation and rendering in section~\ref{sec:background}, explaining the basic principle of NeRF and 3D-GS. We summarize the existing methods and categorize them according to their core ideas in section~\ref{sec:dynamicnerf} and section~\ref{sec:dynamicRepresentation}. We also gather works that focus on volumetric video representation and streaming in section~\ref{sec:streamopt}.
We collect the datasets and benchmarks used in dynamic scene representation and rendering in section~\ref{sec:datasets}, and compare the performance of different methods on these datasets in section~\ref{sec:performance}.
We also discuss the challenges and future directions in dynamic scene reconstruction in section~\ref{sec:future}.

\section{Background}
\label{sec:background}

\subsection{History of Dynamic Scene Reconstruction}
\label{subsec:history}

For the past few decades, the field of dynamic scene reconstruction has seen rapid development. We summarize the representative works of dynamic scene reconstruction from different periods in Figure~\ref{fig:timeline}.

Initial research on dynamic scene reconstruction demands a large number of closely posed cameras to capture the scene from different viewpoints to get complete reconstruction of dynamic shapes~\cite{larsenTemporallyConsistentReconstruction2007,lei2009new,furukawaAccurateDenseRobust2010}.
SfM and its extension, NRSfM, are also a feasible way to capture motion with multiple static cameras or a single moving camera~\cite{breglerRecoveringNonrigid3D2000a,paladini2012optimal,garg2013dense,akhterTrajectorySpaceDual2011,dai2014simple,zhuComplexNonRigidMotion2014,5396339,torresani2008nonrigid,russellEnergyBasedMultiple2011,russellVideoPopupMonocular2014}.
This kind of method determines the spatial and geometric relationships of the target through the movement of regular RGB cameras to obtain motion~\cite{taneja2011modeling} and deformation information. Then it performs point correspondence in multiple frames of images to represent the 3D shape varying over time as a linear combination of a low-rank shape basis~\cite{russellEnergyBasedMultiple2011}.
Although early work is only capable of capturing the motion trajectory or smooth surface of the object~\cite{huang1992dynamic,zhang2003spacetime}, the appearance of the object is not well preserved~\cite{breglerRecoveringNonrigid3D2000a}.
Later works improve the quality of the reconstruction by introducing the appearance information of the object~\cite{russellVideoPopupMonocular2014,wangDynamicSfMDetecting2015}.

Due to hardware limitations, early rendering methods for geometry appearance are based on offline MVS, the goal of which is to reconstruct a complete 3D object model from a collection of images taken from known camera viewpoints~\cite{seitzComparisonEvaluationMultiView2006}.
Early work~\citet{CompleteMultiviewReconstruction} adopts MVS in each frame and reconstructs the dynamic scene probabilistically, fusing narrow and wide baseline stereo features using videos captured by 14 cameras, which introduces temporal cues to recover the places occluded in some frames and improves the quality of the reconstruction. Work at that time can be viewed as static scene reconstruction of different frames~\cite{leroyMultiViewDynamicShape2017}.

Another line of per-frame dynamic scene reconstruction methods can be categorized as SfT approaches~\cite{yuDirectDenseDeformable2015,allenSpaceHumanBody2003,bartoliTemplatebasedReconstructionSingle2012,suwajanakorn2014total,salzmann2008local,ostlund2012laplacian}, which require a preprocessed 3D mesh model containing the surface of target objects for reference and needs to update this model with new frames for point correspondences~\cite{liuTemplateBased3DReconstruction2018}.
~\citet{liRobustSingleviewGeometry2009} presents a robust framework and algorithms for reconstructing the geometry and motion of complex deforming shapes from single-view scans. The method leverages a smooth template to approximate the scanned object coarsely, acting as a geometric and topological prior for reconstruction. It employs a novel space-time adaptive non-rigid registration method to recover large-scale motion and an efficient linear mesh deformation algorithm to synthesize fine-scale details like wrinkles and folds.
The topology change of the object is also a challenge for template-based methods, as the template is usually a fixed mesh model and cannot adapt to the change of the shape of the object~\cite{douFusion4DRealtimePerformance2016}.

\begin{figure*}[ht]
    \centering
    \label{fig:timeline}
    \begin{tikzpicture}
        \node[anchor=south west, inner sep=0]
            (image)
            at
            (0,0)
            {\includegraphics[width=\textwidth]{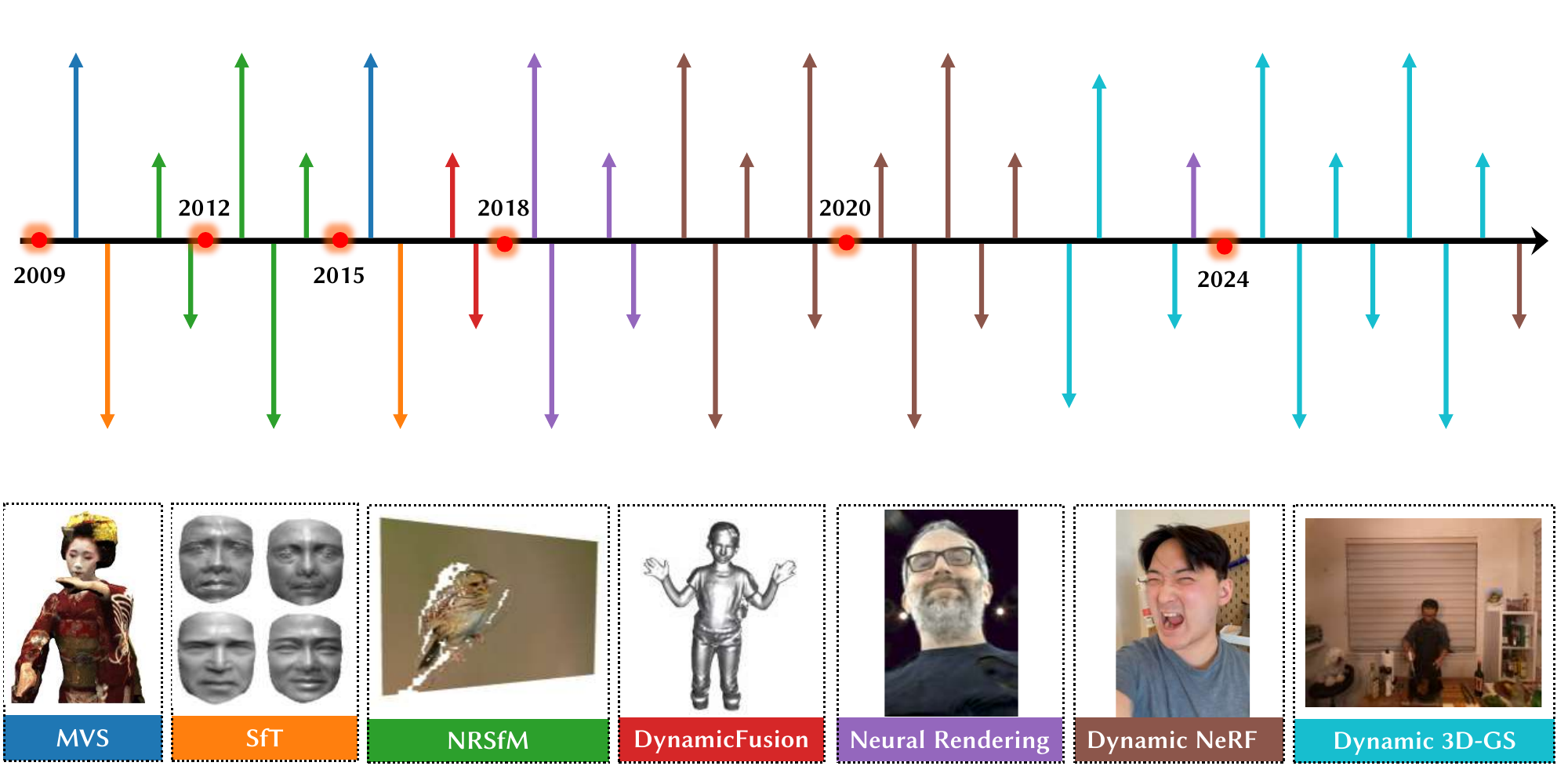}};
        \begin{scope}[x={(image.south east)}, y={(image.north west)}]
            \node at (0.047, 0.98) { \textcolor[HTML]{1f77b4}{ PNWBS}};
            \node at
                (0.047, 0.94)
                { \textcolor[HTML]{1f77b4}{ (\cite{CompleteMultiviewReconstruction})} };
            \node at (0.071, 0.42) { \textcolor[HTML]{ff7f0e}{ TSDR}};
            \node at
                (0.071, 0.38)
                { \textcolor[HTML]{ff7f0e}{ (\cite{liRobustSingleviewGeometry2009})} };
            \node at (0.102, 0.854) { \textcolor[HTML]{2ca02c}{ TS-NRSfM}};
            \node at
                (0.102, 0.814)
                { \textcolor[HTML]{2ca02c}{ (\cite{akhterTrajectorySpaceDual2011})} };
            \node at (0.12, 0.55) { \textcolor[HTML]{2ca02c}{ MSfM}};
            \node at
                (0.12, 0.51)
                { \textcolor[HTML]{2ca02c}{ (\cite{5396339})} };
            \node at (0.153, 0.98) { \textcolor[HTML]{2ca02c}{ DVNRS}};
            \node at
                (0.153, 0.94)
                { \textcolor[HTML]{2ca02c}{ (\cite{garg2013dense})} };
            \node at (0.174, 0.42) { \textcolor[HTML]{2ca02c}{ US-NRSfM}};
            \node at
                (0.174, 0.38)
                { \textcolor[HTML]{2ca02c}{ (\cite{zhuComplexNonRigidMotion2014})} };
            \node at (0.195, 0.854) { \textcolor[HTML]{2ca02c}{ VPop-up}};
            \node at
                (0.195, 0.814)
                { \textcolor[HTML]{2ca02c}{ (\cite{russellVideoPopupMonocular2014})} };
            \node at (0.237, 0.98) { \textcolor[HTML]{1f77b4}{ GDSR}};
            \node at
                (0.237, 0.94)
                { \textcolor[HTML]{1f77b4}{ (\cite{mustafaGeneralDynamicScene2015a})} };
            \node at (0.255, 0.42) { \textcolor[HTML]{ff7f0e}{ DDD}};
            \node at
                (0.255, 0.38)
                { \textcolor[HTML]{ff7f0e}{ (\cite{yuDirectDenseDeformable2015})} };
            \node at (0.289, 0.884) { \textcolor[HTML]{d62728}{ Dynamic-}};
            \node at (0.289, 0.854) { \textcolor[HTML]{d62728}{ -Fusion}};
            \node at
                (0.289, 0.814)
                { \textcolor[HTML]{d62728}{ (\cite{newcombeDynamicFusionReconstructionTracking2015})} };
            \node at (0.302, 0.55) { \textcolor[HTML]{d62728}{ Fusion4D}};
            \node at (0.302, 0.51) { \textcolor[HTML]{d62728}{ (\cite{douFusion4DRealtimePerformance2016})} };
            \node at (0.341, 0.98) { \textcolor[HTML]{9467bd}{ DSVC}};
            \node at
                (0.341, 0.94)
                { \textcolor[HTML]{9467bd}{ (\cite{huangDeepVolumetricVideo2018a})} };
            \node at (0.349, 0.42) { \textcolor[HTML]{9467bd}{ DAM}};
            \node at
                (0.349, 0.38)
                { \textcolor[HTML]{9467bd}{ (\cite{lombardiDeepAppearanceModels2018a})} };
            \node at (0.387, 0.884) { \textcolor[HTML]{9467bd}{ Neural}};
            \node at (0.387, 0.854) { \textcolor[HTML]{9467bd}{ Volumes} };
            \node at
                (0.387, 0.814)
                { \textcolor[HTML]{9467bd}{ (\cite{lombardiNeuralVolumesLearning2019b})} };

            \node at (0.404, 0.55) { \textcolor[HTML]{9467bd}{ OFlow}};
            \node at
                (0.403, 0.51)
                { \textcolor[HTML]{9467bd}{ (\cite{niemeyerOccupancyFlow4D2019})} };
            \node at (0.433, 0.98) { \textcolor[HTML]{8c564b}{ D-NeRF}};
            \node at
                (0.433, 0.94)
                { \textcolor[HTML]{8c564b}{ (\cite{pumarola2020d})} };
            \node at (0.456, 0.42) { \textcolor[HTML]{8c564b}{ NSFF}};
            \node at
                (0.456, 0.38)
                { \textcolor[HTML]{8c564b}{ (\cite{liNeuralSceneFlow2021})} };
            \node at (0.475, 0.854) { \textcolor[HTML]{8c564b}{ Nerfies}};
            \node at
                (0.475, 0.814)
                { \textcolor[HTML]{8c564b}{ (\cite{Park_2021_ICCV})} };
            \node at (0.517, 0.98) { \textcolor[HTML]{8c564b}{ NR-NeRF}};
            \node at
                (0.517, 0.94)
                { \textcolor[HTML]{8c564b}{ (\cite{tretschkNonRigidNeuralRadiance2021a})} };
            \node at (0.52, 0.55) { \textcolor[HTML]{8c564b}{ Video-Nerf}};
            \node at
                (0.52, 0.51)
                { \textcolor[HTML]{8c564b}{ (\cite{xian2021space})} };
            \node at (0.563, 0.854) { \textcolor[HTML]{8c564b}{ FPO}};
            \node at
                (0.563, 0.814)
                { \textcolor[HTML]{8c564b}{ (\cite{wangFourierPlenOctreesDynamic2022})} };

            \node at (0.582, 0.42) { \textcolor[HTML]{8c564b}{ TiNeuVox}};
            \node at
                (0.582, 0.38)
                { \textcolor[HTML]{8c564b}{ (\cite{fangFastDynamicRadiance2022})}};

            \node at (0.603, 0.98) { \textcolor[HTML]{8c564b}{ DyNerf}};
            \node at
                (0.603, 0.94)
                { \textcolor[HTML]{8c564b}{ (\cite{liNeural3DVideo2022a})}};

            \node at (0.626, 0.55) { \textcolor[HTML]{8c564b}{ K-Planes}};
            \node at
                (0.626, 0.51)
                { \textcolor[HTML]{8c564b}{ (\cite{fridovich-keilKPlanesExplicitRadiance2023a})}};

            \node at (0.648, 0.854) { \textcolor[HTML]{8c564b}{ HexPlane}};
            \node at
                (0.648, 0.814)
                { \textcolor[HTML]{8c564b}{ (\cite{caoHexPlaneFastRepresentation2023})}};

            \node at (0.682, 0.45) { \textcolor[HTML]{17becf}{ Deformable}};
            \node at (0.682, 0.42) { \textcolor[HTML]{17becf}{ 3DGS}};
            \node at
                (0.682, 0.38)
                { \textcolor[HTML]{17becf}{ (\cite{yangDeformable3DGaussians2023})}};

            \node at (0.699, 0.98) { \textcolor[HTML]{17becf}{ Dynamic}};
            \node at (0.699, 0.95) { \textcolor[HTML]{17becf}{ 3D-GS}};
            \node at
                (0.699, 0.91)
                { \textcolor[HTML]{17becf}{ (\cite{luitenDynamic3DGaussians2023})}};

            \node at (0.75, 0.55) { \textcolor[HTML]{17becf}{ RealTime4DGS }};
            \node at
                (0.75, 0.51)
                { \textcolor[HTML]{17becf}{ (\cite{yangRealtimePhotorealisticDynamic2023})}};


            \node at (0.76, 0.854) { \textcolor[HTML]{9467bd}{ DynIBaR }};
            \node at
                (0.76, 0.814)
                { \textcolor[HTML]{9467bd}{ (\cite{liDynIBaRNeuralDynamic2023})} };

            \node at (0.804, 0.98) { \textcolor[HTML]{17becf}{ GaussianFlow}};
            \node at
                (0.804, 0.94)
                { \textcolor[HTML]{17becf}{ (\cite{linGaussianFlow4DReconstruction2024})}};

            \node at
                (0.8277142857142857, 0.42)
                { \textcolor[HTML]{17becf}{4D-Rotor GS}};
            \node at
                (0.8277142857142857, 0.38)
                { \textcolor[HTML]{17becf}{ (\cite{duan4DRotorGaussianSplatting2024})}};

            \node at
                (0.8514285714285714, 0.854)
                { \textcolor[HTML]{17becf}{4D-GS}};
            \node at
                (0.8514285714285714, 0.814)
                { \textcolor[HTML]{17becf}{ (\cite{wu4DGaussianSplatting2024})}};

            \node at
                (0.8751428571428571, 0.55)
                { \textcolor[HTML]{17becf}{ E-D3DGS}};
            \node at
                (0.8751428571428571, 0.51)
                { \textcolor[HTML]{17becf}{ (\cite{baeGaussianEmbeddingBasedDeformation2024})}};

            \node at
                (0.897, 0.98)
                { \textcolor[HTML]{17becf}{ GaGS}};
            \node at
                (0.897, 0.94)
                { \textcolor[HTML]{17becf}{ (\cite{lu3DGeometryawareDeformable2024})}};

            \node at
                (0.9225714285714286, 0.42)
                { \textcolor[HTML]{17becf}{ STG}};
            \node at
                (0.9225714285714286, 0.38)
                { \textcolor[HTML]{17becf}{ (\cite{liSpacetimeGaussianFeature2024})}};

            \node at
                (0.9455, 0.854)
                { \textcolor[HTML]{17becf}{ NPGs}};
            \node at
                (0.9455, 0.814)
                { \textcolor[HTML]{17becf}{ (\cite{dasNeuralParametricGaussians2024})}};

            \node at (0.97, 0.55) { \textcolor[HTML]{8c564b}{ V4D }};
            \node at
                (0.97, 0.51)
                { \textcolor[HTML]{8c564b}{ (\cite{ganV4DVoxel4D2024})}};
        \end{scope}
    \end{tikzpicture}

    \caption{\textbf{General timeline of dynamic scene reconstruction methods.}
    The bottom images are adapted from \citet{CompleteMultiviewReconstruction}, \citet{suwajanakornTotalMovingFace2014}, \citet{russellVideoPopupMonocular2014}, \citet{newcombeDynamicFusionReconstructionTracking2015}, \citet{lombardiNeuralVolumesLearning2019b}, \citet{parkNerfiesDeformableNeural2021} and \citet{sun3DGStreamFlyTraining}.}
    \label{fig:timeline}
\end{figure*}

As kinect cameras and depth sensors are introduced to scene reconstruction, dynamic 3D reconstruction method based on RGB-D data source can obtain the depth information of the surface of the scene to be reconstructed directly from the depth sensor, so the estimation of depth information is reduced and the reconstruction speed is significantly improved.
DynamicFusion~\cite{newcombeDynamicFusionReconstructionTracking2015} is the milestone work for 3D reconstruction of dynamic real-time scenes and tackles some inherent challenges in template-based reconstruction techniques.
It is a real-time dense tracking and reconstruction system that can reconstruct the 3D model of a dynamic scene from a single Kinect sensor in real time.
By twisting the reference volume non-rigidly to each incoming frame and fusing the depth samples into the model, the method is able to incrementally update the reference surface model.
Nevertheless, it does not produce a complete reconstruction of the scene's visual information because it does not recover the scene's appearance and lighting information~\cite{newcombeDynamicFusionReconstructionTracking2015}.
Concurrent work~\citet{mustafaGeneralDynamicScene2015a} details a method capable of reconstructing dynamic scenes captured by multiple moving cameras without any prior knowledge of the scene's structure or appearance. This is particularly innovative as it does not require a controlled environment or a pre-existing model of the scene, which is a common requirement for many techniques at that time.
Fusion4D~\cite{douFusion4DRealtimePerformance2016} employs a fully parallelized non-rigid registration framework and a learning-based correspondence matching strategy, integrating volumetric fusion with smooth deformation field estimation in RGBD views. This facilitates incremental reconstruction and effectively handles large inter-frame motions and topology changes.

Recent years have seen a surge of interest in dynamic scene reconstruction and rendering with learning-based or neural rendering methods~\cite{songHDRNetFusionRealtime3D2021, meschederOccupancyNetworksLearning2019, zhao2021spk2imgnet,Ost_2021_CVPR}.
And the first work that combines learning-based method with dynamic scene reconstruction can be traced back to~\citet{delageDynamicBayesianNetwork2006}.
This paper presents a dynamic Bayesian network model capable of autonomous 3D reconstruction from a single indoor image, leveraging prior knowledge about indoor scenes to resolve ambiguities inherent in monocular vision and recover 3D information. 
~\citet{zhengSparseDynamic3D2015} develops a compressed sensing approach that formulates the estimation of 3D structure as a dictionary learning problem, where the dictionary is defined by the temporally varying 3D structure, and local sequencing information is captured through sparse coefficients that describe a locally linear 3D structural interpolation.
It is worth noting that NRSfM methods and SfT methods have achieved substantial progress through neural rendering innovations\cite{Novotny_2019_ICCV,kairandaFSfTShapeFromTemplatePhysicsBased2022}.

Deep learning-based methods have also been widely used in dynamic scene reconstruction in recent years.
~\citet{lombardiNeuralVolumesLearning2019b} employs a learning-based technique that utilizes an encoder-decoder network to transform 2D images into a 3D volumetric representation without the need for explicit object reconstruction or tracking. This approach leverages a differentiable ray-marching operation for end-to-end training, enabling the creation of dynamic and renderable volumes that can generalize well to novel viewpoints. The paper also introduces a dynamic irregular grid structure, implemented through a warp field during ray marching, to enhance resolution and reduce artifacts associated with regular grid structures.
HDR-Net-Fusion~\cite{songHDRNetFusionRealtime3D2021} addresses the challenges of noise and erroneous observations from data capturing devices and the inherently ill-posed nature of non-rigid registration with insufficient information. The system learns to simultaneously reconstruct and refine geometry on the fly with a sparse embedded deformation graph of surfels, using a hierarchical deep reinforcement network.~\citet{huangDeepVolumetricVideo2018a} introduces a novel multi-view Convolutional Neural Network that maps 2D images to a 3D volumetric field, which encodes the probabilistic distribution of surface points of the captured subject. Trained solely on synthetic data, the network generalizes to real footage, demonstrating significantly more robust and accurate reconstruction of dynamic scenes.

NeRF~\cite{mildenhall2020nerfrepresentingscenesneural} and 3D-GS~\cite{kerbl20233d} are two milestone works in the field of neural rendering, which are the two most popular foundations for dynamic scene reconstruction in recent years. NeRF uses an MLP to represent the scene implicitly as a continuous 5D function and renders the scene by querying the MLP at each pixel, allowing for the synthesis of new views that were not present in the training data. 3D-GS is a real-time and photorealistic rendering method for static scenes, which uses 3D Gaussians to represent the scene and renders the scene by projecting the 3D Gaussians to the 2D image plane, achieving high-quality and high-speed rendering of complex scenes.

The development of dynamic NeRF methods has shown a diverse trend. Initially, methods such as Video-Nerf~\cite{liSpacetimeGaussianFeature2024}, Neural Scene Flow Fields~\cite{liNeuralSceneFlow2021}, DynamicNeRF~\cite{Gao-ICCV-DynNeRF}, NeRFlow~\cite{duNeuralRadianceFlow2021}, DCT-NeRF~\cite{wangNeuralTrajectoryFields2021}, and DyBluRF~\cite{sunDyBluRFDynamicNeural2024} emerged, using time as an additional dimension input to the MLP to model scene dynamics, but faced some challenges such as data sparsity and lack of supervision. Subsequently, D-NeRF~\cite{pumarola2020d} pioneered the introduction of a deformation field to model dynamic scenes in canonical space, followed by Nerfies~\cite{Park_2021_ICCV}, HyperNerf~\cite{parkHyperNeRFHigherDimensionalRepresentation2021}, NR-Nerf~\cite{tretschkNonRigidNeuralRadiance2021a}, TiNeuVox~\cite{fangFastDynamicRadiance2022}, DyNerf~\cite{liNeural3DVideo2022a}, Nerf-DS~\cite{yanNeRFDSNeuralRadiance2023}, and DE-NeRF~\cite{maDeformableNeuralRadiance2023}, which further refined this approach. Hybrid methods, including K-Planes~\cite{fridovich-keilKPlanesExplicitRadiance2023a}, HexPlane~\cite{caoHexPlaneFastRepresentation2023}, D-TensoRF~\cite{jangDTensoRFTensorialRadiance2022}, DaReNeRF~\cite{louDaReNeRFDirectionawareRepresentation2024}, Nerf-Player~\cite{10049689}, Hy-DNeRF~\cite{chenMultiResolutionHybridExplicit2024}, DMRF~\cite{qiaoDynamicMeshAwareRadiance2023}, V4D~\cite{ganV4DVoxel4D2024}, and MSTH~\cite{NEURIPS2023_df311263}, integrating explicit geometric structures with implicit neural representations, improving inference speed and interpretability. Additionally, methods such as FPO~\cite{wangFourierPlenOctreesDynamic2022} and Sync-NeRF~\cite{kimSyncNeRFGeneralizingDynamic2024} have focused on specific aspects, such as fast video generation and handling unsynchronized videos, respectively, further expanding the capabilities of Dynamic NeRF.

Dynamic 3D Gaussian representation methods have emerged to address the limitations of 3D-GS in capturing dynamic scenes, and they can be generally categorized into three groups.
Deformation field based methods, such as Deformable 3D-GS~\cite{yangDeformable3DGaussians2023}, GauFR~\cite{liangGauFReGaussianDeformation2023}, and SC-GS~\cite{huangSCGSSparseControlledGaussian2023},Gaussian-Flow~\cite{linGaussianFlow4DReconstruction2024}, DynMF~\cite{kratimenosDynMFNeuralMotion2023}, and SP-GS~\cite{wan2024superpointgaussiansplattingrealtime} aim to deform 3D Gaussians across frames to fit dynamic scenes. They often use an MLP as the deformation network, assuming that only the position, scale, and rotation of the gaussians change. 
4D primitive based methods, including RealTime4DGS~\cite{yangRealtimePhotorealisticDynamic2023}, 4D-Rotor GS~\cite{duan4DRotorGaussianSplatting2024}, and STG~\cite{liSpacetimeGaussianFeature2024}, extend the 3D-GS formulation by integrating the time dimension.
Per-frame training methods, such as Dynamic 3D-GS~\cite{luitenDynamic3DGaussians2023} and 3DGStream~\cite{sun3DGStreamFlyTraining}, initialize 3D Gaussians at certain frames and train them at other frames. Unlike deformation field based methods, they directly train the Gaussians at each frame. 

As the focus of this survey, more details of these two methods and their variants will be discussed in the following sections.

\subsection{Neural Radiance Fields}
\label{sec:nerf}
The core principle of NeRF is to represent the scene as a continuous volumetric function, which encodes the density and color information at all points in space. NeRF achieves this by training a neural network, typically an MLP, on a set of input images. 
\begin{equation}
    F_{\theta} : (\mathbf{x}, \mathbf{d}) \to (\mathbf{c},\sigma) ,
\end{equation}
where $\mathbf{x} = (x,y,z)$ is the location of 3D point in the scene, $\mathbf{d}=(\theta, \phi)$ is the viewing direction, $\mathbf{c}= (r,g,b)$ is the color of the point, and $\sigma$ is the density of the point (or it can be interpreted as the probability of a ray terminating at an infinitesimal particle at that position.). Once the neural radiance field of the scene is obtained, the volume rendering~\cite{10.1145/964965.808594} can be performed to render the color of the scene at each pixel in a certain view direction.
\begin{equation}
\begin{aligned}
    C &=\int_{t_n}^{t_f}T(t)\sigma(\boldsymbol{r}(t))\boldsymbol{c}(\boldsymbol{r}(t),\boldsymbol{d})dt , \\
    T(t)&=\exp(-\int_{t_n}^t\sigma(\boldsymbol{r}(s))ds) ,
\end{aligned}
\end{equation}
where $C$ is the color of the pixel, $T(t)$ is the transmittance of the ray, $\sigma$ is the density of the point, $\boldsymbol{c}$ is the color of the point, $\boldsymbol{r}(t)$ is the position of the ray at time $t$, and $\boldsymbol{d}$ is the viewing direction.

To estimate this continuous integral, NeRF divides the interval $[t_n, t_f]$ into $N$ equal-width bins and randomly selects one sample from each bin, with the quadrature rule by~\citet{468400}.
\begin{equation}
    \begin{aligned}
    \hat{C}(\mathbf{r})&=\sum_{i=1}^N T_i (1-\exp(-\sigma_i \delta_i)) \mathbf{c}_i ,\\ 
    \text{where } T_i &=\exp\left(-\sum_{j=1}^{i-1} \sigma_j \delta_j\right).
\end{aligned}
\end{equation}
With the input image as the supervision and the loss function as the difference between the rendered image and the ground-truth image, the neural network can be optimized by the backpropagation algorithm directly. However, this is not sufficient for achieving state-of-the-art quality, which will be addressed by the two improvements.

As the neural network tends to learn low-frequency functions~\cite{pmlr-v97-rahaman19a}, the high-frequency details of the image are not well preserved. To address this issue, the first improvement is the positional encoding of the input, which encodes input position with a Fourier feature map for the network to learn the complex spatial patterns of the scene.

\begin{equation}
    \scalebox{0.9}{$\displaystyle 
        \gamma(p) = \sin(2^{0} \pi p), \cos(2^{0} \pi p), \cdots, \sin(2^{L-1} \pi p), \cos(2^{L-1} \pi p) 
    $}
\end{equation}

The second improvement is the volume rendering with hierarchical sampling~\cite{10.1145/78964.78965}, which addresses the trade-off between the rendering quality and the number of samples. Two networks are trained jointly to predict the density and color of the scene at different levels of detail; the ``coarse" network predicts the density and color at a lower resolution, and the ``fine" network predicts the density and color based on the output of the ``coarse" network at a higher resolution. The final color of the scene is the alpha composition of the color predicted by the two networks.

\subsection{3D Gaussian Splatting}
\label{sec:3dgs}
3D-GS satisfies the needs for real-time and photorealistic rendering of static scenes. Unlike traditional mesh-based or volume-based rendering methods that require a large amount of memory and computation, 3D-GS takes advantage of the physical properties of the Gaussians and achieves a balance between rendering quality and rendering speed.

A 3D Gaussian has four necessary attributes: the center position $u$, the covariance matrix $\Sigma$, the amplitude $\alpha$, and the color $c$, where $c$ is represented by spherical harmonics (SH) coefficients. All of these attributes are learnable and can be optimized by backward propagation.
A 3D Gaussian is defined as:
\begin{equation}
    G(x) = e^{-\frac{1}{2}(x^T \Sigma^{-1} x)} ,
\end{equation}
where $x$ is the position of the point in world space.

As the covariance matrix $\Sigma$ is symmetric and positive definite, it can be decomposed as:
\begin{equation}
    \Sigma = R S S^T R^T,
\end{equation}
where $R$ is the rotation matrix and $S$ is the diagonal matrix of eigenvalues. $R$ is stored in the form of the quaternion $q$ and $S$ is represented by the 3D scale vector $s$ to allow independent optimization.

For each 3D Gaussian in a scene, the rendering process is to project all the 3D Gaussians to the 2D image plane, just like splatting snowballs onto the wall:
\begin{equation}
    \Sigma' = J W \Sigma W^T J^T ,
\end{equation}
where $J$ is the Jacobian matrix of the affine approximation of the projective transformation, and $W$ is a viewing transformation matrix.

The projected 2D Gaussian inherits the center position and color of the 3D Gaussian, and the covariance matrix is updated to $\Sigma'$. Each projection of the 3D Gaussian will contribute to certain pixel tiles of the image, and the final color of the pixel is the alpha blending of all the contributions. The order depends on the depth of the 3D Gaussians, and the depth is calculated by the distance between the pixel and the center of the 3D Gaussian. Once the sorting is done, the color of the pixel can be calculated by:
\begin{equation}
    C = \sum_{i=1}^{n} \alpha_i c_i \prod_{j=1}^{i-1}(1-a_j) ,
\end{equation}
where $C$ is the final color of the pixel, $\alpha_i$ is the amplitude of the $i$-th 3D gaussian multiplied by equation 1, and $c_i$ represents the computed color.

The optimization of the 3D Gaussian is done by the comparison between the rendered image and the ground-truth image with the fast rasterization. The loss function is defined as:
\begin{equation}
    \mathcal{L} = (1-\lambda) \mathcal{L}_{1} + \lambda \mathcal{L}_{D - SSIM} ,
\end{equation}

\begin{figure*}[ht] 
  \centering
  \centering
  \includegraphics[width=0.95\textwidth]{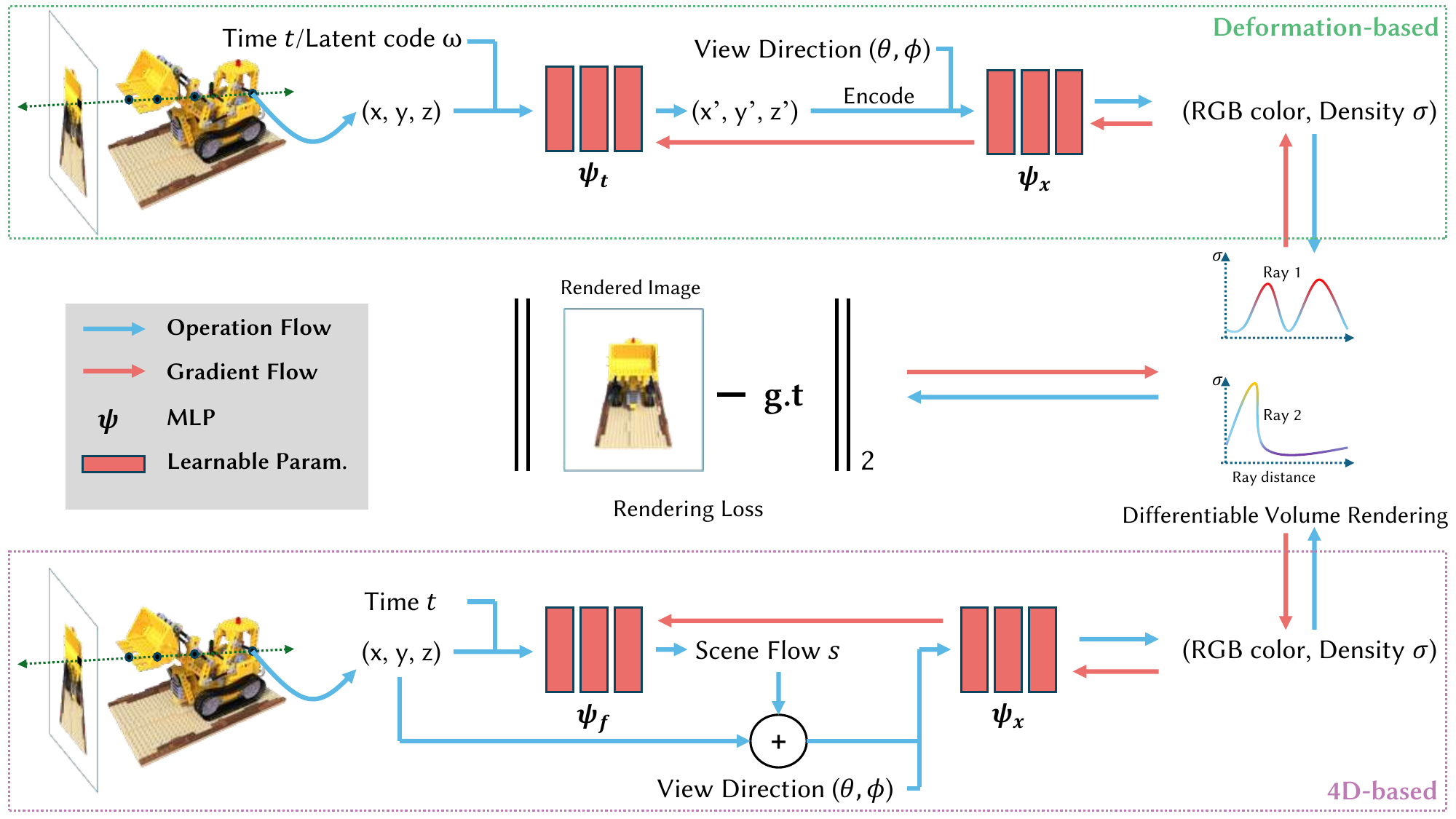}
  \caption{\textbf{General Pipeline of dynamic NeRF methods.} The figure is adapted from~\citet{mildenhall2020nerfrepresentingscenesneural},~\citet{pumarola2020d} and~\citet{Park_2021_ICCV}.} \label{fig:dynamicnerfpipeline}
\end{figure*}

\section{Dynamic Neural Radiance Fields}
\label{sec:dynamicnerf}
Expanding vanilla NeRF to handle dynamic scenes is a challenging task, rather than just adding a new time dimension. The main difficulty arises from the fact that the geometry and appearance of the scene can change over time and objects move within dynamic scenes~\cite{zhangRateawareCompressionNeRFbased2024},
which requires the model to be able to capture these changes. 
In this section, we review the recent work that has tackled this problem by extending NeRF to handle dynamic scenes. 
We refer to these methods as Dynamic Neural Radiance Fields (Dynamic NeRF).

Generally, dynamic NeRF methods can be categorized into three groups:
1) methods that employ time as an additional dimension input to MLP to model the dynamics of the scene.
2) methods that introduce a deformation field to model the dynamic scene in canonical space.
3) methods that integrate explicit geometric structures with implicit neural representation.

The pipeline of the first two groups is shown in Figure~\ref{fig:dynamicnerfpipeline}.

There also exist methods that focus on compressing the dynamic NeRF model to reduce the computational cost or improve the training efficiency of the model, which we will discuss in the following section.

\subsection{4D-based Dynamic NeRF}
\label{subsec:4dnerf}
The first group of dynamic NeRF methods is to use time as an additional dimension input to the MLP.
Although directly introducing an additional time dimension to the 5D input of NeRF is non-trivial~\cite{10049689}, these methods model the dynamics of the scene by encoding the time information into the input of the MLP. Their main idea is to concatenate the time information with the input coordinates and view direction, and feed the concatenated input to the MLP to predict the radiance and volume density.
Thus, the model fuses the spatial and temporal information to model the dynamic scenes.
However, the main challenge of these methods is the sparsity of the data~\cite{10049689} and lacks supervision and priors between frames~\cite{zhanKFDNeRFRethinkingDynamic2024}.

\textbf{Video-NeRF}~\cite{xian2021space} leverages monocular video depth estimation to constrain the time-varying geometry within a globally consistent representation, addressing the motion-appearance ambiguity inherent in single-view videos. This method innovatively models scene dynamics by encoding time information into the MLP, concatenating temporal data with spatial coordinates and view direction. By aggregating frame-wise 2.5D representations and propagating scene contents across time using a static scene loss, the method enables high-quality free-viewpoint video rendering while preserving motion and texture details.

\textbf{Neural Scene Flow Fields}~\cite{liNeuralSceneFlow2021} combines a static scene representation with a dynamic one, allowing the model to leverage multi-view constraints in static regions for higher-quality rendering. Uniquely, it predicts forward and backward 3D scene flow(essentially 3D offset vectors that indicate the position at the subsequent time), denoting the position of a point at adjacent time instances, allowing for sophisticated representation of sharp motion discontinuities and enabling interpolation along both spatial and temporal dimensions. It also introduces data-driven priors such as geometric consistency and single-view depth terms, to learn correct scene geometry, especially for dynamic regions.

\textbf{DynamicNeRF}~\cite{Gao-ICCV-DynNeRF} employs both a static NeRF and a time-varying dynamic NeRF to capture the complexities of dynamic scenes from a single monocular video. The static NeRF is utilized to model the background and non-moving elements of the scene, allowing for accurate reconstruction of the static structure and appearance without interference from moving objects. In contrast, the dynamic NeRF is designed to handle the moving objects within the scene, considering both the 3D position and the time to predict the 3D scene flow. It predicts the volume density and color at a given spatiotemporal location, along with the forward and backward 3D scene flow vectors. These two vectors are pivotal as they describe the displacement of a point in 3D space from one time step to the next, capturing the dynamic essence of the scene. The model uses these flow vectors to warp the radiance fields from adjacent time instances, creating a temporally consistent representation.

\textbf{NeRFlow}~\cite{duNeuralRadianceFlow2021} introduces two critical components: a radiance field modeling the appearance and density of the scene and a flow field enabling their propagation across time. The input and output of radiance field are as same as common 4D-based dynamic NeRF methods, with the addition of a flow field that predicts the 3D scene flow, which is represented by a 4D function that takes in spatial coordinates and time as input and outputs the instantaneous velocity vector, which describes how each point in space is moving over time. Using a neural ODE~\cite{chen2018neural} for continuous modeling and integration of the flow field, the model can generate consistent 4D view synthesis in both space and time.

\textbf{DCT-NeRF}~\cite{wangNeuralTrajectoryFields2021} learns smooth and stable trajectories for each point in space in an input video sequence, allowing for high-quality reconstruction, especially in dynamic regions. By parameterizing trajectories with the Discrete Cosine Transform (DCT), DCT-NeRF enables consistent motion extrapolation beyond the last available frame. The method handles temporal occlusions and reflects changes in reflectance over time, rendering dynamic scenes from novel viewpoints with high fidelity.

\textbf{DyBluRF}~\cite{sunDyBluRFDynamicNeural2024} captures the camera trajectory and object motion by discretizing the exposure time into multiple timestamps and learning camera poses at these timestamps~\cite{wang2023bad}. It uses an MLP to predict the global DCT trajectories of scene objects~\cite{valmadre2012general}, simulating motion blur along the exposure time. To enhance temporal consistency, a cross-time rendering approach is adopted, which models scene correlations across multiple input views and integrates scene information from other frames into the target frame. DyBluRF also introduces data-driven priors to mitigate inaccuracies in depth and optical flow from blurry images~\cite{teed2020raft,ranftl2020towards}, using \textit{Extreme Value Constraints} for depth and optical flow maps.
\subsection{Deformation-based Dynamic NeRF}
\label{subsec:deformnerf}

The second group introduces a deformation field to model dynamic scenes in canonical space.
These methods model the dynamics of the scene by introducing a deformation field that warps the canonical space to the dynamic scene space~\cite{Gao-ICCV-DynNeRF}, and feeds the warped coordinates at each timestamp to the MLP to predict the radiance and volume density for each frame.
In conclusion, the key to these methods is to obtain a canonical space representation of the scene and use the deformation field to warp the canonical space to the dynamic scene space to model dynamic scenes~\cite{pumarola2020d}.

\textbf{D-NeRF}~\cite{pumarola2020d} is the pioneer work that introduces a deformation field to NeRF to model dynamic scenes in canonical space. D-NeRF considers time as an additional input and splits the learning process into two stages: encoding the scene into a canonical space and mapping this representation to the deformed scene at a specific time with a deformation field. This deformation field captures the transformation between the scene at a specific time and a canonical scene configuration, allowing it to effectively model and render dynamic changes in the scene.

\textbf{Nerfies}~\cite{Park_2021_ICCV} optimizes an additional continuous volumetric deformation field that warps each observed point into a canonical 5D NeRF, enabling the representation of non-rigidly deforming scenes. The deformation field is implemented as an MLP and is optimized concurrently with the canonical NeRF model. The paper proposes an elastic regularization for the deformation field, inspired by principles of geometry processing and physical simulation~\cite{zollhofer2014real,douFusion4DRealtimePerformance2016,newcombeDynamicFusionReconstructionTracking2015}, to improve optimization robustness.

\textbf{HyperNerf}~\cite{parkHyperNeRFHigherDimensionalRepresentation2021} addresses discontinuities in the deformation field to capture topological changes by embedding  the scene in a higher-dimensional space, treating the 5D radiance field of each input image as a slice through this ``hyper-space". Inspired by level set methods~\cite{osher1988fronts}, HyperNeRF enables the reconstruction and rendering of scenes with varying topologies, which were previously challenging for continuous deformation field based approaches.

\textbf{NR-NeRF}~\cite{tretschkNonRigidNeuralRadiance2021a} decomposes a dynamic scene into a canonical volume and its corresponding deformation, implementing scene deformation as ray bending. A novel rigidity network is proposed to better constrain rigid regions of the scene, which is used to assign a rigidity score to every point in the canonical volume and helps to distinguish between rigid and non-rigid regions within the scene. By doing so, it allows the deformation field to apply transformations only to the non-rigid parts of the scene while keeping the rigid regions unchanged. The rigidity network is jointly trained with the ray bending network and the canonical neural radiance field.

\textbf{TiNeuVox}~\cite{fangFastDynamicRadiance2022} employs a compact deformation network to encode coarse motion trajectories and enhances temporal information within a radiance network. It proposes a multi-distance interpolation method to model both small and large motions effectively, sweeping movements by interpolating features from voxels at various distances.

\textbf{DyNerf}~\cite{liNeural3DVideo2022a} represents scenes using a compact and expressive dynamic neural radiance field parameterized by temporal latent codes. Rather than directly using time as input, the model encodes scene motion and appearance changes through these latent codes, which are learned to capture the intricate details of moving objects and textures. It significantly boosts training speed and image quality through a hierarchical training scheme and ray importance sampling, enabling the representation of complex and dynamic scenes.

\textbf{NeRF-DS}~\cite{yanNeRFDSNeuralRadiance2023} introduces a surface-aware dynamic NeRF that conditions the radiance field on surface position and orientation in the observation space, aiming at addressing changing reflections on moving specular surfaces in dynamic scenes. This innovation allows the model to maintain different reflected colors for specular surfaces when mapped to a common canonical space. Furthermore, NeRF-DS utilizes a moving object mask to guide the deformation field, providing a stable reference despite color variations on specular surfaces in motion.
  
\textbf{DE-NeRF}~\cite{maDeformableNeuralRadiance2023} unlocks new possibilities for 3D visual modeling of rapidly moving and deforming scenes. It combines RGB and event cameras to model rapidly moving and deforming objects. DE-NeRF leverages the high-speed, asynchronous nature of event cameras to capture visual changes and integrates this data with sparse RGB frames to model dynamic scenes effectively. The method jointly optimizes camera poses and radiance fields, utilizing a neural network to map event timestamps to camera poses.

\begin{figure*}[ht] 
  \centering
  \centering
  \captionsetup{justification=centering}
  \includegraphics[width=0.95\textwidth]{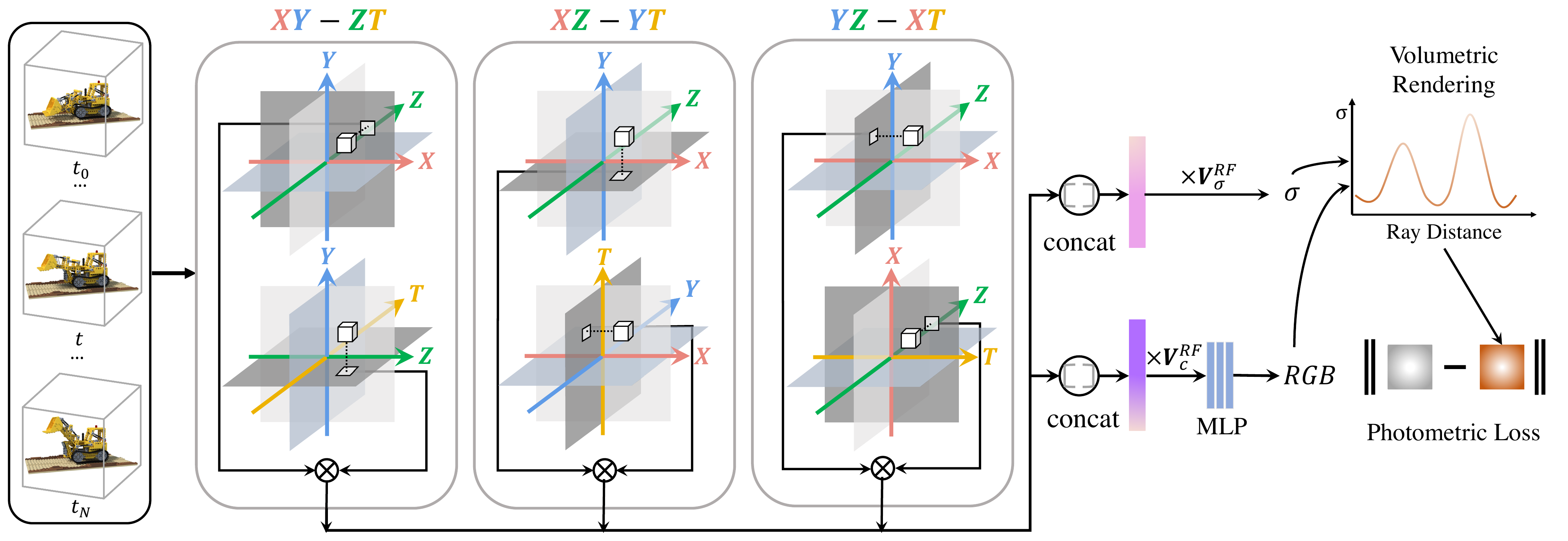}
  \caption{\textbf{Pipeline of HexPlane.} 
The image is courtesy of~\cite{caoHexPlaneFastRepresentation2023}.
  } \label{fig:hexplane}
\end{figure*}

\subsection{Hybrid Dynamic NeRF}
\label{subsec:hybridnerf}
Hybrid Dynamic NeRF methods refer to a data representation technique that integrates explicit geometric structures with implicit neural representations, by performing efficient trilinear interpolation~\cite{liCompressingVolumetricRadiance2023} equipped with lightweight neural networks, which can accelerate the inference process~\cite{chenMultiResolutionHybridExplicit2024} and improve the interpretability of the model~\cite{fridovich-keilKPlanesExplicitRadiance2023a}.
These representations maintain a grid of coordinate feature volumes for rapid rendering~\cite{10049689}, often achieved through spatial decomposition techniques~\cite{fridovich-keilKPlanesExplicitRadiance2023a} that compress geometric data, as this kind of representation requires significant storage overhead~\cite{chenMultiResolutionHybridExplicit2024}.
This kind of work can be viewed as one of the promising directions for improving dynamic NeRF~\cite{liCompressingVolumetricRadiance2023}.

\textbf{K-Planes}~\cite{fridovich-keilKPlanesExplicitRadiance2023a} provides the first white-box, interpretable model for radiance fields that it decomposes scenes into space-only planes and space-time planes, enhancing interpretability and enabling dimension-specific priors. It uses a linear feature decoder with a learned color basis~\cite{Wizadwongsa_2021_CVPR}, offering performance similar to non-linear black-box MLP decoders but with greater transparency and adaptability. It also incorporates a global appearance code per training image, allowing the model to account for changes in lighting or other appearance variations without affecting the geometry.
  
\textbf{HexPlane}~\cite{caoHexPlaneFastRepresentation2023} addresses the computational inefficiency of NeRF by using an explicit factorization approach. It decomposes the 4D spacetime into six feature planes, as shown in Figure~\ref{fig:hexplane}. To compute the features for a point in spacetime, HexPlane projects the point onto each of these planes, extracts feature vectors, and then fuses them through element-wise multiplication and concatenation, which is highly efficient and reduces the need for numerous MLP evaluations. The resulting fused feature vector is passed to a small MLP to predict the color of the point. Compared to K-Planes, the simplicity and specialization of HexPlane for 4D volumes make it particularly well-suited for applications where the rapid rendering of dynamic content is crucial.
 
\textbf{D-TensoRF}~\cite{jangDTensoRFTensorialRadiance2022} extends Tensorial Radiance Fields~\cite{chen2022tensorf} for dynamic scenes, representing the radiance field of a dynamic scene as a 5D tensor, with each axis corresponding to X, Y, Z coordinates and time, and each element featuring multi-channel features. The method decomposes this tensor into lower-rank components using both CANDECOMP/PARAFAC decomposition and a newly proposed Matrix-Matrix decomposition, with the latter aiming to reduce computational overhead. 

\textbf{DaReNeRF}~\cite{louDaReNeRFDirectionawareRepresentation2024} extends plane-based methods~\cite{caoHexPlaneFastRepresentation2023} with direction-aware representation and captures scene dynamics from six different directions using an inverse dual-tree complex wavelet transformation~\cite{selesnick2005dual} to recover plane-based information, combined with a small MLP for color regression and volume rendering in training. The method addresses storage issues with a trainable masking approach and achieves a twofold reduction in training time compared to prior work.

\textbf{NeRF-Player}~\cite{10049689} takes a hybrid representation and leverages the decomposition of 4D spatiotemporal space into three distinct categories: static, deforming, and new areas. This decomposition is accomplished through a self-supervised learning process that predicts the probability that each point in the scene belongs to one of these three categories without manual annotation. The static field represents time-invariant areas, simplifying the modeling task for unchanging background elements. The deforming field captures areas with changing surfaces, such as moving objects, ensuring their consistency across different frames. The newness field handles areas that emerge over time, such as new objects entering the scene. NeRFPlayer employs a sliding window scheme for streaming feature channels of InstantNGP~\cite{muller2022instant} and TensoRF~\cite{chen2022tensorf}, enabling efficient modeling of these neural fields and facilitating real-time rendering. 

\textbf{DMRF}~\cite{qiaoDynamicMeshAwareRadiance2023} establishes a two-way coupling between meshes and NeRF, allowing the updating of radiance and throughput along a cast ray with an arbitrary number of bounces. It addresses the color space discrepancy between path tracing and standard NeRF by training NeRF with high dynamic range (HDR) images, enabling more accurate light transport simulation. The paper also presents a strategy for estimating light sources and casting shadows on NeRF.

\textbf{V4D}~\cite{ganV4DVoxel4D2024} employs a dual voxel format for density and texture fields, processed by a small MLP, and incorporates a lookup table refinement module for pixel-level enhancements with minimal computational overhead. It also introduces conditional positional encoding to improve performance with negligible computational costs~\cite{shechtman2015phase}.

\textbf{MSTH}~\cite{NEURIPS2023_df311263} represents dynamic scenes as a weighted combination of 3D and 4D hash encodings~\cite{muller2022instant}, with weights determined by a learnable mask guided by uncertainty estimation~\cite{kendall2017uncertainties,martin2021nerf}. This method reduces hash collision rates and enables compact representation, allowing rapid optimization and convergence.

\subsection{Other Methods for Dynamic NeRF Representation}
\label{subsec:othernerf}
Apart from the three main categories of dynamic NeRF methods, there also exist other methods that focus on improving the representation of dynamic scenes in NeRF.

\textbf{FPO}~\cite{wangFourierPlenOctreesDynamic2022} integrates generalized NeRF with PlenOctree representation~\cite{yu2021plenoctrees}, volumetric fusion, and Fourier transform to facilitate fast generation and rendering of free-viewpoint videos. By integrating PlenOctree representation with Fourier transform, FPO effectively captures the temporal dynamics of scenes, allowing for the compression of time-varying information into the frequency domain. This approach results in a significant reduction in memory usage and accelerated fine-tuning. It proposes a coarse-to-fine fusion scheme to enhance the speed of scene construction, which rapidly generates the PlenOctree structure using a generalizable technique.

\textbf{Sync-NeRF}~\cite{kimSyncNeRFGeneralizingDynamic2024} extends the capabilities of dynamic neural radiance fields to unsynchronized videos by incorporating per-camera time offsets. These offsets enable the model to account for temporal discrepancies between different viewpoints, which is critical in scenarios where the same frame index across videos does not correspond to the same moment in time. Utilizing an implicit neural representation for temporal embeddings and adjusting grid sampling for time offsets, this approach can be applied to various dynamic neural radiance field architectures.

\begin{figure*}[ht] 
    \centering
    \centering
    \includegraphics[width=0.95\textwidth]{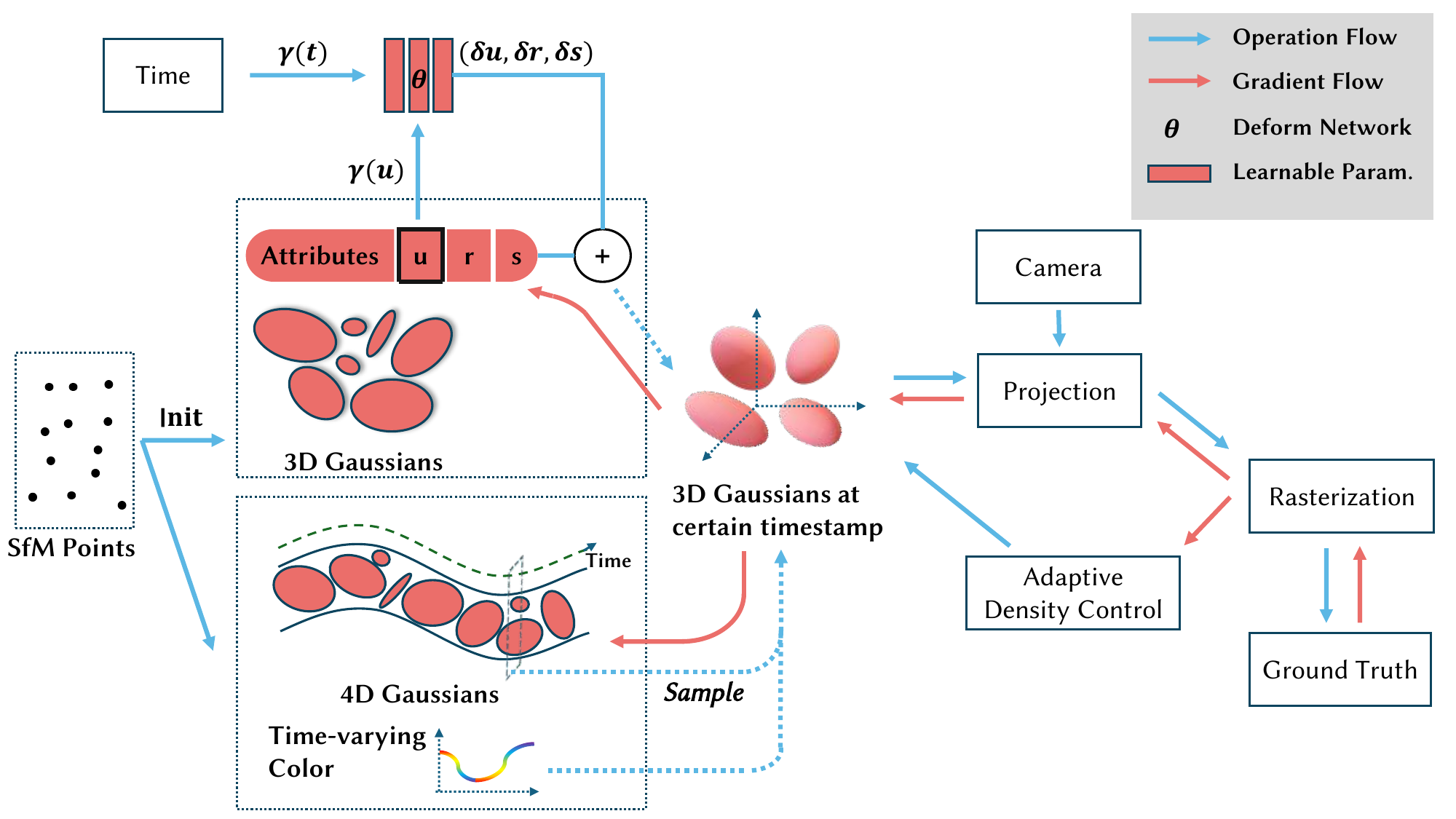}
    \caption{\textbf{General Pipeline of dynamic 3D-GS methods.} 
    For deformable based methods, MLP as a deformation network is inserted into the 3D-GS pipeline to predict the deformation of 3D Gaussians across frames. 4D primitive based methods extend the 3D case by incorporating the time dimension, sampling 4D Gaussians into 3D Gaussians at certain timestamps and incorporating regularizations to ensure the temporal consistency. The figure is adapted from~\citet{kerbl20233d} and~\citet{yangDeformable3DGaussians2023}.} 
    \label{fig:dynamic3dgspipeline}
  \end{figure*}
      
\section{Dynamic 3D Gaussian Splatting}
\label{sec:dynamicRepresentation}
Although 3D-GS is a powerful representation for static 3D scenes, it fails to capture the dynamic nature of the scene. 
To address this limitation, recent research has proposed several dynamic 3D Gaussian representation methods. 
These methods can generally be categorized into three groups:
1) methods that implement a deformation field to modify Gaussians across frames~\cite{liangGauFReGaussianDeformation2023};
2) methods that extend 3D-GS to 4D-GS to represent dynamic scenes~\cite{wu4DGaussianSplatting2024,duan4DRotorGaussianSplatting2024};
3) methods that initialize Gaussians at a certain frame and propagate them to other frames~\cite{sun3DGStreamFlyTraining}.

The pipeline of the first two groups is shown in Figure~\ref{fig:dynamic3dgspipeline}.
In this section, we provide an overview of dynamic 3D Gaussian representation methods.

\subsection{Deformation Field based Methods}
\label{sec:deformation field}
The key idea of deformation field based methods is to deform a set of 3D Gaussians across frames to fit the dynamic scene.
These methods often use an MLP to serve as the deformation network to predict the change of Gaussians.
Many of them assume that only the position ,scale and rotation of the Gaussians change across frames and remain other parameters constant.

\textbf{Deformable 3D-GS}~\cite{yangDeformable3DGaussians2023} is the first to introduce the deformation field to 3D-GS. It extends the 3D-GS to canonical space and leverages a deformation MLP network to learn the offsets of the canonical 3D-GS to the final 3D shape. The deformation network takes timestamps and 3D Gaussian positions as input and predicts the deformation of position, scale, and rotation of 3D Gaussians across frames. The method also uses an Annealing Smooth Training strategy to smooth the rendering jitters caused by inconsistent scenes across frames. 

\textbf{GauFR}~\cite{liangGauFReGaussianDeformation2023} also leverages deformation field. By separating the point clouds into static and dynamic parts through a learnable process, the method only deforms the Gaussians in dynamic part which significantly reduces the computational cost of the deformation network and leads to higher quality overall dynamic view synthesis.
  
\textbf{SC-GS}~\cite{huangSCGSSparseControlledGaussian2023} proposes a spatially constrained 3D-GS representation that extends Gaussians in the canonical space and models the motion through a set of control points. The method leverages an MLP to predict the 6DoF transformation of the control points and deforms the Gaussians according to the computed interpolation weights of neighbouring control points. The method also uses an \textit{ARAP} loss~\cite{10.5555/1281991.1282006} to encourage the control points to be as rigid as possible.

\textbf{Gaussian-Flow}~\cite{linGaussianFlow4DReconstruction2024} introduces a \textit{Dual-Domain Deformation Model} instead of neural network to directly model the dynamics of 3D Gaussians, which integrates both the time domain polynomials and the frequency domain Fourier series. Without any intricate neural field structure, the model can achieve extremely fast training and rendering speed. 

\textbf{DynMF}~\cite{kratimenosDynMFNeuralMotion2023} uses a set of learnable motion basis trajectories to represent the mean position and quaternion of the 3D Gaussians across frames. A small MLP is used to predict a few compact basis trajectories. The method also employs L1 regularization, proposes sparsity loss to force Gaussians to choose fewer trajectories and rigid loss to enhance rigidity.

\textbf{SP-GS}~\cite{wan2024superpointgaussiansplattingrealtime} groups the 3D Gaussians into superpoints \textit{As Rigid As Possible}, using a carefully designed \textit{Property Reconstruction Loss}. Then, the deformation network takes the canonical position of superpoints and time as input to predict the deformation and apply it to the Gaussians within the superpoints. 

\textbf{4D-GS}~\cite{wu4DGaussianSplatting2024} adopts a 4D K-Planes~\cite{fridovichkeil2023kplanesexplicitradiancefields} and a tiny MLP to encode voxel feature from 3D Gaussians to aggregate nearby spatial and temporal Gaussians that share similar spatial and temporal information, and then uses a multi-head Gaussian deformation decoder to compute the change in position, scale and rotation.

\textbf{GPS-Gaussian}~\cite{zhengGPSGaussianGeneralizablePixelwise2023} combines the depth map with two encoders and a Gaussian parameter decoder to predict the 3D Gaussian points, adopting a joint training mechanism to optimize the depth estimation and the Gaussian parameters prediction simultaneously. The method first selects two adjacent source views by conducting a dot product of all input views vectors and the novel view vector. Then, these two views are inputted into a shared image encoder to extract feature maps and conduct binocular depth estimation to get the depth map. The depth map is fed into a depth encoder and fused with image features and the spatial features to predict the scaling map, rotation map, and opacity map, which are used to generate the 3D Gaussians. 

\textbf{D-MiSo}~\cite{waczyńska2024dmiso} takes advantage of Triangle Soup~\cite{1290060} to control the two divided types of 3D Gaussians: CoreGaussians and Sub-Gaussians, where the former encompasses the latter. The CoreGaussians act like the control point of SC-GS that controls the general transformation, while sub-Gaussians contribute more in rendering and modification. The method uses two MLPs to separately predict the general deformation of CoreGaussians and local deformation of the Sub-Gaussians attached to that CoreGaussian. 

\textbf{Video-3DGS}~\cite{shin2024enhancing} focuses on video reconstruction and video editing. For reconstruction, the method first leverages an off-the-shelf open-vocabulary video object segmentation network to conduct \textit{Spatial Decomposition} to extract the foreground moving objects, which are represented by \textit{Frg-3DGS}, and the background points are represented by \textit{Bkg-3DGS}. Then, it conducts \textit{Temporal Decomposition} to divide the video sequence into several clips to ensure that the motion of foreground objects is minimized for the convenience of COLMAP~\cite{7780814} process management. After these two decompositions, the method introduces \textit{MC-COLMAP} to estimate point clouds from videos that recursively invokes COLMAP function on certain clips in order of increasing frames until the return value is ``success". Then, the \textit{Bkg-3DGS} and \textit{Frg-3DGS} can be initialized and deformed by distinct deformation networks. The frame is rendered by the learnable combination of these two kinds of Gaussians.

\textbf{GauMesh}~\cite{xiaoBridging3DGaussian2024} combines 3D-GS with triangle mesh to represent dynamic scenes. It separately initializes the 3D Gaussians and the triangle mesh at the first frame and merges them with alpha composition. Then, the method uses a deformation network to predict the deformation of the 3D Gaussians and the triangle mesh across frames. For dynamic representation, the method applies a multi-resolution Hexplane~\cite{caoHexPlaneFastRepresentation2023} and a shallow MLP to get the features for 3D Gaussians and deforms each vertex in the mesh using a displacement vector to translate it to a new position.

\textbf{E-D3DGS}~\cite{baeGaussianEmbeddingBasedDeformation2024} introduces a novel representation of embedding between the Gaussian band and the MLP for deformation prediction. It assigns each Gaussian a latent learnable embedding to present the unique characteristics of the Gaussian and uses typical temporal embeddings to represent different frames. Then, the MLP takes the latent embedding and temporal embedding at certain frames to predict the change of Gaussian's attributes like position, scale, and rotation. The method also separately conducts deformation for coarse and fine motions~\cite{Feichtenhofer_2019_ICCV}, leveraging a 1D feature grid for N frames and interpolating temporally to decode the embedding for fine deformation. For coarse motions, it just downsamples the feature grid and gets the embedding for coarse deformation. This design helps it better capture the fine details of the dynamic scenes. For regularization, the method proposes \textit{Local Smoothness Regularization} to encourage nearby Gaussians to conduct similar deformation, allowing for accurate capture of textures and fine details.

\textbf{GaussianPrediction}~\cite{2405.19745} intergrates 3D-GS with dynamic scene modeling and future scenario synthesis. For the deformation part, the method encodes the position at a certain frequency and feeds it with a time coefficient into an MLP to predict the deformation of the Gaussians. It also leverages another MLP to derive the lifespan of a Gaussian indirectly to improve rendering quality. For the prediction part, it designs a key point driven framework to deform the whole Gaussians to a certain timestamp without losing original geometric properties. The first step is to initialize the key points that the method takes a clustering techniques to get the key points that drive certain class of Gaussians with motion similarity and spatial proximity. Then, it needs to add key points by localizing the place with a greater Gaussian gradient norm. The last step is to learn the weight that each key point contributes to Gaussians to conduct deformation, which is a time-independent learning process. Then a Graph Convolution Network is used to extract relational features between key points across frames, which passes through a single-layer MLP to predict the future motion and calculate the predicted Gaussians.

\textbf{GaGS}~\cite{lu3DGeometryawareDeformable2024} proposes a geometry-aware feature extraction network to better capture the geometric information of the scene. After initializing the 3D Gaussians at the first frame, the method extracts the features in two branches, the first is to use the 3D coordinates of the 3D Gaussians as point clouds to extract the positional features through an MLP and the second is to transform the point clouds into voxels and extract the local geometric features using a 3D U-Net. Then, these two kinds of features are concatenated and fed into an MLP to form the geometry-aware features, which are later used to predict the deformation of the 3D Gaussians at certain timestamps using the MLP. It is worth noting that this method changes the quaternion representation of rotation of 3D Gaussians to 6D rotation representation, helping the neural networks to learn smooth rotation variation.

\subsection{4D Primitive based Methods}
\label{sec:4D primitive}
4D primitive based methods extend the formulation of 3D-GS and integrate the time dimension to represent dynamic scenes.
For rendering, these methods often sample the 4D Gaussians into 3D Gaussians at certain timestamps to take advantage of the tile-based rasterizer.

\textbf{RealTime4DGS}~\cite{yangRealtimePhotorealisticDynamic2023} is the first to reformulate 3D-GS into 4D-GS. The method treats time dimension and space dimension equally, using a 4D scaling matrix and a pair of isotropic rotations~\cite{cayley1894collected} to represent 4D rotation. Then, the 4D Gaussians can be conditioned on time multiplied by the marginal distribution to render the dynamic scene.

\textbf{4D-Rotor GS}~\cite{duan4DRotorGaussianSplatting2024} introduces 4D rotors~\cite{bosch2020n} to characterize the 4D rotation, which offers a well-defined, interpretable rotation representation that can model both static and dynamic scenes. For sliced 3D Gaussians, the method adds a temporal decay term to adapt to challenging dynamics. Moreover, the method introduces an entropy loss to encourage the opacity of Gaussians to approach either one or zero, and a 4D spatiotemporal consistency loss to enforce consistent motion among neighboring Gaussians.

\textbf{STG}~\cite{liSpacetimeGaussianFeature2024} proposes a temporal radial basis function to model the temporal dynamics of 3D Gaussians. The method uses a polynomial to model the position and rotation of space Gaussians. Instead of using spherical harmonics coefficients to represent color, STG employs a two-layer MLP to predict the temporal color features.

\subsection{Per-frame Training Methods}
\label{sec:per-frame training}
This category of methods initializes 3D Gaussians at certain frames and trains 3D Gaussians at other frames based on the former frames. 
This may be a little similar to the deformation field based methods, but the deformation field based methods often predict the deformation of Gaussians across frames, while the per-frame training methods directly train the Gaussians at each frame.
Methods in this category demand multiple cameras as input to provide the 3D information of the scene~\cite{yangDeformable3DGaussians2023}. 

\textbf{Dynamic 3D-GS}~\cite{luitenDynamic3DGaussians2023} initializes the 3D Gaussians at the first timestamp and conducts static reconstruction of them. The 3D Gaussians have fixed parameters except for their position and rotation, which enables it to track the points across frames. The optimization is conducted by regularization without leveraging neural network. 

\textbf{3DGStream}~\cite{sun3DGStreamFlyTraining} initializates 3D Gaussians at the first frame and creates a Neural Transformation Cache (NTC) to conduct the transformation of 3D Gaussians across frames. NTC is a shallow, fully fused MLP that utilizes multi-resolution hash encoding to enhance compactness, efficiency, adaptability, and prior incorporation. The method also introduces an adaptive 3DG addition strategy that dynamically adds new 3D Gaussians for emerging objects while regulating their overall quantity.

\textbf{DualGS}~\cite{jiang2024robust} focuses on Human-centric Volumetric Videos, using a set of motion-aware joint Gaussians to represent global motions and a larger number of appearance-aware skin Gaussians for visual representation. It employs a coarse-to-fine optimization strategy to separately optimize the joint Gaussians and skin Gaussians. The first stage focuses on the motion and uses a locally as-rigid-as-possible regularizer to regulate the joint Gaussians. Then, it conducts the fine-grained optimization to jointly optimize the joint Gaussians and skin Gaussians to balance the rendering quality and temporal consistency. 

\textbf{V\textsuperscript{3}}~\cite{wang2024v3viewingvolumetricvideos} centralizes on the compactness of 3D Gaussians on mobile devices. For representation, the method uses 2D Gaussian Video to represent the attributes like scale, rotation, and position of 3D Gaussians. During training, the method partitions the frames into groups to facilitate topological transformations and handle the infinite length of the dynamic scene. Then, it initializes the 3D Gaussians at the first frame with NeuS2~\cite{wang2023neus2fastlearningneural} and point cloud. For regularization, the method introduces \textit{Residual Entropy Loss} to encourage repeated values and \textit{Temporal Loss} to enhance the temporal consistency. The paper also provides a player for mobile devices to view the volumetric videos.

\textbf{NPGs}~\cite{dasNeuralParametricGaussians2024} is an object reconstruction method that adopts a two-stage approach to simplify the reconstruction process. In the first stage, the method initializes a neural parametric point model frame by frame to capture the general geometric structure and deformation of the object. The model is represented as the sum of a learnable point basis multiplied by low-rank coefficients produced by MLP.  The second stage is to use the model to drive the Gaussians to conduct reconstruction. The Gaussians are stored in local volumes, which are local unstructured point sets to reflect the local non-rigid deformation of the object. Thus, Gaussians only need to follow the motion of the point basis to reconstruct the object.

\section{Volumetric Video Representation and Optimization of Streaming Transmission}
\label{sec:streamopt}
Volumetric video representation, or Free-Viewpoint Video(FVV), is a promising technology that captures the 3D scene from multiple viewpoints and enables the viewer to navigate the scene in real time, which is viewed as the next significant advancement in media production~\cite{EISERT2023289}.
It differs from common dynamic 3D scene representations in its critical need for reducing the storage size and transmission bandwidth~\cite{zhangRateawareCompressionNeRFbased2024}.
The main goal of this kind of work is to balance bandwidth efficiency and high visual quality.~\cite{liuNextgenerationVolumetricVideo2023}.
Going a step further, the ultimate goal is to make producing and viewing FVVs as seamless as streaming regular 2D videos.~\cite{wangNeuralResidualRadiance2023}.
The difficulty in volumetric video rendering and streaming lies in the following aspects:
1)Efficient encoding and compression strategy: the reconstructed 3D scenes for volumetric video are usually large in size, posing a challenge for how to encode and compress the data efficiently and requiring certain streaming techniques.
2)High-quality rendering: for XR applications, immersive experience is crucial, which requires high-quality rendering of the scenes.
In this section, we will introduce the recent works on the framework of volumetric video representation and transmission, especially focusing on the dynamic NeRF based methods and dynamic 3D-GS based methods.

\begin{figure}[!t]\centering
	\includegraphics[width=\columnwidth]{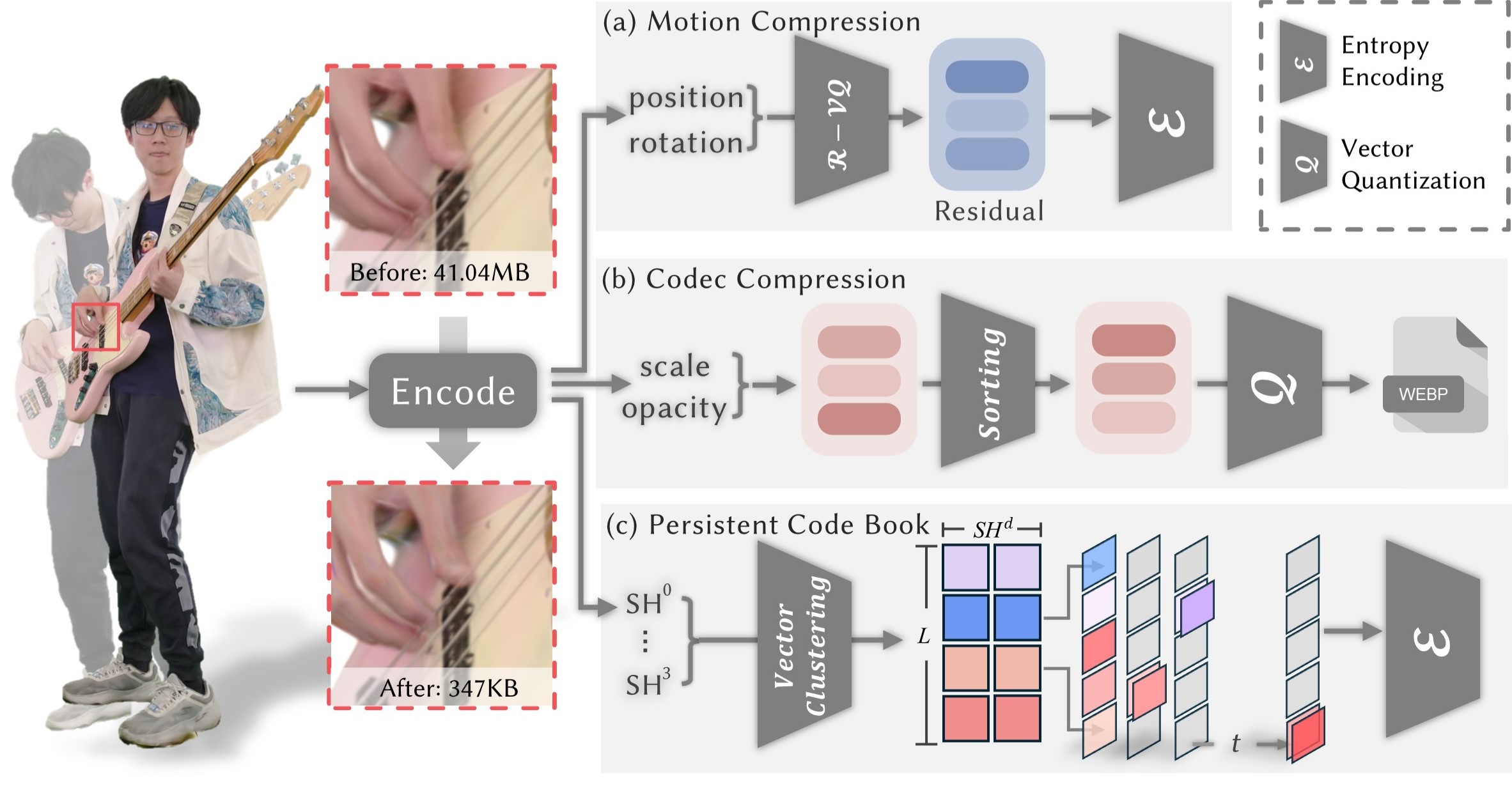}
	\caption{\textbf{Hybrid compression strategy of DualGS.} The image is courtesy of~\citet{jiang2024robust}.}
    \label{fig:compression_dualgs}
\end{figure}

\subsection{Optimization of Dynamic NeRF Modeling}
\label{subsec:streamnerf}
In the realm of volumetric video research, multiple works have focused on enhancing the NeRF representation to achieve efficient compression and high-quality rendering. 

\citet{zhangEfficientDynamicNeRFBased2024a} decomposes the NeRF representation into coefficient fields and basis fields. 
It incrementally updates the basis fields temporally for dynamic modeling and introduces end-to-end joint optimization with simulated quantization and rate estimation, thereby enhancing the rate-distortion performance. 

\textbf{ReRF}~\cite{wangNeuralResidualRadiance2023} employs a tiny global MLP along with compact motion and residual grids to model dynamic scenes, enabling efficient volumetric video rendering with high compression rates. 

\citet{zhangRateawareCompressionNeRFbased2024} develops a rate-aware compression framework. By incorporating adaptive quantization and a spatial-temporal implicit entropy model for accurate bitrate estimation, it optimizes the compact representation and achieves improved compression.

\textbf{HumanRF}~\cite{10.1145/3592415} presents a spatio-temporal radiance field achieved through a low-rank decomposition and adaptive temporal partitioning, allowing for efficient reconstruction of human motions.

\textbf{JointRF}~\cite{zhengJointRFEndtoEndJoint2024} uses a compact residual feature grid and coefficient feature grid to represent dynamic NeRF. It introduces sequential feature compression and end-to-end joint optimization to reduce redundancy and enhance rate-distortion performance.

\textbf{Compactvv}~\cite{guoCompactNeuralVolumetric2023} addresses the high storage cost of volumetric videos. It proposes a dynamic codebook representation, which reduces feature redundancy in time and space through codebook compression and compensates for rendering quality loss with dynamic codes, achieving high compression rates while maintaining quality. 

\textbf{TeTriRF}~\cite{wuTeTriRFTemporalTriPlane2024} achieves efficient volumetric video generation and rendering with reduced storage requirements using a hybrid representation of tri-planes and voxel grids, along with a grouping training scheme and a specific compression pipeline including value quantization, removal of empty spaces, conversion into 2D serialization, and subsequent video encoding.

\textbf{VideoRF}~\cite{wangVideoRFRenderingDynamic2024} converts the 4D feature volumes of dynamic scenes into 2D feature image streams. With a specialized rendering pipeline, training scheme, and developed player, it enables real-time streaming and rendering of dynamic radiance fields on mobile devices, enhancing the accessibility and quality of volumetric video experiences.
 
These works have collectively made significant contributions to advancing the state-of-the-art in NeRF-based volumetric video compression and rendering.

\subsection{Optimization of Dynamic 3D-GS Modeling}
\label{subsec:stream3dgs}
For 3D-GS methods, the explicit representation of the 3D scenes makes it easier to be encoded and compressed, and its high-quality and real-time rendering capability makes it a promising candidate for volumetric video rendering and streaming. 
However, the storage and bandwidth requirement of the 3D-GS methods are higher due to this explicit representation. All of the current 3D-GS based volumetric video rendering and streaming methods employ the per-frame 3D reconstruction strategy.

\textbf{GauMesh}~\cite{xiaoBridging3DGaussian2024} combines 3D Gaussian and triangle mesh primitives. After initializing with a tracked mesh, it jointly optimizes multiple aspects and fuses the two representations through a specific rendering process, achieving high-quality view synthesis of dynamic scenes and adapting well to different scene parts.

\textbf{DualGS}~\cite{jiang2024robust} uses a dual-Gaussian representation, initializes and anchors the Gaussians, and employs a coarse-to-fine training strategy. Additionally, a compression strategy is designed to achieve high-fidelity rendering and a 120-fold compression ratio for immersive experiences in VR environments, as shown in Fig.~\ref{fig:compression_dualgs}.

\textbf{MGA}~\cite{sunMultiframeBitrateAllocation2024a} addresses the bitrate allocation problem for streaming dynamic 3D Gaussians. It identifies four encoding parameters and proposes the MGA algorithm to search for optimal parameters across multiple frames, along with the adaptive MGAA algorithm to reduce computational time.

\textbf{3DGStream}~\cite{sun3DGStreamFlyTraining} focuses on efficient FVV streaming. It uses an NTC to model the transformations of 3D Gaussians, along with an adaptive 3DG addition strategy. Through a two-stage pipeline, it enables on-the-fly dynamic scene reconstruction and real-time rendering with moderate storage requirements.

\textbf{V\textsuperscript{3}}~\cite{wang2024v3viewingvolumetricvideos} transforms dynamic 3D Gaussians into a 2D video format compatible with hardware video codecs, enabling efficient streaming and decoding. It employs a two-stage training strategy: first, using hash encoding and shallow MLP to learn motion and prune Gaussians for streaming; then, fine-tuning to enhance temporal continuity.

\textbf{TGH}~\cite{10.1145/3687919} addresses the challenge of memory footprint and computational complexity by introducing a hierarchical structure that models different scales of motion and dynamics in a scene using 4D Gaussian primitives. This approach allows for adaptive sharing of Gaussian primitives across various temporal segments, reducing the overall model size. The paper also proposes a \textit{Compact Appearance Model} that leverages sparse spherical harmonics to capture view-dependent effects with reduced storage.

\section{Datasets for dynamic scenes}
\label{sec:datasets}
\subsection{Monocular Datasets}
\label{sec:monocular_datasets}
For monocular datasets, D-NeRF~\cite{pumarola2020d}, HyperNerf~\cite{park2021hypernerf} and NeRF-DS~\cite{yan2023nerf} datasets 
are three widely used public datasets for dynamic scenes.

\textbf{D-NeRF} dataset contains 8 synthetic scenes captured in different camera poses at the resolution of 800x800, each with a certain dynamic object. The scenes involve large motion and rigidly deforming objects, such as a lego, balls, and jumping characters.

\textbf{HyperNerf} dataset includes 17 dynamic scenes captured by one or two cameras with consistent camera motions, focusing on topologically varying scenes with rigid and non-rigidly deforming objects without exaggerated frame-to-frame motion. The dataset provides the ground truth depth maps and camera poses for each frame. The most widely used scenes in this dataset are the four \textit{vrig} scenes exhibiting topological changes.

\textbf{NeRF-DS} is a dynamic specular scene dataset containing 8 scenes with various types of moving or rigidly deforming specular objects. 
The scenes are captured by two rigidly mounted cameras and one camera is used for training and the other for testing.

\begin{table}
    \caption{\textbf{Performance comparison of NVS task on D-NeRF~\cite{pumarola2020d}
    datasets.} $^{\star}$denotes the dataset is rendered at the resolution of 800
    $\times$ 800. }
    \label{tab:dnerf}
    
    \centering
    \begin{tabular}{llll}
        \toprule Method                                                    & PSNR↑          & SSIM↑          & LPIPS↓         \\
        \midrule SC-GS~\cite{huangSCGSSparseControlledGaussian2023}        & \textbf{43.31} & \textbf{0.997} & \textbf{0.006} \\
        $^{\star}$Deformable 3D-GS~\cite{yangDeformable3DGaussians2023}    & 39.51          & 0.991          & 0.012          \\
        $^{\star}$SP-GS~\cite{wan2024superpointgaussiansplattingrealtime}  & 37.98          & 0.988          & 0.019          \\
        DynMF~\cite{kratimenosDynMFNeuralMotion2023}                       & 36.89          & 0.983          & 0.020          \\
        GauFRe~\cite{liangGauFReGaussianDeformation2023}                   & 34.80          & 0.982          & 0.020          \\
        $^{\star}$4D-GS~\cite{wu4DGaussianSplatting2024}                   & 34.05          & 0.980          & 0.020          \\
        $^{\star}$ReaTime4DGS~\cite{yangRealtimePhotorealisticDynamic2023} & 34.09          & 0.980          & 0.020          \\
        4D Rotor GS~\cite{duan4DRotorGaussianSplatting2024}                & 34.26          & 0.970          & 0.030          \\
        Gaussian-Flow~\cite{linGaussianFlow4DReconstruction2024}           & 34.27          & 0.980          & 0.030          \\
        V4D~\cite{ganV4DVoxel4D2024}                                       & 33.72          & 0.980          & 0.020          \\
        Hy-DNeRF~\cite{chenMultiResolutionHybridExplicit2024}              & 33.71          & 0.976          & 0.019          \\
        $^{\star}$TiNeuVox-B~\cite{fangFastDynamicRadiance2022}            & 32.67          & 0.970          & 0.040          \\
        K-Planes~\cite{fridovich-keilKPlanesExplicitRadiance2023a}         & 32.32          & 0.973          & 0.038          \\
        HexPlane~\cite{caoHexPlaneFastRepresentation2023}                  & 31.04          & 0.970          & 0.040          \\
        D-NeRF~\cite{pumarola2020d}                                        & 31.69          & 0.975          & 0.058          \\
        D-TensoRF~\cite{jangDTensoRFTensorialRadiance2022}                 & 30.68          & 0.960          & 0.030          \\
        \bottomrule
    \end{tabular}
\end{table}

\subsection{Multi-view Datasets}
\label{sec:multi-view_datasets}
For multi-view datasets, there exist a variety of datasets for dynamic scenes, among which the Plenoptic video dataset~\cite{li2022neural3dvideosynthesis} is the most widely used dataset. 

\textbf{Plenoptic} video dataset contains 9 real-world dynamic scenes focusing on different video synthesis challenges, including~\cite{li2022neural3dvideosynthesis}:
1)High Reflectivity, Translucency, and Transparency
2)Dynamic Scene Changes and Motions
3)Self-Cast Moving Shadows
4)Volumetric Effects
5)Complex View-Dependent Effects
6)Variable Lighting Conditions
7)Multiple People Moving in Open Spaces, Viewed Through Transparent Windows Under Dim Indoor Lighting.
The videos are captured by 15-21 cameras facing the central character in the scene in a time-synchronized manner at the fps of 30.

\textbf{ReRF} dataset~\cite{wangNeuralResidualRadiance2023} contains 3 released scenes with 74 cameras around the character in the center. The dataset is challenging in the human motion.

\textbf{Actors-HQ}~\cite{10.1145/3592415} is distinguished by its capture of high-fidelity dynamic human motion at an unprecedented resolution of 12MP using a rig of 160 cameras. The dataset encompasses a diverse range of actions performed by 4 female and 4 male actors clad in everyday attire, which adds variability to the cloth dynamics.

\textbf{Google Immersive} dataset~\cite{broxton2020immersive} is another multi-view dataset for dynamic scenes that contains 15 scenes with 46 cameras or fewer.
The dataset also provides detailed camera calibration information produced by SfM algorithms~\cite{7780814}.

\textbf{Meet Room} dataset~\cite{li2022streaming} includes 4 scenes with 13 cameras in total. The dataset is composed of videos at 30 fps with the resolution of 1280x720, which last for 10 seconds.

\textbf{HiFi4G} dataset~\cite{Jiang_2024_CVPR} is an informative multi-view dataset focusing on human motions. The dataset contains 6 scenes with 81 cameras, covering a wide range of complex human motions, such as \textit{playing instruments}, \textit{dancing}, and \textit{changing clothes}.

\textbf{Particle-NeRF} dataset~\cite{abou2022particlenerf} is a synthetic dataset using Blender to generate dynamic scenes. It publishes 6 scenes with one or multiple objects captured by 20 cameras distributed on the top half of the sphere.

\begin{table}
    \caption{\textbf{Performance comparison of NVS task on NeRF-DS~\cite{yanNeRFDSNeuralRadiance2023}
    dataset.}$^{\star}$ Deformable 3D-GS~\cite{yangDeformable3DGaussians2023}
    and SP-GS~\cite{wan2024superpointgaussiansplattingrealtime} use metric SSIM.
    $^{\dagger}$ denotes that they employ LPIPS with VGG.}

    \centering
    \label{tab:nerfds}
    \begin{tabular}{llll}
        \toprule Method                                                   & PSNR↑          & MS-SSIM↑        & LPIPS↓            \\
        \midrule Deformable 3D-GS~\cite{yangDeformable3DGaussians2023}    & \textbf{24.10} & 0.852$^{\star}$ & 0.177$^{\dagger}$ \\
        GauFRe~\cite{liangGauFReGaussianDeformation2023}                  & 23.80          & 0.887           & 0.144             \\
        SC-GS~\cite{huangSCGSSparseControlledGaussian2023}                & \textbf{24.10} & 0.891           & 0.140             \\
        Nerfies~\cite{Park_2021_ICCV}                                     & 20.10          & 0.707           & 0.349             \\
        HyperNeRF~\cite{parkHyperNeRFHigherDimensionalRepresentation2021} & 22.50          & 0.827           & 0.206             \\
        NeRF-DS~\cite{yanNeRFDSNeuralRadiance2023}                        & 23.90          & \textbf{0.898}  & \textbf{0.127}    \\
        TiNeuVox-B~\cite{fangFastDynamicRadiance2022}                     & 21.70          & 0.818           & 0.219             \\
        SP-GS~\cite{wan2024superpointgaussiansplattingrealtime}           & 23.15          & 0.834$^{\star}$ & 0.206$^{\dagger}$ \\
        \bottomrule
    \end{tabular}
\end{table}

\begin{table}
    \caption{\textbf{Performance comparison of NVS task on HyperNeRF~\cite{parkHyperNeRFHigherDimensionalRepresentation2021}
    dataset.}$^{\star}$DynMF~\cite{kratimenosDynMFNeuralMotion2023} use metric SSIM.
    }

    \centering
    \label{tab:hypernerf}
    \begin{tabular}{lll}
        \toprule Method                                                   & PSNR↑          & MS-SSIM↑        \\
        \midrule Gaussian-Flow~\cite{linGaussianFlow4DReconstruction2024}  & 26.3           & 0.862           \\
        4D-GS~\cite{wu4DGaussianSplatting2024}                            & 25.20          & 0.845           \\
        DynMF~\cite{kratimenosDynMFNeuralMotion2023}                      & 25.27          & 0.750$^{\star}$ \\
        HyperNeRF~\cite{parkHyperNeRFHigherDimensionalRepresentation2021} & 23.47          & 0.856           \\
        Nerfies~\cite{Park_2021_ICCV}                                     & 22.10          & 0.803           \\
        NeRFPlayer~\cite{10049689}                                        & 23.70          & 0.803           \\
        TiNeuVox-B~\cite{fangFastDynamicRadiance2022}                     & 24.30          & 0.837           \\
        V4D~\cite{ganV4DVoxel4D2024}                                      & 24.80          & 0.832           \\
        Neural Scene Flow Fields~\cite{liNeuralSceneFlow2021}             & 26.40          & 0.917           \\
        SP-GS~\cite{wan2024superpointgaussiansplattingrealtime}           & \textbf{26.78} & \textbf{0.892}  \\
        Hy-DNeRF~\cite{chenMultiResolutionHybridExplicit2024}             & 22.90          & 0.718           \\
        \bottomrule
    \end{tabular}
\end{table}

\begin{table*}
    [tp]
    \centering
    \caption{\textbf{Performance comparison of NVS task on Plenoptic Video
    dataset~\cite{li2022neural3dvideosynthesis}.} $^{\star}$HexPlane~\cite{caoHexPlaneFastRepresentation2023}
    reports average results excluding the coffee-martini scene and is rendered
    at the resolution of 1024 $\times$ 768.}
    \label{tab:plenoptic}
    \begin{tabular}{lllllll}
        \toprule Scenes                        & Method                                                      & PSNR↑          & D-SSIM↓        & LPIPS↓         & FPS↑            \\
        \midrule \multirow{9}{*}{Flame Salmon} & 4D-GS~\cite{wu4DGaussianSplatting2024}                      & 31.02          & 0.030          & 0.150          & 36.00           \\
                                               & RealTime4DGS~\cite{yangRealtimePhotorealisticDynamic2023}   & \textbf{32.01}          & \textbf{0.014} & \textbf{0.055}          & 114.00          \\
                                               & 4D-Rotor GS~\cite{duan4DRotorGaussianSplatting2024}         & 31.62          & 0.030          & 0.140          & \textbf{277.47} \\
                                               & DyNeRF~\cite{liNeural3DVideo2022a}                          & 29.58          & 0.020          & 0.083          & -               \\
                                               & HexPlane~\cite{caoHexPlaneFastRepresentation2023}           & 29.47          & 0.018          & 0.078          & -               \\
                                               & STG~\cite{liSpacetimeGaussianFeature2024}                   & 29.48          & 0.022          & 0.063          & 103.00          \\
                                               & DynMF~\cite{kratimenosDynMFNeuralMotion2023}                & 31.70          & 0.020          & 0.110          & 130.00          \\
                                               & MSTH~\cite{NEURIPS2023_df311263}                            & 29.92          & 0.020          & 0.063          & -               \\
        \midrule \multirow{8}{*}{All Scenes}   & HexPlane$^{\star}$~\cite{caoHexPlaneFastRepresentation2023} & 31.71          & \textbf{0.014} & 0.075          & 0.20            \\
                                               & Gaussian-Flow~\cite{linGaussianFlow4DReconstruction2024}     & 32.00          & 0.015          & -              & -               \\
                                               & K-Planes~\cite{fridovich-keilKPlanesExplicitRadiance2023a}  & 31.63          & 0.018          & -              & 0.30            \\
                                               & 4D-GS~\cite{wu4DGaussianSplatting2024}                      & 31.15          & 0.016          & \textbf{0.049} & \textbf{30.00}           \\
                                               & MSTH~\cite{NEURIPS2023_df311263}                            & \textbf{32.37} & 0.015          & 0.056          & 2.00            \\
                                               & NerfPlayer~\cite{10049689}                                  & 30.69          & 0.034          & 0.111          & 0.05            \\
                                               & TeTriRF~\cite{wuTeTriRFTemporalTriPlane2024}                & 30.43          & 0.047          & 0.248          & -               \\
        \bottomrule
    \end{tabular}
\end{table*}

\section{Performance Comparison}
\label{sec:performance}
\subsection{Metrics}
To evaluate the performance of dynamic 3D-GS methods, there exist several
commonly used metrics including the following: 1) PSNR 2) SSIM 3) LPIPS 4) FPS 5)
Training Time.

\textbf{PSNR} (Peak Signal-to-Noise Ratio) is a commonly used metric to quantify
the difference between the quality of the reconstructed 3D scenes and the ground
truth image~\cite{wiki:Peak_signal-to-noise_ratio}. A higher PSNR value indicates better quality of the reconstructed 3D scenes.

\begin{equation}
    \begin{aligned}
        {\mathit{MSE}} & ={\frac{1}{m\,n}}\sum_{i=0}^{m-1}\sum_{j=0}^{n-1}[I(i,j)-K(i,j)]^{2}.        \\
        \mathit{PSNR}  & =10\cdot \log_{10}\left({\frac{{\mathit{MAX}}_{I}^{2}}{\mathit{MSE}}}\right) \\
                       & =20\cdot \log_{10}({\mathit{MAX}}_{I})-10\cdot \log_{10}({\mathit{MSE}}).
    \end{aligned}
\end{equation}
where $\mathit{MSE}$ is the mean squared error between the ground truth image $I$
and the reconstructed image $K$, and $\mathit{MAX}_{I}$ is the maximum possible
pixel value of the image.

\textbf{SSIM} (Structural Similarity) is a metric that measures the perceived similarity
between the ground truth image and the reconstructed image. SSIM ranges from -1 to
1, where 1 indicates perfect similarity between the two images. SSIM has several
variants, including Multiscale SSIM (MS-SSIM)~\cite{wang2003multiscale} and Structural
dissimilarity (DSSIM). The SSIM of two images $x$ and $y$ is defined as follows:
\begin{equation}
    { \displaystyle {\hbox{SSIM}}(x,y)={\frac{(2\mu_{x}\mu_{y}+c_{1})(2\sigma_{xy}+c_{2})}{(\mu_{x}^{2}+\mu_{y}^{2}+c_{1})(\sigma_{x}^{2}+\sigma_{y}^{2}+c_{2})}}}
    ,
\end{equation}
where $\mu_{x}$ and $\mu_{y}$ are the average pixel values of images $x$ and $y$,
$\sigma_{x}^{2}$ and $\sigma_{y}^{2}$ are the variances of images $x$ and $y$, $\sigma
_{xy}$ is the covariance of images $x$ and $y$, and $c_{1}$ and $c_{2}$ are
constants to stabilize the division with weak denominator~\cite{wiki:Structural_similarity_index_measure}.

\textbf{LPIPS} (Learned Perceptual Image Patch Similarity) is a metric that measures
the perceptual similarity between the ground truth image and the reconstructed image
through a deep neural network, which can better capture the human perception of
image quality than PSNR and SSIM~\cite{zhang2018unreasonable}.

\textbf{VMAF} (Video Multi-method Assessment Fusion) is a metric that measures the
quality of the reconstructed 3D scenes by fusing the results of multiple quality
metrics using Support Vector Machine.

\begin{table}[tp]
    \caption{\textbf{Performance comparison of volumetric video rendering on
    ReRF~\cite{wangNeuralResidualRadiance2023} dataset.}}

    \centering
    \label{tab:ReRFfvv}
    \resizebox{\linewidth}{!}{
    \begin{tabular}{llllll}
        \toprule Method                                               & PSNR↑          & SSIM↑          & LPIPS↓         & Train(min)↓   & Storage(MB)↓   \\
        \midrule 3DGStream~\cite{sun3DGStreamFlyTraining}             & 27.26          & 0.960          & -              & \textbf{0.12} & 7.644          \\
        ReRF~\cite{wangNeuralResidualRadiance2023}                    & 31.84          & 0.974          & 0.042          & -             & 0.645          \\
        VideoRF~\cite{wangVideoRFRenderingDynamic2024}                & 32.01          & 0.976          & \textbf{0.023} & 20            & 0.658          \\
        V\textsuperscript{3}~\cite{wang2024v3viewingvolumetricvideos} & \textbf{32.97} & \textbf{0.983} & -              & 0.82          & 0.532          \\
        JointRF~\cite{zhengJointRFEndtoEndJoint2024}                  & 33.23          & 0.980          & -              & -             & 0.470          \\
        TeTriRF~\cite{wuTeTriRFTemporalTriPlane2024}                  & 30.18          & 0.962          & 0.056          & 0.58          & \textbf{0.070} \\
        HumanRF~\cite{10.1145/3592415}                                & 28.82          & 0.900          & 0.050          & -             & 2.800          \\
        \bottomrule
    \end{tabular}
    }
\end{table}

\subsection{Scene Reconstruction and Novel View Synthesis}
\label{sec:scene_reconstruction} Scene reconstruction is a fundamental problem
in computer vision and graphics. It is the process of creating a 3D model of a scene
to demonstrate the geometry, motion and visual appearance of the objects
involved in the scenes~\cite{chen2024survey3dgaussiansplatting} from a set of
images or videos. As for Novel View Synthesis (NVS), it is the process of
synthesizing the appointed view of a scene from the given images or videos. Due to
the explicit representation of 3D-GS~\cite{kerbl20233d}, in dynamic 3D-GS, the scene
reconstruction and NVS tasks are usually combined together that scene reconstruction
is the foundation of NVS.

For monocular dynamic scene reconstruction, we choose the D-Nerf dataset~\cite{pumarola2020d},
HyperNerf Dataset~\cite{yanNeRFDSNeuralRadiance2023} and NeRF-DS dataset~\cite{parkHyperNeRFHigherDimensionalRepresentation2021}
to evaluate the performance of dynamic 3D-GS methods. We collect the PSNR, SSIM,
and LPIPS of the methods from the papers and summarize them in Table~\ref{tab:dnerf},
\ref{tab:nerfds}, \ref{tab:hypernerf}. It is worth noting that not all the methods
are rendering at the same resolution, which may affect the performance of these
metrics. Also, the training and rendering process on different GPUs will affect
the training time and FPS. Due to less focus on storage of the monocular dynamic
scene reconstruction, we do not include this metric in the table.

On the D-NeRF dataset, most of the methods are evaluated at the resolution of 800
$\times$ 800, while Deformable 3D-GS~\cite{yangDeformable3DGaussians2023}, SP-GS~\cite{wan2024superpointgaussiansplattingrealtime},
4D-GS~\cite{wu4DGaussianSplatting2024}, and ReaTime4DGS~\cite{yangRealtimePhotorealisticDynamic2023}
are evaluated at the resolution of 800 $\times$ 800.

SC-GS~\cite{wan2024superpointgaussiansplattingrealtime} demonstrates superior
performance with the highest PSNR and SSIM, and the lowest LPIPS, indicating its
ability to generate high-quality novel views with minimal perceptual loss. In contrast,
Deformable 3D-GS~\cite{yangDeformable3DGaussians2023} and SP-GS~\cite{wan2024superpointgaussiansplattingrealtime},
while still competitive, show slightly higher LPIPS values, suggesting a greater
perceptual difference from the original scenes, though at a higher resolution.

Shifting to the NeRF-DS dataset, Deformable 3D-GS~\cite{yangDeformable3DGaussians2023}
and SC-GS~\cite{huangSCGSSparseControlledGaussian2023} again emerge as top performers,
with SC-GS maintaining its lead in PSNR and SSIM, and Deformable 3D-GS showing a
significant LPIPS score when employing VGG, which indicates a trade-off in
perceptual quality. The focus of this dataset on real-world scenes with dynamic specular objects presents unique challenges, which are reflected in the performance
metrics.

On the HyperNeRF~\cite{parkHyperNeRFHigherDimensionalRepresentation2021} dataset,
all the methods are evaluated at the resolution of 540 $\times$ 960 or 536 $\times$
960. Among them, the method SP-GS~\cite{wan2024superpointgaussiansplattingrealtime}
stands out with the highest PSNR and MS-SSIM, showcasing its robustness in handling
higher-dimensional representations and generating accurate novel views. The
performance of Gaussian-Flow~\cite{linGaussianFlow4DReconstruction2024} and
Neural Scene Flow Fields~\cite{liNeuralSceneFlow2021} is also noteworthy, with Gaussian-Flow
achieving a competitive PSNR and MS-SSIM, and Neural Scene Flow Fields leading
in MS-SSIM, highlighting the effectiveness of these methods in capturing scene dynamics.

For multi-view dynamic scene reconstruction, we collect the PSNR, D-SSIM, SSIM, LPIPS,
FPS, on the Plenoptic Video~\cite{li2022neural3dvideosynthesis} dataset of the methods
from the papers and summarize in Table~\ref{tab:plenoptic}, the performance
of various methods is assessed across different scenes at resolutions of 540
$\times$ 960 or 536 $\times$ 960. In the \textit{Flame Salmon} scene,
RealTime4DGS~\cite{yangRealtimePhotorealisticDynamic2023} stands out with the
highest FPS, showcasing its real-time rendering capability. 4D-Rotor GS~\cite{duan4DRotorGaussianSplatting2024}
also demonstrates robust performance with a notable FPS, highlighting its efficiency
in managing higher-dimensional representations. In terms of image quality
metrics, 4D-GS~\cite{wu4DGaussianSplatting2024} achieves a competitive standing with
strong PSNR values. And for methods evaluated at all the scenes of the dataset, MSTH~\cite{NEURIPS2023_df311263}
excels with the highest PSNR and competitive LPIPS scores, indicating its
ability to generate high-fidelity imagery across various scenes. 

These findings underscore the importance of advanced NVS techniques in producing accurate and dynamic novel views, especially in the context of complex scene representations.

\subsection{Volumetric Video Rendering and Streaming}
\label{sec:fvv} Volumetric video rendering and streaming is sensitive to the
quality of the reconstructed 3D scenes, the compression and encoding strategy,
the rendering speed, and the storage requirement. We collect the PSNR, SSIM,
LPIPS, FPS, Training time, and per-frame Storage of the methods and their test datasets
from the papers and summarize them in Table~\ref{tab:ReRFfvv} and Table~\ref{Actors-HQfvv}.
The task is evaluated on ReRF~\cite{wangNeuralResidualRadiance2023} dataset and
Actors-HQ~\cite{10.1145/3592415} dataset.

On the ReRF dataset, the performance comparison for free-view video rendering reveals
that V\textsuperscript{3}~\cite{wang2024v3viewingvolumetricvideos} excels in PSNR
and SSIM metrics, while JointRF~\cite{zhengJointRFEndtoEndJoint2024} stands out
for its efficiency in storage requirements. ReRF~\cite{wangNeuralResidualRadiance2023}
performs best in the LPIPS metric, indicating its strength in perceptual
similarity. Although TeTriRF~\cite{wuTeTriRFTemporalTriPlane2024} lags slightly in
LPIPS, it has the lowest storage demand per frame, highlighting its efficiency in
storage management.

\begin{table}[tp]
    \caption{\textbf{Performance comparison of volumetric video rendering on
    Actors-HQ~\cite{10.1145/3592415} dataset.}}
    \centering
    \label{Actors-HQfvv} \resizebox{\linewidth}{!}{
    \begin{tabular}{llllll}
        \toprule Method                                               & PSNR↑          & SSIM↑          & LPIPS↓ & Train(min)↓ & Storage(MB)↓ \\
        \midrule ReRF~\cite{wangNeuralResidualRadiance2023}           & 28.33          & 0.836          & 0.296  & -           & 0.554        \\
        VideoRF~\cite{wangVideoRFRenderingDynamic2024}                & 28.46          & 0.838          & 0.278  & 20.00         & 0.426        \\
        HumanNeRF~\cite{10.1145/3592415}                              & 30.14          & \textbf{0.966} & -      & 1.74        & 8.225        \\
        V\textsuperscript{3}~\cite{wang2024v3viewingvolumetricvideos} & \textbf{32.28} & 0.946          & -      & 0.89        & \textbf{0.513}        \\
        3DGStream~\cite{sun3DGStreamFlyTraining}                      & 27.34          & 0.856          & -      & \textbf{0.19}        & 7.629        \\
        \bottomrule
    \end{tabular}
    }
\end{table}

On the Actors-HQ dataset, V\textsuperscript{3}~\cite{wang2024v3viewingvolumetricvideos}
once again achieves the highest scores in PSNR and SSIM, while HumanNeRF~\cite{10.1145/3592415}
has the highest LPIPS score, which may suggest a larger perceptual difference between
its generated images and the real images. ReRF~\cite{wangNeuralResidualRadiance2023}
and VideoRF~\cite{wangVideoRFRenderingDynamic2024} show close performance in
PSNR and SSIM, but ReRF performs better in LPIPS.
\section{Future Directions}
\label{sec:future}

\textbf{Large scale unbounded dynamic scene rendering.}
Although current research~\cite{huang2024s3gaussian,Liu_2023_ICCV,zhouDrivingGaussianCompositeGaussian2024,Turki_2023_CVPR,yanStreetGaussiansModeling2024,liVDGVisionOnlyDynamic2024,khanAutoSplatConstrainedGaussian2024} has made strides in handling dynamic urban scenes, further efforts to scale up and handle even more extensive and complex environments are needed. On the one hand, it is crucial to improve the adaptability of models to complex dynamic scenes, such as handling intense object motions and dealing with more complex scene changes. This requires further optimization of scene decomposition~\cite{jiang2024robust} and dynamic object modeling methods to achieve more accurate reconstruction and rendering. However, improving the efficiency of the novel representation~\cite{park2024point,xiaoBridging3DGaussian2024,10747249} to meet real-time requirements is a key development trend. Exploring more efficient training strategies~\cite{zhengGPSGaussianGeneralizablePixelwise2023}, introducing feasible geometry priors~\cite{ganV4DVoxel4D2024}, and improving rendering algorithms will help to achieve faster processing speeds while ensuring rendering quality.

\textbf{Reconstruction from sparse views and incomplete camera poses.} Reconstructing 3D scenes from sparse views and incomplete camera poses presents a formidable challenge. Such scenarios are common in resource-limited conditions, such as on mobile devices~\cite{wang2024v3viewingvolumetricvideos} or when deploying a large array of cameras is impractical and costly. Current 3D-GS based methods depend on SfM points heavily~\cite{wan2024superpointgaussiansplattingrealtime,liangGauFReGaussianDeformation2023}, but software like~\citet{schonberger2016structure} and \citet{pan2025global} designed for static scenes may not be suitable for dynamic scenes. As it only initializes the point cloud at certain frames and needs strategies~\cite{sun3DGStreamFlyTraining} to handle objects newly entering the scene~\cite{luiten2024dynamic}.

\textbf{On-the-fly training.}
The ability to train models on the fly is of paramount importance~\cite{sun3DGStreamFlyTraining}. Despite the remarkable progress in dynamic scene reconstruction, many existing representations, while achieving fast rendering speeds, still face limitations in online training capabilities~\cite{liSpacetimeGaussianFeature2024}. This deficiency restricts their application in various streaming scenarios~\cite{liSpacetimeGaussianFeature2024} where real-time adaptation is essential. Although some methods~\cite{sun3DGStreamFlyTraining,yan2023od,liu2024dynamics} have made progress in this direction, there is still much room for improvement.

\textbf{Data structure and novel primitive for dynamic scenes.}
Current methods face challenges in accurately representing complex geometries~\cite{li2023dynibar,duan4DRotorGaussianSplatting2024}, handling acute dynamic changes~\cite{sunDyBluRFDynamicNeural2024,mihajlovicSplatFieldsNeuralGaussian2025}, and optimizing storage and computational resources~\cite{NEURIPS2023_df311263,liu2024lgs}. Against this backdrop, future direction in data structure~\cite{fischer2024multi,louDaReNeRFDirectionawareRepresentation2024,jiang2024robust,zhang2025efficient} and novel primitive~\cite{guo2024tetsphere} for dynamic scenes hold great potential for further advancement.

\section{Conclusion}
In this survey, we provide a comprehensive overview of the state-of-the-art in dynamic scene reconstruction and rendering. We have discussed the evolution of dynamic scene representation from traditional methods to the latest 3D-GS based approaches. We collect and categorize the most representative methods in the literature and compare them in terms of rendering quality, efficiency, and robustness. 

In addition, we highlight the challenges and limitations of current methods and identify future research directions. We believe that the insights provided in this survey will help researchers and practitioners better understand the current landscape of dynamic scene reconstruction and inspire new ideas for future research.

\small
\bibliographystyle{IEEEtranN}
\bibliography{ref}

\begin{thebibliography}{176}
\providecommand{\natexlab}[1]{#1}
\providecommand{\url}[1]{#1}
\csname url@samestyle\endcsname
\providecommand{\newblock}{\relax}
\providecommand{\bibinfo}[2]{#2}
\providecommand{\BIBentrySTDinterwordspacing}{\spaceskip=0pt\relax}
\providecommand{\BIBentryALTinterwordstretchfactor}{4}
\providecommand{\BIBentryALTinterwordspacing}{\spaceskip=\fontdimen2\font plus
\BIBentryALTinterwordstretchfactor\fontdimen3\font minus
  \fontdimen4\font\relax}
\providecommand{\BIBforeignlanguage}[2]{{%
\expandafter\ifx\csname l@#1\endcsname\relax
\typeout{** WARNING: IEEEtranN.bst: No hyphenation pattern has been}%
\typeout{** loaded for the language `#1'. Using the pattern for}%
\typeout{** the default language instead.}%
\else
\language=\csname l@#1\endcsname
\fi
#2}}
\providecommand{\BIBdecl}{\relax}
\BIBdecl

\bibitem[Li et~al.(2024{\natexlab{a}})Li, Chen, Li, and
  Xu]{liSpacetimeGaussianFeature2024}
Z.~Li, Z.~Chen, Z.~Li, and Y.~Xu, ``Spacetime gaussian feature splatting for
  real-time dynamic view synthesis,'' in \emph{Proceedings of the IEEE/CVF
  Conference on Computer Vision and Pattern Recognition (CVPR)}, June 2024, pp.
  8508--8520.

\bibitem[Li et~al.(2021)Li, Niklaus, Snavely, and Wang]{liNeuralSceneFlow2021}
Z.~Li, S.~Niklaus, N.~Snavely, and O.~Wang, ``Neural {{Scene Flow Fields}} for
  {{Space-Time View Synthesis}} of {{Dynamic Scenes}},'' in \emph{Proceedings
  of the {{IEEE}}/{{CVF Conference}} on {{Computer Vision}} and {{Pattern
  Recognition}}}, 2021, pp. 6498--6508.

\bibitem[Gao et~al.(2021)Gao, Saraf, Kopf, and Huang]{Gao-ICCV-DynNeRF}
C.~Gao, A.~Saraf, J.~Kopf, and J.-B. Huang, ``Dynamic view synthesis from
  dynamic monocular video,'' in \emph{Proceedings of the IEEE International
  Conference on Computer Vision}, 2021.

\bibitem[Du et~al.(2021)Du, Zhang, Yu, Tenenbaum, and
  Wu]{duNeuralRadianceFlow2021}
Y.~Du, Y.~Zhang, H.-X. Yu, J.~B. Tenenbaum, and J.~Wu, ``Neural {{Radiance
  Flow}} for {{4D View Synthesis}} and {{Video Processing}},'' in \emph{2021
  {{IEEE}}/{{CVF International Conference}} on {{Computer Vision}}
  ({{ICCV}})}.\hskip 1em plus 0.5em minus 0.4em\relax IEEE Computer Society,
  2021, pp. 14\,304--14\,314.

\bibitem[Wang et~al.(2021)Wang, Eckart, Lucey, and
  Gallo]{wangNeuralTrajectoryFields2021}
C.~Wang, B.~Eckart, S.~Lucey, and O.~Gallo, ``Neural {{Trajectory Fields}} for
  {{Dynamic Novel View Synthesis}},'' 2021.

\bibitem[Sun et~al.(2024{\natexlab{a}})Sun, Li, Shen, Ye, Xian, and
  Cao]{sunDyBluRFDynamicNeural2024}
H.~Sun, X.~Li, L.~Shen, X.~Ye, K.~Xian, and Z.~Cao, ``{{DyBluRF}}: {{Dynamic
  Neural Radiance Fields}} from {{Blurry Monocular Video}},'' in
  \emph{Proceedings of the {{IEEE}}/{{CVF Conference}} on {{Computer Vision}}
  and {{Pattern Recognition}}}, 2024, pp. 7517--7527.

\bibitem[Pumarola et~al.(2021)Pumarola, Corona, Pons-Moll, and
  Moreno-Noguer]{pumarola2020d}
A.~Pumarola, E.~Corona, G.~Pons-Moll, and F.~Moreno-Noguer, ``D-nerf: Neural
  radiance fields for dynamic scenes,'' in \emph{Proceedings of the IEEE/CVF
  Conference on Computer Vision and Pattern Recognition (CVPR)}, June 2021, pp.
  10\,318--10\,327.

\bibitem[Park et~al.(2021{\natexlab{a}})Park, Sinha, Barron, Bouaziz, Goldman,
  Seitz, and Martin-Brualla]{Park_2021_ICCV}
K.~Park, U.~Sinha, J.~T. Barron, S.~Bouaziz, D.~B. Goldman, S.~M. Seitz, and
  R.~Martin-Brualla, ``Nerfies: Deformable neural radiance fields,'' in
  \emph{Proceedings of the IEEE/CVF International Conference on Computer Vision
  (ICCV)}, October 2021, pp. 5865--5874.

\bibitem[Park et~al.(2021{\natexlab{b}})Park, Sinha, Hedman, Barron, Bouaziz,
  Goldman, Martin-Brualla, and
  Seitz]{parkHyperNeRFHigherDimensionalRepresentation2021}
K.~Park, U.~Sinha, P.~Hedman, J.~T. Barron, S.~Bouaziz, D.~B. Goldman,
  R.~Martin-Brualla, and S.~M. Seitz, ``Hypernerf: a higher-dimensional
  representation for topologically varying neural radiance fields,'' vol.~40,
  no.~6, Dec. 2021.

\bibitem[Tretschk et~al.(2021)Tretschk, Tewari, Golyanik, Zollh{\"o}fer,
  Lassner, and Theobalt]{tretschkNonRigidNeuralRadiance2021a}
E.~Tretschk, A.~Tewari, V.~Golyanik, M.~Zollh{\"o}fer, C.~Lassner, and
  C.~Theobalt, ``Non-{{Rigid Neural Radiance Fields}}: {{Reconstruction}} and
  {{Novel View Synthesis}} of a {{Dynamic Scene From Monocular Video}},'' in
  \emph{Proceedings of the {{IEEE}}/{{CVF International Conference}} on
  {{Computer Vision}}}, 2021, pp. 12\,959--12\,970.

\bibitem[Fang et~al.(2022)Fang, Yi, Wang, Xie, Zhang, Liu, Nie{\ss}ner, and
  Tian]{fangFastDynamicRadiance2022}
J.~Fang, T.~Yi, X.~Wang, L.~Xie, X.~Zhang, W.~Liu, M.~Nie{\ss}ner, and Q.~Tian,
  ``Fast {{Dynamic Radiance Fields}} with {{Time-Aware Neural Voxels}},'' in
  \emph{{{SIGGRAPH Asia}} 2022 {{Conference Papers}}}, ser. {{SA}} '22.\hskip
  1em plus 0.5em minus 0.4em\relax New York, NY, USA: Association for Computing
  Machinery, 2022, pp. 1--9.

\bibitem[Li et~al.(2022{\natexlab{a}})Li, Slavcheva, Zollh{\"o}fer, Green,
  Lassner, Kim, Schmidt, Lovegrove, Goesele, Newcombe, and
  Lv]{liNeural3DVideo2022a}
T.~Li, M.~Slavcheva, M.~Zollh{\"o}fer, S.~Green, C.~Lassner, C.~Kim,
  T.~Schmidt, S.~Lovegrove, M.~Goesele, R.~Newcombe, and Z.~Lv, ``Neural {{3D
  Video Synthesis From Multi-View Video}},'' in \emph{Proceedings of the
  {{IEEE}}/{{CVF Conference}} on {{Computer Vision}} and {{Pattern
  Recognition}}}, 2022, pp. 5521--5531.

\bibitem[Yan et~al.(2023{\natexlab{a}})Yan, Li, and
  Lee]{yanNeRFDSNeuralRadiance2023}
Z.~Yan, C.~Li, and G.~H. Lee, ``{{NeRF-DS}}: {{Neural Radiance Fields}} for
  {{Dynamic Specular Objects}},'' in \emph{Proceedings of the {{IEEE}}/{{CVF
  Conference}} on {{Computer Vision}} and {{Pattern Recognition}}}, 2023, pp.
  8285--8295.

\bibitem[Ma et~al.(2023)Ma, Paudel, Chhatkuli, and
  Van~Gool]{maDeformableNeuralRadiance2023}
Q.~Ma, D.~P. Paudel, A.~Chhatkuli, and L.~Van~Gool, ``Deformable {{Neural
  Radiance Fields}} using {{RGB}} and {{Event Cameras}},'' in \emph{Proceedings
  of the {{IEEE}}/{{CVF International Conference}} on {{Computer Vision}}},
  2023, pp. 3590--3600.

\bibitem[{Fridovich-Keil} et~al.(2023){Fridovich-Keil}, Meanti, Warburg, Recht,
  and Kanazawa]{fridovich-keilKPlanesExplicitRadiance2023a}
S.~{Fridovich-Keil}, G.~Meanti, F.~R. Warburg, B.~Recht, and A.~Kanazawa,
  ``K-{{Planes}}: {{Explicit Radiance Fields}} in {{Space}}, {{Time}}, and
  {{Appearance}},'' in \emph{Proceedings of the {{IEEE}}/{{CVF Conference}} on
  {{Computer Vision}} and {{Pattern Recognition}}}, 2023, pp. 12\,479--12\,488.

\bibitem[Cao and Johnson(2023)]{caoHexPlaneFastRepresentation2023}
A.~Cao and J.~Johnson, ``{{HexPlane}}: {{A Fast Representation}} for {{Dynamic
  Scenes}},'' in \emph{Proceedings of the {{IEEE}}/{{CVF Conference}} on
  {{Computer Vision}} and {{Pattern Recognition}}}, 2023, pp. 130--141.

\bibitem[Jang and Kim(2022)]{jangDTensoRFTensorialRadiance2022}
H.~Jang and D.~Kim, ``D-{{TensoRF}}: {{Tensorial Radiance Fields}} for
  {{Dynamic Scenes}},'' 2022.

\bibitem[Lou et~al.(2024)Lou, Planche, Gao, Li, Luan, Ding, Chen, Noble, and
  Wu]{louDaReNeRFDirectionawareRepresentation2024}
A.~Lou, B.~Planche, Z.~Gao, Y.~Li, T.~Luan, H.~Ding, T.~Chen, J.~Noble, and
  Z.~Wu, ``{{DaReNeRF}}: {{Direction-aware Representation}} for {{Dynamic
  Scenes}},'' in \emph{Proceedings of the {{IEEE}}/{{CVF Conference}} on
  {{Computer Vision}} and {{Pattern Recognition}}}, 2024, pp. 5031--5042.

\bibitem[Song et~al.(2023)Song, Chen, Li, Chen, Chen, Yuan, Xu, and
  Geiger]{10049689}
L.~Song, A.~Chen, Z.~Li, Z.~Chen, L.~Chen, J.~Yuan, Y.~Xu, and A.~Geiger,
  ``Nerfplayer: A streamable dynamic scene representation with decomposed
  neural radiance fields,'' \emph{IEEE Transactions on Visualization and
  Computer Graphics}, vol.~29, no.~5, pp. 2732--2742, 2023.

\bibitem[Chen et~al.(2024{\natexlab{a}})Chen, Yan, Yue, and
  Zhou]{chenMultiResolutionHybridExplicit2024}
Y.~Chen, W.~Yan, G.~Yue, and W.~Zhou, ``Multi-{{Resolution Hybrid Explicit
  Representation}} for {{Novel View Synthesis}} of {{Dynamic Scenes}},''
  \emph{IEEE Transactions on Intelligent Vehicles}, pp. 1--11, 2024.

\bibitem[Qiao et~al.(2023)Qiao, Gao, Xu, Feng, Huang, and
  Lin]{qiaoDynamicMeshAwareRadiance2023}
Y.-L. Qiao, A.~Gao, Y.~Xu, Y.~Feng, J.-B. Huang, and M.~C. Lin, ``Dynamic
  {{Mesh-Aware Radiance Fields}},'' in \emph{Proceedings of the {{IEEE}}/{{CVF
  International Conference}} on {{Computer Vision}}}, 2023, pp. 385--396.

\bibitem[Gan et~al.(2024)Gan, Xu, Huang, Chen, and Yokoya]{ganV4DVoxel4D2024}
W.~Gan, H.~Xu, Y.~Huang, S.~Chen, and N.~Yokoya, ``{{V4D}}: {{Voxel}} for {{4D
  Novel View Synthesis}},'' \emph{IEEE Transactions on Visualization and
  Computer Graphics}, vol.~30, no.~2, pp. 1579--1591, 2024.

\bibitem[Wang et~al.(2023{\natexlab{a}})Wang, Chen, Wang, Song, and
  Liu]{NEURIPS2023_df311263}
F.~Wang, Z.~Chen, G.~Wang, Y.~Song, and H.~Liu, ``Masked space-time hash
  encoding for efficient dynamic scene reconstruction,'' in \emph{Advances in
  Neural Information Processing Systems}, A.~Oh, T.~Naumann, A.~Globerson,
  K.~Saenko, M.~Hardt, and S.~Levine, Eds., vol.~36.\hskip 1em plus 0.5em minus
  0.4em\relax Curran Associates, Inc., 2023, pp. 70\,497--70\,510.

\bibitem[Yang et~al.(2024)Yang, Gao, Zhou, Jiao, Zhang, and
  Jin]{yangDeformable3DGaussians2023}
Z.~Yang, X.~Gao, W.~Zhou, S.~Jiao, Y.~Zhang, and X.~Jin, ``Deformable 3d
  gaussians for high-fidelity monocular dynamic scene reconstruction,'' in
  \emph{Proceedings of the IEEE/CVF Conference on Computer Vision and Pattern
  Recognition (CVPR)}, June 2024, pp. 20\,331--20\,341.

\bibitem[Liang et~al.(2023)Liang, Khan, Li, {Nguyen-Phuoc}, Lanman, Tompkin,
  and Xiao]{liangGauFReGaussianDeformation2023}
Y.~Liang, N.~Khan, Z.~Li, T.~{Nguyen-Phuoc}, D.~Lanman, J.~Tompkin, and
  L.~Xiao, ``{{GauFRe}}: {{Gaussian Deformation Fields}} for {{Real-time
  Dynamic Novel View Synthesis}},'' https://arxiv.org/abs/2312.11458v1, 2023.

\bibitem[Huang et~al.(2024{\natexlab{a}})Huang, Sun, Yang, Lyu, Cao, and
  Qi]{huangSCGSSparseControlledGaussian2023}
Y.-H. Huang, Y.-T. Sun, Z.~Yang, X.~Lyu, Y.-P. Cao, and X.~Qi, ``Sc-gs:
  Sparse-controlled gaussian splatting for editable dynamic scenes,'' in
  \emph{Proceedings of the IEEE/CVF Conference on Computer Vision and Pattern
  Recognition (CVPR)}, June 2024, pp. 4220--4230.

\bibitem[Lin et~al.(2024)Lin, Dai, Zhu, and
  Yao]{linGaussianFlow4DReconstruction2024}
Y.~Lin, Z.~Dai, S.~Zhu, and Y.~Yao, ``Gaussian-{{Flow}}: {{4D Reconstruction}}
  with {{Dynamic 3D Gaussian Particle}},'' in \emph{Proceedings of the
  {{IEEE}}/{{CVF Conference}} on {{Computer Vision}} and {{Pattern
  Recognition}}}, 2024, pp. 21\,136--21\,145.

\bibitem[Kratimenos et~al.(2024)Kratimenos, Lei, and
  Daniilidis]{kratimenosDynMFNeuralMotion2023}
A.~Kratimenos, J.~Lei, and K.~Daniilidis, ``Dynmf: Neural motion factorization
  for real-time dynamic view synthesis with 3d gaussian splatting,'' in
  \emph{European Conference on Computer Vision}.\hskip 1em plus 0.5em minus
  0.4em\relax Springer, 2024, pp. 252--269.

\bibitem[Diwen~Wan(2024)]{wan2024superpointgaussiansplattingrealtime}
G.~Z. Diwen~Wan, Ruijie~Lu, ``Superpoint gaussian splatting for real-time
  high-fidelity dynamic scene reconstruction,'' in \emph{Forty-first
  International Conference on Machine Learning}, 2024.

\bibitem[Wu et~al.(2024{\natexlab{a}})Wu, Yi, Fang, Xie, Zhang, Wei, Liu, Tian,
  and Wang]{wu4DGaussianSplatting2024}
G.~Wu, T.~Yi, J.~Fang, L.~Xie, X.~Zhang, W.~Wei, W.~Liu, Q.~Tian, and X.~Wang,
  ``4d gaussian splatting for real-time dynamic scene rendering,'' in
  \emph{Proceedings of the IEEE/CVF Conference on Computer Vision and Pattern
  Recognition (CVPR)}, June 2024, pp. 20\,310--20\,320.

\bibitem[Zheng et~al.(2024{\natexlab{a}})Zheng, Zhou, Shao, Liu, Zhang, Nie,
  and Liu]{zhengGPSGaussianGeneralizablePixelwise2023}
S.~Zheng, B.~Zhou, R.~Shao, B.~Liu, S.~Zhang, L.~Nie, and Y.~Liu,
  ``Gps-gaussian: Generalizable pixel-wise 3d gaussian splatting for real-time
  human novel view synthesis,'' in \emph{Proceedings of the IEEE/CVF Conference
  on Computer Vision and Pattern Recognition (CVPR)}, June 2024, pp.
  19\,680--19\,690.

\bibitem[Waczynska et~al.(2025)Waczynska, Borycki, Kaleta, Tadeja, and
  Spurek]{waczyńska2024dmiso}
J.~Waczynska, P.~Borycki, J.~Kaleta, S.~Tadeja, and P.~Spurek, ``D-miso:
  Editing dynamic 3d scenes using multi-gaussians soup,'' \emph{Advances in
  Neural Information Processing Systems}, vol.~37, pp. 107\,865--107\,889,
  2025.

\bibitem[Schönberger and Frahm(2016)]{7780814}
J.~L. Schönberger and J.-M. Frahm, ``Structure-from-motion revisited,'' in
  \emph{2016 IEEE Conference on Computer Vision and Pattern Recognition
  (CVPR)}, 2016, pp. 4104--4113.

\bibitem[Shin et~al.(2024)Shin, Yu, Shen, Kweon, Yoon, and
  Chen]{shin2024enhancing}
I.~Shin, Q.~Yu, X.~Shen, I.~S. Kweon, K.-J. Yoon, and L.-C. Chen, ``Enhancing
  temporal consistency in video editing by reconstructing videos with 3d
  gaussian splatting,'' \emph{arXiv preprint arXiv:2406.02541}, 2024.

\bibitem[Yang et~al.(2023)Yang, Yang, Pan, and
  Zhang]{yangRealtimePhotorealisticDynamic2023}
Z.~Yang, H.~Yang, Z.~Pan, and L.~Zhang, ``Real-time {{Photorealistic Dynamic
  Scene Representation}} and {{Rendering}} with {{4D Gaussian Splatting}},''
  https://arxiv.org/abs/2310.10642v3, 2023.

\bibitem[Duan et~al.(2024)Duan, Wei, Dai, He, Chen, and
  Chen]{duan4DRotorGaussianSplatting2024}
Y.~Duan, F.~Wei, Q.~Dai, Y.~He, W.~Chen, and B.~Chen, ``4d-rotor gaussian
  splatting: Towards efficient novel view synthesis for dynamic scenes,'' in
  \emph{ACM SIGGRAPH 2024 Conference Papers}, ser. SIGGRAPH '24.\hskip 1em plus
  0.5em minus 0.4em\relax New York, NY, USA: Association for Computing
  Machinery, 2024.

\bibitem[Luiten et~al.(2024{\natexlab{a}})Luiten, Kopanas, Leibe, and
  Ramanan]{luitenDynamic3DGaussians2023}
J.~Luiten, G.~Kopanas, B.~Leibe, and D.~Ramanan, ``Dynamic 3d gaussians:
  Tracking by persistent dynamic view synthesis,'' in \emph{2024 International
  Conference on 3D Vision (3DV)}.\hskip 1em plus 0.5em minus 0.4em\relax IEEE,
  2024, pp. 800--809.

\bibitem[Sun et~al.(2024{\natexlab{b}})Sun, Jiao, Li, Zhang, Zhao, and
  Xing]{sun3DGStreamFlyTraining}
J.~Sun, H.~Jiao, G.~Li, Z.~Zhang, L.~Zhao, and W.~Xing, ``3dgstream: On-the-fly
  training of 3d gaussians for efficient streaming of photo-realistic
  free-viewpoint videos,'' in \emph{Proceedings of the IEEE/CVF Conference on
  Computer Vision and Pattern Recognition (CVPR)}, June 2024, pp.
  20\,675--20\,685.

\bibitem[Jiang et~al.(2024{\natexlab{a}})Jiang, Shen, Hong, Guo, Wu, Zhang, Yu,
  and Xu]{jiang2024robust}
Y.~Jiang, Z.~Shen, Y.~Hong, C.~Guo, Y.~Wu, Y.~Zhang, J.~Yu, and L.~Xu, ``Robust
  dual gaussian splatting for immersive human-centric volumetric videos,''
  \emph{ACM Trans. Graph.}, vol.~43, no.~6, Nov. 2024.

\bibitem[Das et~al.(2024)Das, Wewer, Yunus, Ilg, and
  Lenssen]{dasNeuralParametricGaussians2024}
D.~Das, C.~Wewer, R.~Yunus, E.~Ilg, and J.~E. Lenssen, ``Neural parametric
  gaussians for monocular non-rigid object reconstruction,'' in
  \emph{Proceedings of the IEEE/CVF Conference on Computer Vision and Pattern
  Recognition (CVPR)}, June 2024, pp. 10\,715--10\,725.

\bibitem[Wang et~al.(2024{\natexlab{a}})Wang, Zhang, Wang, Yao, Xie, Yu, Wu,
  and Xu]{wang2024v3viewingvolumetricvideos}
P.~Wang, Z.~Zhang, L.~Wang, K.~Yao, S.~Xie, J.~Yu, M.~Wu, and L.~Xu, ``V\^{} 3:
  Viewing volumetric videos on mobiles via streamable 2d dynamic gaussians,''
  \emph{ACM Transactions on Graphics (TOG)}, vol.~43, no.~6, pp. 1--13, 2024.

\bibitem[Zhang et~al.(2024{\natexlab{a}})Zhang, Lu, Liang, Tang, Hu, and
  Song]{zhangEfficientDynamicNeRFBased2024a}
Z.~Zhang, G.~Lu, H.~Liang, A.~Tang, Q.~Hu, and L.~Song, ``Efficient
  dynamic-nerf based volumetric video coding with rate distortion
  optimization,'' in \emph{2024 IEEE International Conference on Multimedia and
  Expo (ICME)}, 2024, pp. 1--6.

\bibitem[Wang et~al.(2023{\natexlab{b}})Wang, Hu, He, Wang, Yu, Tuytelaars, Xu,
  and Wu]{wangNeuralResidualRadiance2023}
L.~Wang, Q.~Hu, Q.~He, Z.~Wang, J.~Yu, T.~Tuytelaars, L.~Xu, and M.~Wu,
  ``Neural {{Residual Radiance Fields}} for {{Streamably Free-Viewpoint
  Videos}},'' in \emph{Proceedings of the {{IEEE}}/{{CVF Conference}} on
  {{Computer Vision}} and {{Pattern Recognition}}}, 2023, pp. 76--87.

\bibitem[Zhang et~al.(2024{\natexlab{b}})Zhang, Lu, Liang, Cheng, Tang, and
  Song]{zhangRateawareCompressionNeRFbased2024}
Z.~Zhang, G.~Lu, H.~Liang, Z.~Cheng, A.~Tang, and L.~Song, ``Rate-aware
  {{Compression}} for {{NeRF-based Volumetric Video}},'' in \emph{Proceedings
  of the 32nd {{ACM International Conference}} on {{Multimedia}}}, ser. {{MM}}
  '24.\hskip 1em plus 0.5em minus 0.4em\relax New York, NY, USA: Association
  for Computing Machinery, 2024, pp. 3974--3983.

\bibitem[I\c{s}\i{}k et~al.(2023)I\c{s}\i{}k, R\"{u}nz, Georgopoulos,
  Khakhulin, Starck, Agapito, and Nie\ss{}ner]{10.1145/3592415}
M.~I\c{s}\i{}k, M.~R\"{u}nz, M.~Georgopoulos, T.~Khakhulin, J.~Starck,
  L.~Agapito, and M.~Nie\ss{}ner, ``Humanrf: High-fidelity neural radiance
  fields for humans in motion,'' vol.~42, no.~4, Jul. 2023.

\bibitem[Zheng et~al.(2024{\natexlab{b}})Zheng, Zhong, Hu, Zhang, Song, Zhang,
  and Wang]{zhengJointRFEndtoEndJoint2024}
Z.~Zheng, H.~Zhong, Q.~Hu, X.~Zhang, L.~Song, Y.~Zhang, and Y.~Wang, ``Jointrf:
  End-to-end joint optimization for dynamic neural radiance field
  representation and compression,'' in \emph{2024 IEEE International Conference
  on Image Processing (ICIP)}, 2024, pp. 3292--3298.

\bibitem[Guo et~al.(2023)Guo, Peng, Yan, Mou, Shen, Bao, and
  Zhou]{guoCompactNeuralVolumetric2023}
H.~Guo, S.~Peng, Y.~Yan, L.~Mou, Y.~Shen, H.~Bao, and X.~Zhou, ``Compact
  {{Neural Volumetric Video Representations}} with {{Dynamic Codebooks}},''
  \emph{Advances in Neural Information Processing Systems}, vol.~36, pp.
  75\,884--75\,895, 2023.

\bibitem[Wu et~al.(2024{\natexlab{b}})Wu, Wang, Kouros, and
  Tuytelaars]{wuTeTriRFTemporalTriPlane2024}
M.~Wu, Z.~Wang, G.~Kouros, and T.~Tuytelaars, ``{{TeTriRF}}: {{Temporal
  Tri-Plane Radiance Fields}} for {{Efficient Free-Viewpoint Video}},'' in
  \emph{Proceedings of the {{IEEE}}/{{CVF Conference}} on {{Computer Vision}}
  and {{Pattern Recognition}}}, 2024, pp. 6487--6496.

\bibitem[Wang et~al.(2024{\natexlab{b}})Wang, Yao, Guo, Zhang, Hu, Yu, Xu, and
  Wu]{wangVideoRFRenderingDynamic2024}
L.~Wang, K.~Yao, C.~Guo, Z.~Zhang, Q.~Hu, J.~Yu, L.~Xu, and M.~Wu,
  ``{{VideoRF}}: {{Rendering Dynamic Radiance Fields}} as {{2D Feature Video
  Streams}},'' in \emph{Proceedings of the {{IEEE}}/{{CVF Conference}} on
  {{Computer Vision}} and {{Pattern Recognition}}}, 2024, pp. 470--481.

\bibitem[Xiao et~al.(2024)Xiao, Wang, Li, Cai, Fan, Xue, Yang, Shen, and
  Gao]{xiaoBridging3DGaussian2024}
Y.~Xiao, X.~Wang, J.~Li, H.~Cai, Y.~Fan, N.~Xue, M.~Yang, Y.~Shen, and S.~Gao,
  ``Bridging {{3D Gaussian}} and {{Mesh}} for {{Freeview Video Rendering}},''
  2024.

\bibitem[Sun et~al.(2024{\natexlab{c}})Sun, Shi, Ooi, Huang, and
  Hsu]{sunMultiframeBitrateAllocation2024a}
Y.-C. Sun, Y.~Shi, W.~T. Ooi, C.-Y. Huang, and C.-H. Hsu, ``Multi-frame
  {{Bitrate Allocation}} of {{Dynamic 3D Gaussian Splatting Streaming Over
  Dynamic Networks}},'' in \emph{Proceedings of the 2024 {{SIGCOMM Workshop}}
  on {{Emerging Multimedia Systems}}}, ser. {{EMS}} '24.\hskip 1em plus 0.5em
  minus 0.4em\relax New York, NY, USA: Association for Computing Machinery,
  2024, pp. 1--7.

\bibitem[Xu et~al.(2024)Xu, Xu, Yu, Peng, Sun, Bao, and Zhou]{10.1145/3687919}
Z.~Xu, Y.~Xu, Z.~Yu, S.~Peng, J.~Sun, H.~Bao, and X.~Zhou, ``Representing long
  volumetric video with temporal gaussian hierarchy,'' vol.~43, no.~6, Nov.
  2024.

\bibitem[Park et~al.(2021{\natexlab{c}})Park, Sinha, Hedman, Barron, Bouaziz,
  Goldman, Martin-Brualla, and Seitz]{park2021hypernerf}
K.~Park, U.~Sinha, P.~Hedman, J.~T. Barron, S.~Bouaziz, D.~B. Goldman,
  R.~Martin-Brualla, and S.~M. Seitz, ``Hypernerf: A higher-dimensional
  representation for topologically varying neural radiance fields,'' \emph{ACM
  Trans. Graph.}, vol.~40, no.~6, dec 2021.

\bibitem[Yan et~al.(2023{\natexlab{b}})Yan, Li, and Lee]{yan2023nerf}
Z.~Yan, C.~Li, and G.~H. Lee, ``Nerf-ds: Neural radiance fields for dynamic
  specular objects,'' in \emph{Proceedings of the IEEE/CVF Conference on
  Computer Vision and Pattern Recognition}, 2023, pp. 8285--8295.

\bibitem[Li et~al.(2022{\natexlab{b}})Li, Slavcheva, Zollh\"ofer, Green,
  Lassner, Kim, Schmidt, Lovegrove, Goesele, Newcombe, and
  Lv]{li2022neural3dvideosynthesis}
T.~Li, M.~Slavcheva, M.~Zollh\"ofer, S.~Green, C.~Lassner, C.~Kim, T.~Schmidt,
  S.~Lovegrove, M.~Goesele, R.~Newcombe, and Z.~Lv, ``Neural 3d video synthesis
  from multi-view video,'' in \emph{Proceedings of the IEEE/CVF Conference on
  Computer Vision and Pattern Recognition (CVPR)}, June 2022, pp. 5521--5531.

\bibitem[Broxton et~al.(2020)Broxton, Flynn, Overbeck, Erickson, Hedman,
  DuVall, Dourgarian, Busch, Whalen, and Debevec]{broxton2020immersive}
M.~Broxton, J.~Flynn, R.~Overbeck, D.~Erickson, P.~Hedman, M.~DuVall,
  J.~Dourgarian, J.~Busch, M.~Whalen, and P.~Debevec, ``Immersive light field
  video with a layered mesh representation,'' vol.~39, no.~4, pp. 86:1--86:15,
  2020.

\bibitem[Li et~al.(2022{\natexlab{c}})Li, Shen, Wang, Shen, and
  Tan]{li2022streaming}
L.~Li, Z.~Shen, Z.~Wang, L.~Shen, and P.~Tan, ``Streaming radiance fields for
  3d video synthesis,'' \emph{Advances in Neural Information Processing
  Systems}, vol.~35, pp. 13\,485--13\,498, 2022.

\bibitem[Jiang et~al.(2024{\natexlab{b}})Jiang, Shen, Wang, Su, Hong, Zhang,
  Yu, and Xu]{Jiang_2024_CVPR}
Y.~Jiang, Z.~Shen, P.~Wang, Z.~Su, Y.~Hong, Y.~Zhang, J.~Yu, and L.~Xu,
  ``Hifi4g: High-fidelity human performance rendering via compact gaussian
  splatting,'' in \emph{Proceedings of the IEEE/CVF Conference on Computer
  Vision and Pattern Recognition (CVPR)}, June 2024, pp. 19\,734--19\,745.

\bibitem[Abou-Chakra et~al.(2024)Abou-Chakra, Dayoub, and
  S\"underhauf]{abou2022particlenerf}
J.~Abou-Chakra, F.~Dayoub, and N.~S\"underhauf, ``Particlenerf: A
  particle-based encoding for online neural radiance fields,'' in
  \emph{Proceedings of the IEEE/CVF Winter Conference on Applications of
  Computer Vision (WACV)}, January 2024, pp. 5975--5984.

\bibitem[Wu et~al.(2024{\natexlab{c}})Wu, Dai, Deng, Chen, Wu, Cao, Shan, and
  Qi]{wu2024cl}
X.~Wu, P.~Dai, W.~Deng, H.~Chen, Y.~Wu, Y.-P. Cao, Y.~Shan, and X.~Qi,
  ``Cl-nerf: continual learning of neural radiance fields for evolving scene
  representation,'' \emph{Advances in Neural Information Processing Systems},
  vol.~36, 2024.

\bibitem[Xu et~al.(2019)Xu, Cheng, Guo, Han, Liu, and Fang]{xu2019flyfusion}
L.~Xu, W.~Cheng, K.~Guo, L.~Han, Y.~Liu, and L.~Fang, ``Flyfusion: Realtime
  dynamic scene reconstruction using a flying depth camera,'' \emph{IEEE
  transactions on visualization and computer graphics}, vol.~27, no.~1, pp.
  68--82, 2019.

\bibitem[Seitz et~al.(2006)Seitz, Curless, Diebel, Scharstein, and
  Szeliski]{seitzComparisonEvaluationMultiView2006}
S.~Seitz, B.~Curless, J.~Diebel, D.~Scharstein, and R.~Szeliski, ``A
  {{Comparison}} and {{Evaluation}} of {{Multi-View Stereo Reconstruction
  Algorithms}},'' in \emph{2006 {{IEEE Computer Society Conference}} on
  {{Computer Vision}} and {{Pattern Recognition}} ({{CVPR}}'06)}, vol.~1, 2006,
  pp. 519--528.

\bibitem[Leroy et~al.(2017)Leroy, Franco, and
  Boyer]{leroyMultiViewDynamicShape2017}
V.~Leroy, J.-S. Franco, and E.~Boyer, ``Multi-{{View Dynamic Shape Refinement
  Using Local Temporal Integration}},'' in \emph{Proceedings of the {{IEEE
  International Conference}} on {{Computer Vision}}}, 2017, pp. 3094--3103.

\bibitem[Paladini et~al.(2012)Paladini, Del~Bue, Xavier, Agapito,
  Sto{\v{s}}i{\'c}, and Dodig]{paladini2012optimal}
M.~Paladini, A.~Del~Bue, J.~Xavier, L.~Agapito, M.~Sto{\v{s}}i{\'c}, and
  M.~Dodig, ``Optimal metric projections for deformable and articulated
  structure-from-motion,'' \emph{International journal of computer vision},
  vol.~96, pp. 252--276, 2012.

\bibitem[Garg et~al.(2013)Garg, Roussos, and Agapito]{garg2013dense}
R.~Garg, A.~Roussos, and L.~Agapito, ``Dense variational reconstruction of
  non-rigid surfaces from monocular video,'' in \emph{Proceedings of the IEEE
  Conference on computer vision and pattern recognition}, 2013, pp. 1272--1279.

\bibitem[Bartoli et~al.(2012)Bartoli, G{\'e}rard, Chadebecq, and
  Collins]{bartoliTemplatebasedReconstructionSingle2012}
A.~Bartoli, Y.~G{\'e}rard, F.~Chadebecq, and T.~Collins, ``On template-based
  reconstruction from a single view: {{Analytical}} solutions and proofs of
  well-posedness for developable, isometric and conformal surfaces,'' in
  \emph{2012 {{IEEE Conference}} on {{Computer Vision}} and {{Pattern
  Recognition}}}, 2012, pp. 2026--2033.

\bibitem[Li et~al.(2009)Li, Adams, Guibas, and
  Pauly]{liRobustSingleviewGeometry2009}
H.~Li, B.~Adams, L.~J. Guibas, and M.~Pauly, ``Robust single-view geometry and
  motion reconstruction,'' \emph{ACM Trans. Graph.}, vol.~28, no.~5, pp. 1--10,
  2009.

\bibitem[Dou et~al.(2016)Dou, Khamis, Degtyarev, Davidson, Fanello, Kowdle,
  Escolano, Rhemann, Kim, Taylor, Kohli, Tankovich, and
  Izadi]{douFusion4DRealtimePerformance2016}
M.~Dou, S.~Khamis, Y.~Degtyarev, P.~Davidson, S.~R. Fanello, A.~Kowdle, S.~O.
  Escolano, C.~Rhemann, D.~Kim, J.~Taylor, P.~Kohli, V.~Tankovich, and
  S.~Izadi, ``{{Fusion4D}}: Real-time performance capture of challenging
  scenes,'' \emph{ACM Trans. Graph.}, vol.~35, no.~4, pp. 114:1--114:13, 2016.

\bibitem[Newcombe et~al.(2015)Newcombe, Fox, and
  Seitz]{newcombeDynamicFusionReconstructionTracking2015}
R.~A. Newcombe, D.~Fox, and S.~M. Seitz, ``{{DynamicFusion}}:
  {{Reconstruction}} and {{Tracking}} of {{Non-Rigid Scenes}} in
  {{Real-Time}},'' in \emph{Proceedings of the {{IEEE Conference}} on
  {{Computer Vision}} and {{Pattern Recognition}}}, 2015, pp. 343--352.

\bibitem[Mildenhall et~al.(2020)Mildenhall, Srinivasan, Tancik, Barron,
  Ramamoorthi, and Ng]{mildenhall2020nerfrepresentingscenesneural}
B.~Mildenhall, P.~P. Srinivasan, M.~Tancik, J.~T. Barron, R.~Ramamoorthi, and
  R.~Ng, ``Nerf: Representing scenes as neural radiance fields for view
  synthesis,'' in \emph{Computer Vision -- ECCV 2020}, A.~Vedaldi, H.~Bischof,
  T.~Brox, and J.-M. Frahm, Eds.\hskip 1em plus 0.5em minus 0.4em\relax Cham:
  Springer International Publishing, 2020, pp. 405--421.

\bibitem[Zhao et~al.(2023)Zhao, Huang, and Huang]{10296239}
B.~Zhao, W.~Huang, and Y.~Huang, ``A novel hardware accelerator of nerf based
  on xilinx ultrascale and ultrascale+ fpga,'' in \emph{2023 33rd International
  Conference on Field-Programmable Logic and Applications (FPL)}, 2023, pp.
  197--203.

\bibitem[Chen et~al.(2024{\natexlab{b}})Chen, Wu, Harandi, and
  Cai]{chen2024far}
Y.~Chen, Q.~Wu, M.~Harandi, and J.~Cai, ``How far can we compress
  instant-ngp-based nerf?'' in \emph{Proceedings of the IEEE/CVF Conference on
  Computer Vision and Pattern Recognition}, 2024, pp. 20\,321--20\,330.

\bibitem[Kerbl et~al.(2023)Kerbl, Kopanas, Leimk{\"u}hler, and
  Drettakis]{kerbl20233d}
B.~Kerbl, G.~Kopanas, T.~Leimk{\"u}hler, and G.~Drettakis, ``3d gaussian
  splatting for real-time radiance field rendering.'' \emph{ACM Trans. Graph.},
  vol.~42, no.~4, pp. 139--1, 2023.

\bibitem[Chen and Wang(2024)]{chen2024survey3dgaussiansplatting}
G.~Chen and W.~Wang, ``A survey on 3d gaussian splatting,'' 2024.

\bibitem[Ingale and J.(2021)]{ingaleRealtime3DReconstruction2021d}
A.~K. Ingale and D.~U. J., ``Real-time {{3D}} reconstruction techniques applied
  in dynamic scenes: {{A}} systematic literature review,'' \emph{Computer
  Science Review}, vol.~39, p. 100338, 2021.

\bibitem[Tretschk et~al.(2023)Tretschk, Kairanda, BR, Dabral, Kortylewski,
  Egger, Habermann, Fua, Theobalt, and Golyanik]{tretschkStateArtDense2023}
E.~Tretschk, N.~Kairanda, M.~BR, R.~Dabral, A.~Kortylewski, B.~Egger,
  M.~Habermann, P.~Fua, C.~Theobalt, and V.~Golyanik, ``State of the art in
  dense monocular non-rigid 3d reconstruction,'' in \emph{Computer Graphics
  Forum}, vol.~42, no.~2.\hskip 1em plus 0.5em minus 0.4em\relax Wiley Online
  Library, 2023, pp. 485--520.

\bibitem[Yunus et~al.(2024)Yunus, Lenssen, Niemeyer, Liao, Rupprecht, Theobalt,
  {Pons-Moll}, Huang, Golyanik, and Ilg]{yunusRecentTrends3D2024}
R.~Yunus, J.~E. Lenssen, M.~Niemeyer, Y.~Liao, C.~Rupprecht, C.~Theobalt,
  G.~{Pons-Moll}, J.-B. Huang, V.~Golyanik, and E.~Ilg, ``Recent {{Trends}} in
  {{3D Reconstruction}} of {{General Non-Rigid Scenes}},'' \emph{Computer
  Graphics Forum}, vol.~43, no.~2, p. e15062, 2024.

\bibitem[Entezami and Guan(2024)]{entezamiAIDrivenInnovationsVolumetric2024}
E.~Entezami and H.~Guan, ``{{AI-Driven Innovations}} in {{Volumetric Video
  Streaming}}: {{A Review}},'' 2024.

\bibitem[Larsen et~al.(2007)Larsen, Mordohai, Pollefeys, and
  Fuchs]{larsenTemporallyConsistentReconstruction2007}
E.~S. Larsen, P.~Mordohai, M.~Pollefeys, and H.~Fuchs, ``Temporally
  {{Consistent Reconstruction}} from {{Multiple Video Streams Using Enhanced
  Belief Propagation}},'' in \emph{2007 {{IEEE}} 11th {{International
  Conference}} on {{Computer Vision}}}, 2007, pp. 1--8.

\bibitem[Lei et~al.(2009)Lei, Da~Chen, and Yang]{lei2009new}
C.~Lei, X.~Da~Chen, and Y.~H. Yang, ``A new multiview spacetime-consistent
  depth recovery framework for free viewpoint video rendering,'' in \emph{2009
  IEEE 12th International Conference on Computer Vision}.\hskip 1em plus 0.5em
  minus 0.4em\relax IEEE, 2009, pp. 1570--1577.

\bibitem[Furukawa and Ponce(2010)]{furukawaAccurateDenseRobust2010}
Y.~Furukawa and J.~Ponce, ``Accurate, {{Dense}}, and {{Robust Multiview
  Stereopsis}},'' \emph{IEEE Transactions on Pattern Analysis and Machine
  Intelligence}, vol.~32, no.~8, pp. 1362--1376, 2010.

\bibitem[Bregler et~al.(2000)Bregler, Hertzmann, and
  Biermann]{breglerRecoveringNonrigid3D2000a}
C.~Bregler, A.~Hertzmann, and H.~Biermann, ``Recovering non-rigid {{3D}} shape
  from image streams,'' in \emph{Proceedings {{IEEE Conference}} on {{Computer
  Vision}} and {{Pattern Recognition}}. {{CVPR}} 2000 ({{Cat}}.
  {{No}}.{{PR00662}})}, vol.~2, 2000, pp. 690--696 vol.2.

\bibitem[Akhter et~al.(2011)Akhter, Sheikh, Khan, and
  Kanade]{akhterTrajectorySpaceDual2011}
I.~Akhter, Y.~Sheikh, S.~Khan, and T.~Kanade, ``Trajectory {{Space}}: {{A Dual
  Representation}} for {{Nonrigid Structure}} from {{Motion}},'' \emph{IEEE
  Transactions on Pattern Analysis and Machine Intelligence}, vol.~33, no.~7,
  pp. 1442--1456, 2011.

\bibitem[Dai et~al.(2014)Dai, Li, and He]{dai2014simple}
Y.~Dai, H.~Li, and M.~He, ``A simple prior-free method for non-rigid
  structure-from-motion factorizatione,'' \emph{International Journal of
  Computer Vision}, vol. 107, pp. 101--122, 2014.

\bibitem[Zhu et~al.(2014)Zhu, Huang, De~La~Torre, and
  Lucey]{zhuComplexNonRigidMotion2014}
Y.~Zhu, D.~Huang, F.~De~La~Torre, and S.~Lucey, ``Complex {{Non-Rigid Motion 3D
  Reconstruction}} by {{Union}} of {{Subspaces}},'' in \emph{Proceedings of the
  {{IEEE Conference}} on {{Computer Vision}} and {{Pattern Recognition}}},
  2014, pp. 1542--1549.

\bibitem[Ozden et~al.(2010)Ozden, Schindler, and Van~Gool]{5396339}
K.~E. Ozden, K.~Schindler, and L.~Van~Gool, ``Multibody structure-from-motion
  in practice,'' \emph{IEEE Transactions on Pattern Analysis and Machine
  Intelligence}, vol.~32, no.~6, pp. 1134--1141, 2010.

\bibitem[Torresani et~al.(2008)Torresani, Hertzmann, and
  Bregler]{torresani2008nonrigid}
L.~Torresani, A.~Hertzmann, and C.~Bregler, ``Nonrigid structure-from-motion:
  Estimating shape and motion with hierarchical priors,'' \emph{IEEE
  transactions on pattern analysis and machine intelligence}, vol.~30, no.~5,
  pp. 878--892, 2008.

\bibitem[Russell et~al.(2011)Russell, Fayad, and
  Agapito]{russellEnergyBasedMultiple2011}
C.~Russell, J.~Fayad, and L.~Agapito, ``Energy based multiple model fitting for
  non-rigid structure from motion,'' in \emph{{{CVPR}} 2011}, 2011, pp.
  3009--3016.

\bibitem[Russell et~al.(2014)Russell, Yu, and
  Agapito]{russellVideoPopupMonocular2014}
C.~Russell, R.~Yu, and L.~Agapito, ``Video {{Pop-up}}: {{Monocular 3D
  Reconstruction}} of {{Dynamic Scenes}},'' in \emph{Computer {{Vision}} --
  {{ECCV}} 2014}, D.~Fleet, T.~Pajdla, B.~Schiele, and T.~Tuytelaars,
  Eds.\hskip 1em plus 0.5em minus 0.4em\relax Cham: Springer International
  Publishing, 2014, pp. 583--598.

\bibitem[Taneja et~al.(2011)Taneja, Ballan, and Pollefeys]{taneja2011modeling}
A.~Taneja, L.~Ballan, and M.~Pollefeys, ``Modeling dynamic scenes recorded with
  freely moving cameras,'' in \emph{Computer Vision--ACCV 2010: 10th Asian
  Conference on Computer Vision, Queenstown, New Zealand, November 8-12, 2010,
  Revised Selected Papers, Part III 10}.\hskip 1em plus 0.5em minus 0.4em\relax
  Springer, 2011, pp. 613--626.

\bibitem[Huang and Wu(1992)]{huang1992dynamic}
C.-L. Huang and C.-H. Wu, ``Dynamic scene analysis using path and shape
  coherence,'' \emph{Pattern recognition}, vol.~25, no.~5, pp. 445--461, 1992.

\bibitem[Zhang et~al.(2003)Zhang, Curless, and Seitz]{zhang2003spacetime}
L.~Zhang, B.~Curless, and S.~M. Seitz, ``Spacetime stereo: Shape recovery for
  dynamic scenes,'' in \emph{2003 IEEE Computer Society Conference on Computer
  Vision and Pattern Recognition, 2003. Proceedings.}, vol.~2.\hskip 1em plus
  0.5em minus 0.4em\relax IEEE, 2003, pp. II--367.

\bibitem[Wang et~al.(2015)Wang, Kohli, and Mitra]{wangDynamicSfMDetecting2015}
T.~Y. Wang, P.~Kohli, and N.~J. Mitra, ``Dynamic {{SfM}}: {{Detecting Scene
  Changes}} from {{Image Pairs}},'' \emph{Computer Graphics Forum}, vol.~34,
  no.~5, pp. 177--189, 2015.

\bibitem[Tung et~al.(2009)Tung, Nobuhara, and
  Matsuyama]{CompleteMultiviewReconstruction}
T.~Tung, S.~Nobuhara, and T.~Matsuyama, ``Complete multi-view reconstruction of
  dynamic scenes from probabilistic fusion of narrow and wide baseline
  stereo,'' in \emph{2009 IEEE 12th International Conference on Computer
  Vision}, 2009, pp. 1709--1716.

\bibitem[Yu et~al.(2015)Yu, Russell, Campbell, and
  Agapito]{yuDirectDenseDeformable2015}
R.~Yu, C.~Russell, N.~D.~F. Campbell, and L.~Agapito, ``Direct, {{Dense}}, and
  {{Deformable}}: {{Template-Based Non-Rigid 3D Reconstruction From RGB
  Video}},'' in \emph{Proceedings of the {{IEEE International Conference}} on
  {{Computer Vision}}}, 2015, pp. 918--926.

\bibitem[Allen et~al.(2003)Allen, Curless, and
  Popovi{\'c}]{allenSpaceHumanBody2003}
B.~Allen, B.~Curless, and Z.~Popovi{\'c}, ``The space of human body shapes:
  Reconstruction and parameterization from range scans,'' \emph{ACM Trans.
  Graph.}, vol.~22, no.~3, pp. 587--594, 2003.

\bibitem[Suwajanakorn et~al.(2014{\natexlab{a}})Suwajanakorn,
  Kemelmacher-Shlizerman, and Seitz]{suwajanakorn2014total}
S.~Suwajanakorn, I.~Kemelmacher-Shlizerman, and S.~M. Seitz, ``Total moving
  face reconstruction,'' in \emph{Computer Vision--ECCV 2014: 13th European
  Conference, Zurich, Switzerland, September 6-12, 2014, Proceedings, Part IV
  13}.\hskip 1em plus 0.5em minus 0.4em\relax Springer, 2014, pp. 796--812.

\bibitem[Salzmann et~al.(2008)Salzmann, Urtasun, and Fua]{salzmann2008local}
M.~Salzmann, R.~Urtasun, and P.~Fua, ``Local deformation models for monocular
  3d shape recovery,'' in \emph{2008 IEEE conference on computer vision and
  pattern recognition}.\hskip 1em plus 0.5em minus 0.4em\relax IEEE, 2008, pp.
  1--8.

\bibitem[{\"O}stlund et~al.(2012){\"O}stlund, Varol, Ngo, and
  Fua]{ostlund2012laplacian}
J.~{\"O}stlund, A.~Varol, D.~T. Ngo, and P.~Fua, ``Laplacian meshes for
  monocular 3d shape recovery,'' in \emph{Computer Vision--ECCV 2012: 12th
  European Conference on Computer Vision, Florence, Italy, October 7-13, 2012,
  Proceedings, Part III 12}.\hskip 1em plus 0.5em minus 0.4em\relax Springer,
  2012, pp. 412--425.

\bibitem[Liu et~al.(2018)Liu, Peng, Zhou, Liu, and
  Gerndt]{liuTemplateBased3DReconstruction2018}
Y.~Liu, X.~Peng, W.~Zhou, B.~Liu, and A.~Gerndt, ``Template-{{Based 3D
  Reconstruction}} of {{Non-rigid Deformable Object}} from {{Monocular
  Video}},'' \emph{3D Research}, vol.~9, no.~2, p.~21, 2018.

\bibitem[Mustafa et~al.(2015)Mustafa, Kim, Guillemaut, and
  Hilton]{mustafaGeneralDynamicScene2015a}
A.~Mustafa, H.~Kim, J.-Y. Guillemaut, and A.~Hilton, ``General {{Dynamic Scene
  Reconstruction From Multiple View Video}},'' in \emph{Proceedings of the
  {{IEEE International Conference}} on {{Computer Vision}}}, 2015, pp.
  900--908.

\bibitem[Huang et~al.(2018)Huang, Li, Chen, Zhao, Xing, LeGendre, Luo, Ma, and
  Li]{huangDeepVolumetricVideo2018a}
Z.~Huang, T.~Li, W.~Chen, Y.~Zhao, J.~Xing, C.~LeGendre, L.~Luo, C.~Ma, and
  H.~Li, ``Deep {{Volumetric Video From Very Sparse Multi-View Performance
  Capture}},'' in \emph{Proceedings of the {{European Conference}} on
  {{Computer Vision}} ({{ECCV}})}, 2018, pp. 336--354.

\bibitem[Lombardi et~al.(2018)Lombardi, Saragih, Simon, and
  Sheikh]{lombardiDeepAppearanceModels2018a}
S.~Lombardi, J.~Saragih, T.~Simon, and Y.~Sheikh, ``Deep appearance models for
  face rendering,'' \emph{ACM Trans. Graph.}, vol.~37, no.~4, pp. 68:1--68:13,
  2018.

\bibitem[Lombardi et~al.(2019)Lombardi, Simon, Saragih, Schwartz, Lehrmann, and
  Sheikh]{lombardiNeuralVolumesLearning2019b}
S.~Lombardi, T.~Simon, J.~Saragih, G.~Schwartz, A.~Lehrmann, and Y.~Sheikh,
  ``Neural {{Volumes}}: {{Learning Dynamic Renderable Volumes}} from
  {{Images}},'' \emph{ACM Transactions on Graphics}, vol.~38, no.~4, pp. 1--14,
  2019.

\bibitem[Niemeyer et~al.(2019)Niemeyer, Mescheder, Oechsle, and
  Geiger]{niemeyerOccupancyFlow4D2019}
M.~Niemeyer, L.~Mescheder, M.~Oechsle, and A.~Geiger, ``Occupancy {{Flow}}:
  {{4D Reconstruction}} by {{Learning Particle Dynamics}},'' in
  \emph{Proceedings of the {{IEEE}}/{{CVF International Conference}} on
  {{Computer Vision}}}, 2019, pp. 5379--5389.

\bibitem[Xian et~al.(2021)Xian, Huang, Kopf, and Kim]{xian2021space}
W.~Xian, J.-B. Huang, J.~Kopf, and C.~Kim, ``Space-time neural irradiance
  fields for free-viewpoint video,'' in \emph{Proceedings of the IEEE/CVF
  conference on computer vision and pattern recognition}, 2021, pp. 9421--9431.

\bibitem[Wang et~al.(2022)Wang, Zhang, Liu, Zhao, Zhang, Zhang, Wu, Yu, and
  Xu]{wangFourierPlenOctreesDynamic2022}
L.~Wang, J.~Zhang, X.~Liu, F.~Zhao, Y.~Zhang, Y.~Zhang, M.~Wu, J.~Yu, and
  L.~Xu, ``Fourier {{PlenOctrees}} for {{Dynamic Radiance Field Rendering}} in
  {{Real-Time}},'' in \emph{Proceedings of the {{IEEE}}/{{CVF Conference}} on
  {{Computer Vision}} and {{Pattern Recognition}}}, 2022, pp. 13\,524--13\,534.

\bibitem[Li et~al.(2023{\natexlab{a}})Li, Wang, Cole, Tucker, and
  Snavely]{liDynIBaRNeuralDynamic2023}
Z.~Li, Q.~Wang, F.~Cole, R.~Tucker, and N.~Snavely, ``{{DynIBaR}}: {{Neural
  Dynamic Image-Based Rendering}},'' in \emph{Proceedings of the {{IEEE}}/{{CVF
  Conference}} on {{Computer Vision}} and {{Pattern Recognition}}}, 2023, pp.
  4273--4284.

\bibitem[Bae et~al.(2024)Bae, Kim, Yun, Lee, Bang, and
  Uh]{baeGaussianEmbeddingBasedDeformation2024}
J.~Bae, S.~Kim, Y.~Yun, H.~Lee, G.~Bang, and Y.~Uh, ``Per-gaussian
  embedding-based deformation for deformable 3d gaussian splatting,'' in
  \emph{European Conference on Computer Vision}.\hskip 1em plus 0.5em minus
  0.4em\relax Springer, 2024, pp. 321--335.

\bibitem[Lu et~al.(2024)Lu, Guo, Hui, Chen, Yang, Tang, Zhu, and
  Dai]{lu3DGeometryawareDeformable2024}
Z.~Lu, X.~Guo, L.~Hui, T.~Chen, M.~Yang, X.~Tang, F.~Zhu, and Y.~Dai, ``3d
  geometry-aware deformable gaussian splatting for dynamic view synthesis,'' in
  \emph{Proceedings of the IEEE/CVF Conference on Computer Vision and Pattern
  Recognition (CVPR)}, June 2024, pp. 8900--8910.

\bibitem[Suwajanakorn et~al.(2014{\natexlab{b}})Suwajanakorn,
  {Kemelmacher-Shlizerman}, and Seitz]{suwajanakornTotalMovingFace2014}
S.~Suwajanakorn, I.~{Kemelmacher-Shlizerman}, and S.~M. Seitz, ``Total {{Moving
  Face Reconstruction}},'' in \emph{Computer {{Vision}} -- {{ECCV}} 2014},
  D.~Fleet, T.~Pajdla, B.~Schiele, and T.~Tuytelaars, Eds.\hskip 1em plus 0.5em
  minus 0.4em\relax Cham: Springer International Publishing, 2014, pp.
  796--812.

\bibitem[Park et~al.(2021{\natexlab{d}})Park, Sinha, Barron, Bouaziz, Goldman,
  Seitz, and {Martin-Brualla}]{parkNerfiesDeformableNeural2021}
K.~Park, U.~Sinha, J.~T. Barron, S.~Bouaziz, D.~B. Goldman, S.~M. Seitz, and
  R.~{Martin-Brualla}, ``Nerfies: {{Deformable Neural Radiance Fields}},'' in
  \emph{Proceedings of the {{IEEE}}/{{CVF International Conference}} on
  {{Computer Vision}}}, 2021, pp. 5865--5874.

\bibitem[Song et~al.(2021)Song, Huang, Cao, and
  Mu]{songHDRNetFusionRealtime3D2021}
H.~Song, J.~Huang, Y.-P. Cao, and T.-J. Mu, ``{{HDR-Net-Fusion}}: {{Real-time
  3D}} dynamic scene reconstruction with a hierarchical deep reinforcement
  network,'' \emph{Computational Visual Media}, vol.~7, no.~4, pp. 419--435,
  2021.

\bibitem[Mescheder et~al.(2019)Mescheder, Oechsle, Niemeyer, Nowozin, and
  Geiger]{meschederOccupancyNetworksLearning2019}
L.~Mescheder, M.~Oechsle, M.~Niemeyer, S.~Nowozin, and A.~Geiger, ``Occupancy
  {{Networks}}: {{Learning 3D Reconstruction}} in {{Function Space}},'' in
  \emph{Proceedings of the {{IEEE}}/{{CVF Conference}} on {{Computer Vision}}
  and {{Pattern Recognition}}}, 2019, pp. 4460--4470.

\bibitem[Zhao et~al.(2021)Zhao, Xiong, Liu, Zhang, and
  Huang]{zhao2021spk2imgnet}
J.~Zhao, R.~Xiong, H.~Liu, J.~Zhang, and T.~Huang, ``Spk2imgnet: Learning to
  reconstruct dynamic scene from continuous spike stream,'' in
  \emph{Proceedings of the IEEE/CVF Conference on Computer Vision and Pattern
  Recognition}, 2021, pp. 11\,996--12\,005.

\bibitem[Ost et~al.(2021)Ost, Mannan, Thuerey, Knodt, and Heide]{Ost_2021_CVPR}
J.~Ost, F.~Mannan, N.~Thuerey, J.~Knodt, and F.~Heide, ``Neural scene graphs
  for dynamic scenes,'' in \emph{Proceedings of the IEEE/CVF Conference on
  Computer Vision and Pattern Recognition (CVPR)}, June 2021, pp. 2856--2865.

\bibitem[Delage et~al.(2006)Delage, Lee, and
  Ng]{delageDynamicBayesianNetwork2006}
E.~Delage, H.~Lee, and A.~Ng, ``A {{Dynamic Bayesian Network Model}} for
  {{Autonomous 3D Reconstruction}} from a {{Single Indoor Image}},'' in
  \emph{2006 {{IEEE Computer Society Conference}} on {{Computer Vision}} and
  {{Pattern Recognition}} ({{CVPR}}'06)}, vol.~2, 2006, pp. 2418--2428.

\bibitem[Zheng et~al.(2015)Zheng, Ji, Dunn, and
  Frahm]{zhengSparseDynamic3D2015}
E.~Zheng, D.~Ji, E.~Dunn, and J.-M. Frahm, ``Sparse {{Dynamic 3D Reconstruction
  From Unsynchronized Videos}},'' in \emph{Proceedings of the {{IEEE
  International Conference}} on {{Computer Vision}}}, 2015, pp. 4435--4443.

\bibitem[Novotny et~al.(2019)Novotny, Ravi, Graham, Neverova, and
  Vedaldi]{Novotny_2019_ICCV}
D.~Novotny, N.~Ravi, B.~Graham, N.~Neverova, and A.~Vedaldi, ``C3dpo: Canonical
  3d pose networks for non-rigid structure from motion,'' in \emph{Proceedings
  of the IEEE/CVF International Conference on Computer Vision (ICCV)}, October
  2019.

\bibitem[Kairanda et~al.(2022)Kairanda, Tretschk, Elgharib, Theobalt, and
  Golyanik]{kairandaFSfTShapeFromTemplatePhysicsBased2022}
N.~Kairanda, E.~Tretschk, M.~Elgharib, C.~Theobalt, and V.~Golyanik,
  ``F-{{SfT}}: {{Shape-From-Template With}} a {{Physics-Based Deformation
  Model}},'' in \emph{Proceedings of the {{IEEE}}/{{CVF Conference}} on
  {{Computer Vision}} and {{Pattern Recognition}}}, 2022, pp. 3948--3958.

\bibitem[Kim et~al.(2024)Kim, Bae, Yun, Lee, Bang, and
  Uh]{kimSyncNeRFGeneralizingDynamic2024}
S.~Kim, J.~Bae, Y.~Yun, H.~Lee, G.~Bang, and Y.~Uh, ``Sync-{{NeRF}}:
  {{Generalizing Dynamic NeRFs}} to {{Unsynchronized Videos}},''
  \emph{Proceedings of the AAAI Conference on Artificial Intelligence},
  vol.~38, no.~3, pp. 2777--2785, 2024.

\bibitem[Kajiya and Von~Herzen(1984)]{10.1145/964965.808594}
J.~T. Kajiya and B.~P. Von~Herzen, ``Ray tracing volume densities,'' vol.~18,
  no.~3, p. 165–174, Jan. 1984.

\bibitem[Max(1995)]{468400}
N.~Max, ``Optical models for direct volume rendering,'' \emph{IEEE Transactions
  on Visualization and Computer Graphics}, vol.~1, no.~2, pp. 99--108, 1995.

\bibitem[Rahaman et~al.(2019)Rahaman, Baratin, Arpit, Draxler, Lin, Hamprecht,
  Bengio, and Courville]{pmlr-v97-rahaman19a}
N.~Rahaman, A.~Baratin, D.~Arpit, F.~Draxler, M.~Lin, F.~Hamprecht, Y.~Bengio,
  and A.~Courville, ``On the spectral bias of neural networks,'' in
  \emph{Proceedings of the 36th International Conference on Machine Learning},
  ser. Proceedings of Machine Learning Research, K.~Chaudhuri and
  R.~Salakhutdinov, Eds., vol.~97.\hskip 1em plus 0.5em minus 0.4em\relax PMLR,
  09--15 Jun 2019, pp. 5301--5310.

\bibitem[Levoy(1990)]{10.1145/78964.78965}
M.~Levoy, ``Efficient ray tracing of volume data,'' \emph{ACM Trans. Graph.},
  vol.~9, no.~3, p. 245–261, Jul. 1990.

\bibitem[Zhan et~al.(2024)Zhan, Li, Niu, Zhong, Nobuhara, Nishino, and
  Zheng]{zhanKFDNeRFRethinkingDynamic2024}
Y.~Zhan, Z.~Li, M.~Niu, Z.~Zhong, S.~Nobuhara, K.~Nishino, and Y.~Zheng,
  ``Kfd-nerf: Rethinking dynamic nerf with kalman filter,'' in \emph{European
  Conference on Computer Vision}.\hskip 1em plus 0.5em minus 0.4em\relax
  Springer, 2024, pp. 1--18.

\bibitem[Chen et~al.(2018)Chen, Rubanova, Bettencourt, and
  Duvenaud]{chen2018neural}
R.~T. Chen, Y.~Rubanova, J.~Bettencourt, and D.~K. Duvenaud, ``Neural ordinary
  differential equations,'' \emph{Advances in neural information processing
  systems}, vol.~31, 2018.

\bibitem[Wang et~al.(2023{\natexlab{c}})Wang, Zhao, Ma, and Liu]{wang2023bad}
P.~Wang, L.~Zhao, R.~Ma, and P.~Liu, ``Bad-nerf: Bundle adjusted deblur neural
  radiance fields,'' in \emph{Proceedings of the IEEE/CVF Conference on
  Computer Vision and Pattern Recognition}, 2023, pp. 4170--4179.

\bibitem[Valmadre and Lucey(2012)]{valmadre2012general}
J.~Valmadre and S.~Lucey, ``General trajectory prior for non-rigid
  reconstruction,'' in \emph{2012 IEEE Conference on Computer Vision and
  Pattern Recognition}.\hskip 1em plus 0.5em minus 0.4em\relax IEEE, 2012, pp.
  1394--1401.

\bibitem[Teed and Deng(2020)]{teed2020raft}
Z.~Teed and J.~Deng, ``Raft: Recurrent all-pairs field transforms for optical
  flow,'' in \emph{Computer Vision--ECCV 2020: 16th European Conference,
  Glasgow, UK, August 23--28, 2020, Proceedings, Part II 16}.\hskip 1em plus
  0.5em minus 0.4em\relax Springer, 2020, pp. 402--419.

\bibitem[Ranftl et~al.(2020)Ranftl, Lasinger, Hafner, Schindler, and
  Koltun]{ranftl2020towards}
R.~Ranftl, K.~Lasinger, D.~Hafner, K.~Schindler, and V.~Koltun, ``Towards
  robust monocular depth estimation: Mixing datasets for zero-shot
  cross-dataset transfer,'' \emph{IEEE transactions on pattern analysis and
  machine intelligence}, vol.~44, no.~3, pp. 1623--1637, 2020.

\bibitem[Zollh{\"o}fer et~al.(2014)Zollh{\"o}fer, Nie{\ss}ner, Izadi, Rehmann,
  Zach, Fisher, Wu, Fitzgibbon, Loop, Theobalt, et~al.]{zollhofer2014real}
M.~Zollh{\"o}fer, M.~Nie{\ss}ner, S.~Izadi, C.~Rehmann, C.~Zach, M.~Fisher,
  C.~Wu, A.~Fitzgibbon, C.~Loop, C.~Theobalt \emph{et~al.}, ``Real-time
  non-rigid reconstruction using an rgb-d camera,'' \emph{ACM Transactions on
  Graphics (ToG)}, vol.~33, no.~4, pp. 1--12, 2014.

\bibitem[Osher and Sethian(1988)]{osher1988fronts}
S.~Osher and J.~A. Sethian, ``Fronts propagating with curvature-dependent
  speed: Algorithms based on hamilton-jacobi formulations,'' \emph{Journal of
  computational physics}, vol.~79, no.~1, pp. 12--49, 1988.

\bibitem[Li et~al.(2023{\natexlab{b}})Li, Shen, Wang, Shen, and
  Bo]{liCompressingVolumetricRadiance2023}
L.~Li, Z.~Shen, Z.~Wang, L.~Shen, and L.~Bo, ``Compressing {{Volumetric
  Radiance Fields}} to 1 {{MB}},'' in \emph{Proceedings of the {{IEEE}}/{{CVF
  Conference}} on {{Computer Vision}} and {{Pattern Recognition}}}, 2023, pp.
  4222--4231.

\bibitem[Wizadwongsa et~al.(2021)Wizadwongsa, Phongthawee, Yenphraphai, and
  Suwajanakorn]{Wizadwongsa_2021_CVPR}
S.~Wizadwongsa, P.~Phongthawee, J.~Yenphraphai, and S.~Suwajanakorn, ``Nex:
  Real-time view synthesis with neural basis expansion,'' in \emph{Proceedings
  of the IEEE/CVF Conference on Computer Vision and Pattern Recognition
  (CVPR)}, June 2021, pp. 8534--8543.

\bibitem[Chen et~al.(2022)Chen, Xu, Geiger, Yu, and Su]{chen2022tensorf}
A.~Chen, Z.~Xu, A.~Geiger, J.~Yu, and H.~Su, ``Tensorf: Tensorial radiance
  fields,'' in \emph{European conference on computer vision}.\hskip 1em plus
  0.5em minus 0.4em\relax Springer, 2022, pp. 333--350.

\bibitem[Selesnick et~al.(2005)Selesnick, Baraniuk, and
  Kingsbury]{selesnick2005dual}
I.~W. Selesnick, R.~G. Baraniuk, and N.~C. Kingsbury, ``The dual-tree complex
  wavelet transform,'' \emph{IEEE signal processing magazine}, vol.~22, no.~6,
  pp. 123--151, 2005.

\bibitem[M{\"u}ller et~al.(2022)M{\"u}ller, Evans, Schied, and
  Keller]{muller2022instant}
T.~M{\"u}ller, A.~Evans, C.~Schied, and A.~Keller, ``Instant neural graphics
  primitives with a multiresolution hash encoding,'' \emph{ACM transactions on
  graphics (TOG)}, vol.~41, no.~4, pp. 1--15, 2022.

\bibitem[Shechtman et~al.(2015)Shechtman, Eldar, Cohen, Chapman, Miao, and
  Segev]{shechtman2015phase}
Y.~Shechtman, Y.~C. Eldar, O.~Cohen, H.~N. Chapman, J.~Miao, and M.~Segev,
  ``Phase retrieval with application to optical imaging: a contemporary
  overview,'' \emph{IEEE signal processing magazine}, vol.~32, no.~3, pp.
  87--109, 2015.

\bibitem[Kendall and Gal(2017)]{kendall2017uncertainties}
A.~Kendall and Y.~Gal, ``What uncertainties do we need in bayesian deep
  learning for computer vision?'' \emph{Advances in neural information
  processing systems}, vol.~30, 2017.

\bibitem[Martin-Brualla et~al.(2021)Martin-Brualla, Radwan, Sajjadi, Barron,
  Dosovitskiy, and Duckworth]{martin2021nerf}
R.~Martin-Brualla, N.~Radwan, M.~S. Sajjadi, J.~T. Barron, A.~Dosovitskiy, and
  D.~Duckworth, ``Nerf in the wild: Neural radiance fields for unconstrained
  photo collections,'' in \emph{Proceedings of the IEEE/CVF conference on
  computer vision and pattern recognition}, 2021, pp. 7210--7219.

\bibitem[Yu et~al.(2021)Yu, Li, Tancik, Li, Ng, and
  Kanazawa]{yu2021plenoctrees}
A.~Yu, R.~Li, M.~Tancik, H.~Li, R.~Ng, and A.~Kanazawa, ``Plenoctrees for
  real-time rendering of neural radiance fields,'' in \emph{Proceedings of the
  IEEE/CVF International Conference on Computer Vision}, 2021, pp. 5752--5761.

\bibitem[Sorkine and Alexa(2007)]{10.5555/1281991.1282006}
O.~Sorkine and M.~Alexa, ``As-rigid-as-possible surface modeling,'' in
  \emph{Proceedings of the Fifth Eurographics Symposium on Geometry
  Processing}, ser. SGP '07.\hskip 1em plus 0.5em minus 0.4em\relax Goslar,
  DEU: Eurographics Association, 2007, p. 109–116.

\bibitem[Fridovich-Keil et~al.(2023)Fridovich-Keil, Meanti, Warburg, Recht, and
  Kanazawa]{fridovichkeil2023kplanesexplicitradiancefields}
S.~Fridovich-Keil, G.~Meanti, F.~R. Warburg, B.~Recht, and A.~Kanazawa,
  ``K-planes: Explicit radiance fields in space, time, and appearance,'' in
  \emph{Proceedings of the IEEE/CVF Conference on Computer Vision and Pattern
  Recognition (CVPR)}, June 2023, pp. 12\,479--12\,488.

\bibitem[Shin et~al.(2004)Shin, Park, Choi, Chung, and Rhee]{1290060}
H.~Shin, J.~Park, B.~Choi, Y.~Chung, and S.~Rhee, ``Efficient topology
  construction from triangle soup,'' in \emph{Geometric Modeling and
  Processing, 2004. Proceedings}, 2004, pp. 359--364.

\bibitem[Feichtenhofer et~al.(2019)Feichtenhofer, Fan, Malik, and
  He]{Feichtenhofer_2019_ICCV}
C.~Feichtenhofer, H.~Fan, J.~Malik, and K.~He, ``Slowfast networks for video
  recognition,'' in \emph{Proceedings of the IEEE/CVF International Conference
  on Computer Vision (ICCV)}, October 2019.

\bibitem[Zhao et~al.(2024)Zhao, Li, Sun, Zeng, Shen, Ma, Zhang, Bao, and
  Cui]{2405.19745}
B.~Zhao, Y.~Li, Z.~Sun, L.~Zeng, Y.~Shen, R.~Ma, Y.~Zhang, H.~Bao, and Z.~Cui,
  ``Gaussianprediction: Dynamic 3d gaussian prediction for motion extrapolation
  and free view synthesis,'' in \emph{ACM SIGGRAPH 2024 Conference Papers},
  ser. SIGGRAPH '24.\hskip 1em plus 0.5em minus 0.4em\relax New York, NY, USA:
  Association for Computing Machinery, 2024.

\bibitem[Cayley(1894)]{cayley1894collected}
A.~Cayley, \emph{The collected mathematical papers}.\hskip 1em plus 0.5em minus
  0.4em\relax University, 1894, vol.~7.

\bibitem[Bosch(2020)]{bosch2020n}
M.~T. Bosch, ``N-dimensional rigid body dynamics,'' \emph{ACM Transactions on
  Graphics (TOG)}, vol.~39, no.~4, pp. 55--1, 2020.

\bibitem[Wang et~al.(2023{\natexlab{d}})Wang, Han, Habermann, Daniilidis,
  Theobalt, and Liu]{wang2023neus2fastlearningneural}
Y.~Wang, Q.~Han, M.~Habermann, K.~Daniilidis, C.~Theobalt, and L.~Liu, ``Neus2:
  Fast learning of neural implicit surfaces for multi-view reconstruction,'' in
  \emph{Proceedings of the IEEE/CVF International Conference on Computer Vision
  (ICCV)}, October 2023, pp. 3295--3306.

\bibitem[Eisert et~al.(2023)Eisert, Schreer, Feldmann, Hellge, and
  Hilsmann]{EISERT2023289}
P.~Eisert, O.~Schreer, I.~Feldmann, C.~Hellge, and A.~Hilsmann, ``Chapter 11 -
  volumetric video – acquisition, interaction, streaming and rendering,'' in
  \emph{Immersive Video Technologies}, G.~Valenzise, M.~Alain, E.~Zerman, and
  C.~Ozcinar, Eds.\hskip 1em plus 0.5em minus 0.4em\relax Academic Press, 2023,
  pp. 289--326.

\bibitem[Liu et~al.(2023{\natexlab{a}})Liu, Cheng, Wu, and
  Han]{liuNextgenerationVolumetricVideo2023}
K.~Liu, R.~Cheng, N.~Wu, and B.~Han, ``Toward {{Next-generation Volumetric
  Video Streaming}} with {{Neural-based Content Representations}},'' in
  \emph{Proceedings of the 1st {{ACM Workshop}} on {{Mobile Immersive
  Computing}}, {{Networking}}, and {{Systems}}}, ser. {{ImmerCom}} '23.\hskip
  1em plus 0.5em minus 0.4em\relax New York, NY, USA: Association for Computing
  Machinery, 2023, pp. 199--207.

\bibitem[Wikipedia(2024{\natexlab{a}})]{wiki:Peak_signal-to-noise_ratio}
Wikipedia, ``{Peak signal-to-noise ratio} --- {W}ikipedia{,} the free
  encyclopedia,'' 2024, [Online; accessed 03-November-2024].

\bibitem[Wang et~al.(2003)Wang, Simoncelli, and Bovik]{wang2003multiscale}
Z.~Wang, E.~P. Simoncelli, and A.~C. Bovik, ``Multiscale structural similarity
  for image quality assessment,'' in \emph{The Thrity-Seventh Asilomar
  Conference on Signals, Systems \& Computers, 2003}, vol.~2.\hskip 1em plus
  0.5em minus 0.4em\relax Ieee, 2003, pp. 1398--1402.

\bibitem[Wikipedia(2024{\natexlab{b}})]{wiki:Structural_similarity_index_measure}
Wikipedia, ``{Structural similarity index measure} --- {W}ikipedia{,} the free
  encyclopedia,'' 2024, [Online; accessed 03-November-2024].

\bibitem[Zhang et~al.(2018)Zhang, Isola, Efros, Shechtman, and
  Wang]{zhang2018unreasonable}
R.~Zhang, P.~Isola, A.~A. Efros, E.~Shechtman, and O.~Wang, ``The unreasonable
  effectiveness of deep features as a perceptual metric,'' in \emph{Proceedings
  of the IEEE conference on computer vision and pattern recognition}, 2018, pp.
  586--595.

\bibitem[Huang et~al.(2024{\natexlab{b}})Huang, Wei, Zheng, An, Lu, Zhan,
  Tomizuka, Keutzer, and Zhang]{huang2024s3gaussian}
N.~Huang, X.~Wei, W.~Zheng, P.~An, M.~Lu, W.~Zhan, M.~Tomizuka, K.~Keutzer, and
  S.~Zhang, ``S3gaussian: Self-supervised street gaussians for autonomous
  driving,'' \emph{arXiv preprint arXiv:2405.20323}, 2024.

\bibitem[Liu et~al.(2023{\natexlab{b}})Liu, Chen, Yang, Wang, Manivasagam, and
  Urtasun]{Liu_2023_ICCV}
J.~Y. Liu, Y.~Chen, Z.~Yang, J.~Wang, S.~Manivasagam, and R.~Urtasun,
  ``Real-time neural rasterization for large scenes,'' in \emph{Proceedings of
  the IEEE/CVF International Conference on Computer Vision (ICCV)}, October
  2023, pp. 8416--8427.

\bibitem[Zhou et~al.(2024)Zhou, Lin, Shan, Wang, Sun, and
  Yang]{zhouDrivingGaussianCompositeGaussian2024}
X.~Zhou, Z.~Lin, X.~Shan, Y.~Wang, D.~Sun, and M.-H. Yang,
  ``{{DrivingGaussian}}: {{Composite Gaussian Splatting}} for {{Surrounding
  Dynamic Autonomous Driving Scenes}},'' in \emph{Proceedings of the
  {{IEEE}}/{{CVF Conference}} on {{Computer Vision}} and {{Pattern
  Recognition}}}, 2024, pp. 21\,634--21\,643.

\bibitem[Turki et~al.(2023)Turki, Zhang, Ferroni, and Ramanan]{Turki_2023_CVPR}
H.~Turki, J.~Y. Zhang, F.~Ferroni, and D.~Ramanan, ``Suds: Scalable urban
  dynamic scenes,'' in \emph{Proceedings of the IEEE/CVF Conference on Computer
  Vision and Pattern Recognition (CVPR)}, June 2023, pp. 12\,375--12\,385.

\bibitem[Yan et~al.(2024)Yan, Lin, Zhou, Wang, Sun, Zhan, Lang, Zhou, and
  Peng]{yanStreetGaussiansModeling2024}
Y.~Yan, H.~Lin, C.~Zhou, W.~Wang, H.~Sun, K.~Zhan, X.~Lang, X.~Zhou, and
  S.~Peng, ``Street gaussians: Modeling dynamic urban scenes with gaussian
  splatting,'' in \emph{European Conference on Computer Vision}.\hskip 1em plus
  0.5em minus 0.4em\relax Springer, 2024, pp. 156--173.

\bibitem[Li et~al.(2024{\natexlab{b}})Li, Li, Zhang, Wu, Shi, Zhao, Feng, Ding,
  Wang, and Han]{liVDGVisionOnlyDynamic2024}
H.~Li, J.~Li, D.~Zhang, C.~Wu, J.~Shi, C.~Zhao, H.~Feng, E.~Ding, J.~Wang, and
  J.~Han, ``{{VDG}}: {{Vision-Only Dynamic Gaussian}} for {{Driving
  Simulation}},'' 2024.

\bibitem[Khan et~al.(2024)Khan, Fazlali, Sharma, Cao, Bai, Ren, and
  Liu]{khanAutoSplatConstrainedGaussian2024}
M.~Khan, H.~Fazlali, D.~Sharma, T.~Cao, D.~Bai, Y.~Ren, and B.~Liu,
  ``{{AutoSplat}}: {{Constrained Gaussian Splatting}} for {{Autonomous Driving
  Scene Reconstruction}},'' 2024.

\bibitem[Park and Kim(2024)]{park2024point}
B.~Park and C.~Kim, ``Point-dynrf: Point-based dynamic radiance fields from a
  monocular video,'' in \emph{Proceedings of the IEEE/CVF Winter Conference on
  Applications of Computer Vision}, 2024, pp. 3171--3181.

\bibitem[Li et~al.(2024{\natexlab{c}})Li, Yin, Liu, Yang, Chen, Jiang, Yu, and
  Fan]{10747249}
W.~Li, F.~Yin, W.~Liu, Y.~Yang, X.~Chen, B.~Jiang, G.~Yu, and J.~Fan,
  ``Unbounded-gs: Extending 3d gaussian splatting with hybrid representation
  for unbounded large-scale scene reconstruction,'' \emph{IEEE Robotics and
  Automation Letters}, vol.~9, no.~12, pp. 11\,529--11\,536, 2024.

\bibitem[Schonberger and Frahm(2016)]{schonberger2016structure}
J.~L. Schonberger and J.-M. Frahm, ``Structure-from-motion revisited,'' in
  \emph{Proceedings of the IEEE conference on computer vision and pattern
  recognition}, 2016, pp. 4104--4113.

\bibitem[Pan et~al.(2025)Pan, Bar{\'a}th, Pollefeys, and
  Sch{\"o}nberger]{pan2025global}
L.~Pan, D.~Bar{\'a}th, M.~Pollefeys, and J.~L. Sch{\"o}nberger, ``Global
  structure-from-motion revisited,'' in \emph{European Conference on Computer
  Vision}.\hskip 1em plus 0.5em minus 0.4em\relax Springer, 2025, pp. 58--77.

\bibitem[Luiten et~al.(2024{\natexlab{b}})Luiten, Kopanas, Leibe, and
  Ramanan]{luiten2024dynamic}
J.~Luiten, G.~Kopanas, B.~Leibe, and D.~Ramanan, ``Dynamic 3d gaussians:
  Tracking by persistent dynamic view synthesis,'' in \emph{2024 International
  Conference on 3D Vision (3DV)}.\hskip 1em plus 0.5em minus 0.4em\relax IEEE,
  2024, pp. 800--809.

\bibitem[Yan et~al.(2023{\natexlab{c}})Yan, Li, and Lee]{yan2023od}
Z.~Yan, C.~Li, and G.~H. Lee, ``Od-nerf: Efficient training of on-the-fly
  dynamic neural radiance fields,'' \emph{arXiv preprint arXiv:2305.14831},
  2023.

\bibitem[Liu et~al.(2024{\natexlab{a}})Liu, Hu, Zhang, Shao, Lin, and
  Zhang]{liu2024dynamics}
Z.~Liu, Y.~Hu, X.~Zhang, J.~Shao, Z.~Lin, and J.~Zhang, ``Dynamics-aware
  gaussian splatting streaming towards fast on-the-fly training for 4d
  reconstruction,'' \emph{arXiv preprint arXiv:2411.14847}, 2024.

\bibitem[Li et~al.(2023{\natexlab{c}})Li, Wang, Cole, Tucker, and
  Snavely]{li2023dynibar}
Z.~Li, Q.~Wang, F.~Cole, R.~Tucker, and N.~Snavely, ``Dynibar: Neural dynamic
  image-based rendering,'' in \emph{Proceedings of the IEEE/CVF Conference on
  Computer Vision and Pattern Recognition}, 2023, pp. 4273--4284.

\bibitem[Mihajlovic et~al.(2025)Mihajlovic, Prokudin, Tang, Maier, Bogo, Tung,
  and Boyer]{mihajlovicSplatFieldsNeuralGaussian2025}
M.~Mihajlovic, S.~Prokudin, S.~Tang, R.~Maier, F.~Bogo, T.~Tung, and E.~Boyer,
  ``{{SplatFields}}: {{Neural Gaussian Splats}} for~{{Sparse 3D}} and~{{4D
  Reconstruction}},'' in \emph{Computer {{Vision}} -- {{ECCV}} 2024},
  A.~Leonardis, E.~Ricci, S.~Roth, O.~Russakovsky, T.~Sattler, and G.~Varol,
  Eds.\hskip 1em plus 0.5em minus 0.4em\relax Cham: Springer Nature
  Switzerland, 2025, pp. 313--332.

\bibitem[Liu et~al.(2024{\natexlab{b}})Liu, Liu, Li, Li, and Yuan]{liu2024lgs}
H.~Liu, Y.~Liu, C.~Li, W.~Li, and Y.~Yuan, ``Lgs: A light-weight 4d gaussian
  splatting for efficient surgical scene reconstruction,'' in
  \emph{International Conference on Medical Image Computing and
  Computer-Assisted Intervention}.\hskip 1em plus 0.5em minus 0.4em\relax
  Springer, 2024, pp. 660--670.

\bibitem[Fischer et~al.(2024)Fischer, Porzi, Bulo, Pollefeys, and
  Kontschieder]{fischer2024multi}
T.~Fischer, L.~Porzi, S.~R. Bulo, M.~Pollefeys, and P.~Kontschieder,
  ``Multi-level neural scene graphs for dynamic urban environments,'' in
  \emph{Proceedings of the IEEE/CVF Conference on Computer Vision and Pattern
  Recognition}, 2024, pp. 21\,125--21\,135.

\bibitem[Zhang et~al.(2025)Zhang, Chen, and Cui]{zhang2025efficient}
Y.~Zhang, G.~Chen, and S.~Cui, ``Efficient large-scale scene representation
  with a hybrid of high-resolution grid and plane features,'' \emph{Pattern
  Recognition}, vol. 158, p. 111001, 2025.

\bibitem[Guo et~al.(2024)Guo, Wang, He, and Matusik]{guo2024tetsphere}
M.~Guo, B.~Wang, K.~He, and W.~Matusik, ``Tetsphere splatting: Representing
  high-quality geometry with lagrangian volumetric meshes,'' \emph{arXiv
  preprint arXiv:2405.20283}, 2024.

\end{thebibliography}

\end{document}